\documentclass[reprint,amsmath,amssymb,floatfix]{revtex4-2}

\usepackage{geometry}
\usepackage{pstricks-add}
\usepackage{amsmath,amssymb,bbm,bm}
\usepackage{wrapfig}
\usepackage{graphicx}
\usepackage{epstopdf}
\usepackage{color}
\usepackage{hyperref,textcomp}
\usepackage{slashed}

\newcommand{\parea}[1]{\left[#1\right]}

\begin{document}

\title{Probing $H_0$ and resolving AGN disks with ultrafast photon counters}

\author{Neal Dalal}
\email{ndalal@pitp.ca}
\affiliation{Perimeter Institute for Theoretical Physics, 31 Caroline Street N., Waterloo, Ontario, N2L 2Y5, Canada}
\author{Marios Galanis}
\email{mgalanis@pitp.ca}
\affiliation{Perimeter Institute for Theoretical Physics, 31 Caroline Street N., Waterloo, Ontario, N2L 2Y5, Canada}
\author{Charles Gammie}
\affiliation{Illinois Center for Advanced Studies of the Universe, Department of Physics,
University of Illinois at Urbana-Champaign, Champaign, Illinois, USA}
\author{Samuel E.\ Gralla}
\affiliation{Department of Physics, University of Arizona, Tucson, AZ 85721, USA}
\author{Norman Murray}
\affiliation{Canadian Institute for Theoretical Astrophysics, University of Toronto, 60 St.\ George Street, Toronto, ON M5S 3H8, Canada}
\date{\today}

\begin{abstract}
Intensity interferometry is a technique developed many decades ago, that has recently enjoyed a renaissance thanks in part to advances in photodetector technology.  We investigate the potential for long-baseline optical intensity interferometry to observe bright, active galactic nuclei (AGN) associated with rapidly accreting supermassive black holes.  We argue that realistic telescope arrays similar in area to existing Cherenkov arrays, if equipped with modern high-precision single photon detectors, can achieve a sufficiently high signal to noise ratio not only to detect distant AGN, but also to study them in great detail.  We explore the science potential of such observations by considering two examples. First, we find that intensity interferometric observations of bright nearby AGN can allow detailed studies of the central accretion disks powering the AGN, allowing reconstruction of many disk properties like the radial profile.  Next, we argue that intensity interferometers can spatially resolve the broad-line regions of AGN at cosmological distances, and thereby provide a geometric determination of the angular diameter distances to those AGN when combined with reverberation mapping.  Since this measurement can be performed for AGN at distances of hundreds of megaparsecs, this directly measures the Hubble expansion rate $H_0$, with a precision adequate to resolve the recent Hubble tension.  Finally, we speculate on future applications that would be enabled by even larger intensity interferometer arrays.

\end{abstract}

\maketitle

\section{\label{sec:intro}Introduction}

Long-baseline amplitude interferometry is now a well-established method to achieve high angular resolution using modest apertures \cite{BornWolf,ThompsonBook,Lawson2000,Labeyrie.book}. Radio interferometry generally entails recording the received electric field at a high frequency using heterodyne techniques, and then subsequently correlating the recorded electric fields measured at widely separated telescopes.  Spectacular examples of radio interferometry include CHIME observations of fast radio bursts \cite{CHIMEFRB}, Event Horizon Telescope observations of supermassive black holes (SMBH) \cite{EHT}, and ALMA observations of protoplanetary disks \cite{ALMA} and gravitational lenses \cite{Hezaveh2016}.  In contrast, optical and near-infrared interferometry typically involves physically combining the light observed at different telescopes.  Colloquially, radio interferometry may be performed in software, but optical interferometry must be performed using hardware that combines the light from different telescopes.  Because of practical limitations in the distances over which optical light may be transmitted 
while maintaining phase coherence, optical interferometers typically have baselines $< 500$ m, far smaller than the thousand-km baselines used in Very Long Baseline Interferometry (VLBI) radio observations.

An alternative form of optical interferometry that can be performed in software is called intensity interferometry, pioneered in seminal works by \citet{HanburyBrown1954}
(see also \cite{HanburyBrown1956,Purcell1956,1957RSPSA.242..300B,Twiss1957, 1958RSPSA.243..291B,1958RSPSA.248..199B,1958RSPSA.248..222B}).  
This method relies on the super-poissonian statistics of light emitted by any broadband source with nonzero bandwidth $\Delta\nu$. In contrast to a laser which has only shot noise fluctuations whose fractional amplitude decreases with the square-root of the intensity, thermal light has $\mathcal{O}(1)$ intensity fluctuations even in the limit of infinite photons. This is colloquially described as ``bunching'' of photons, so that a large fluctuation from the mean will result in a correlated increase of recorded photons in two photodetectors. These macroscopic fluctuations are random and occur over timescales of order $t\sim\Delta\nu^{-1}$.  This bunching effect allows us to perform interferometry without 
correlating the electric field, but instead by correlating photon counts recorded at different telescopes.  Since this method removes the requirement to combine light from different telescopes, arbitrarily long baselines may in principle be used, similar to radio interferometry.  Early applications of intensity interferometry included the determination of stellar diameters \cite{HanburyBrown1974,Twiss1969}, and over ensuing decades this method has found application across a variety of fields \cite{Baym1998,Loudon2000}. 

Intensity interferometers and more familiar amplitude interferometers are both sensitive to the Fourier transform of the sky brightness, but one significant advantage of amplitude interferometry is that it measures both the amplitude and the phase of the complex visibility (defined below).  In contrast, intensity interferometers can only determine the amplitude of the visibility, i.e.\ they are insensitive to the phase of the visibility at each baseline (although many phase retrieval algorithms exist, e.g.\ \cite{Fienup1982,Dong2023}).  Since phase information is required for imaging, amplitude interferometers have understandably become more commonly used than intensity interferometers.  Another challenge faced by intensity interferometers is reaching a significant signal to noise ratio (SNR) in the photon correlations.  As we shall see below, the SNR in intensity interferometry is a steep function of the number of photons collected, which is why previous attempts focused on extremely bright sources like nearby stars.  With larger collecting areas and more efficient photodetectors, it becomes feasible to study fainter sources, including extragalactic sources.  Another requirement for high SNR is to measure the arrival times of single photons with a precision of order the coherence time of intensity fluctuations, $t \sim \Delta\nu^{-1}$.  Achieving $\Delta t\,\Delta\nu \sim 1$ typically requires both high resolution in timing (small $\Delta t$), and also narrow bandwidths (small $\Delta\nu$).  In their seminal measurement of stellar radii at the Narrabri Observatory \cite{1967MNRAS.137..375H}, Hanbury Brown and Twiss used Photomultiplier Tubes (PMTs) with $\sim 10$ns timing resolution,
which allowed for a 60 MHz bandwidth. 
More recently, the intensity interferometry module of the VERITAS telescopes \cite{kieda2021veritasstellar}, the H.E.S.S. telescopes \cite{Zmija2023}, and the MAGIC telescopes \cite{MAGIC2024} also employed PMTs. 

However, photodetection has evolved tremendously over the past two decades, with new technologies of exquisite precision even becoming commercially available. The technologies with the best time resolution today are Single Photon Avalanche Diodes (SPADs) and Superconducting Nanowire Single Photon Detectors (SNSPDs). Their detection technologies are very different: SNSPDs rely on an electric signal from a local breaking of superconductivity by an impinging photon~\cite{SteinhauerReview,Hadfield2020}, whereas SPADs are diodes with a large reverse-bias current, which produces a sustaining avalanche when triggered by as little as a single photon. Both technologies have demonstrated time-jitters orders of magnitude shorter than PMTs. In particular, single SNSPDs have been demonstrated from the IR to the UV, with time-jitters of 14ps at 1550nm~\cite{SpiropuluShaw2021}, 2.6ps in the optical~\cite{ShawBerggren2020} and 62ps in the UV~\cite{WollmanUV}. Similarly, SPADs have been shown to have time-jitters of $28$ps at 820nm~\cite{Zappa2018} and even $8.7$ps have recently been reported~\cite{becker2022}. 
Note that the quoted time jitter typically refers to the FWHM of the error distribution.  Although the jitter distributions are typically non-Gaussian, for the purposes of the forecasts in this paper, we will approximate them as Gaussian distributions with $\sigma \approx {\rm FWHM}/2.35$.  

Fast photo-detectors are now also beginning to become available in arrays. 
Recently SNSPDs have been demonstrated in arrays with 400,000 pixels and timing jitter below 3ps \cite{Oripov2023}.
Similarly, multiplexing of SPADs is also an established technology, with demonstrated arrays as large as 256$\times$256 pixels, with a 300ps time jitter and  supported data rates on the order of Gbit/s, being considered for use in telecommunications~\cite{LargeFormatAull}. The exquisite time-jitter of SPADs has not escaped astronomers, with the LIDAR landing system of the Europa astrobiology mission using a SPAD array. A 2048$\times$32 array was shown to have a time-jitter of 250ps at 532nm~\cite{EuropaSlides,EuropaLidar}, whereas a smaller array, 1024$\times$32 pixels, achieved 32ps~\cite{SOIspad}.
SPADs have recently been used for intensity interferometric measurements on small arrays, that already have comparable sensitivities to VERITAS~\cite{horch2022observations,Guerin_space}.

The spectacular improvement in fast photodetectors is poised to revolutionize intensity interferometry, enabling an array of new applications going beyond previous studies of individual nearby stars.  
In particular, we argue that it is now feasible to study extragalactic sources using intensity interferometers, including bright active galactic nuclei (AGN). Long-baseline interferometers can spatially resolve AGN emission, a potentially transformative development in AGN science that also provides applications for cosmology and for understanding gravity in the strong-field regime, as we describe below. 

\section{Review of interferometry} \label{sec:review}

In this section, we summarize basic aspects of amplitude interferometry and intensity interferometry \cite{ThompsonBook,Lawson2000,Labeyrie.book,HanburyBrown1956}, and define notation that will be used later.  

Consider a source with specific intensity $I_\nu(\nu, \hat n)$, and specific flux $F_\nu = \int I_\nu\, d^2\hat n$, observed at central frequency $\nu_0$ and bandwidth $\Delta\nu \ll \nu_0$ by a telescope with area $A$ \cite{RybickiLightman}.  Photons from this source are detected at a mean rate $\Gamma = A\,F_\nu(\nu_0) \Delta\nu/(h\nu_0)$.  In time interval $\delta t$, the mean number of photons detected in this band is $\bar N = \Gamma\, \delta t$, and the variance in the number is $\langle \delta N^2\rangle = \bar N$.  

Two different telescopes, each of area $A$, separated by baseline $\bm{B}$, will observe correlations in the light that each receives from this source.  Conventionally, these correlations are expressed in terms of a quantity called the visibility, defined as
\begin{equation} \label{eq:visibility}
V(\nu, \bm{B}, \Delta t) = 
\int I_\nu(\nu,\hat n) 
\exp\left(i [\bm{k} \cdot \bm{B} - \omega\Delta t] \right) d^2 \hat n 
\end{equation}
where $\omega=2\pi\nu$, $\bm{k} = \omega \hat n / c$, and $\Delta t$ is the time lag in the cross-correlation. 
For example, the famous van Cittert-Zernike theorem relates the visibility to correlations of the electric fields measured at two telescopes observing a spatially incoherent source, $\langle E^*(\bm{x},t)E(\bm{x}+\bm{B},t+\Delta t)\rangle \propto V(\bm{B},\Delta t)$.
If we observe a finite bandwidth $\Delta\nu$, then we integrate Eqn.\ \eqref{eq:visibility} over frequencies.  We can write the frequency range as $|\nu-\nu_0| < \Delta\nu/2$, for central frequency $\nu_0$ and bandwidth $\Delta\nu$.  
Let us define the normalized fringe visibility as 
\begin{equation}\label{eq:normvis}
{\cal V}(\nu_0, \Delta\nu, \bm{B}, \Delta t) = \frac
{\int_{\nu_0-\Delta\nu/2}^{\nu_0+\Delta\nu/2} d\nu\, V(\nu, \bm{B}, \Delta t)}
{\int_{\nu_0-\Delta\nu/2}^{\nu_0+\Delta\nu/2}d\nu \int d^2 \hat n \, I_\nu(\nu,\hat n)}.
\end{equation}
This normalized visibility ${\cal V}$ sets the fringe contrast measured by optical amplitude interferometers \cite{Lawson2000}. 

The observable for an intensity interferometer is also closely related to ${\cal V}$.  Intensity interferometry correlates the photon counts $N(t)$ detected at different telescopes.  For a pair of telescopes indexed by subscripts $i$ and $j$, each telescope's photon detection rate fluctuates over time, and since the intensity $I\sim |E|^2$, then we immediately see $\langle I(\bm{x}) I(\bm{x}+\bm{B})\rangle \sim |V(\bm{B})|^2$.  More precisely,
the covariance of the detected counts is given by \cite{Loudon2000}
\begin{eqnarray}
\langle\delta N_i \, \delta N_j\rangle  &=& \frac{\Gamma^2}2 
\int dt_i dt_j\, |{\cal V}(\nu_0, \Delta\nu, \bm{B}, t_i-t_j)|^2 
\nonumber \\
&=& \frac{\Gamma^2 T}2 \int d\Delta t\,
|{\cal V}(\nu_0, \Delta\nu, \bm{B}, \Delta t)|^2,
\label{eq:countscovariance}
\end{eqnarray}
where $\Delta t = t_i - t_j$ and the $t_i$ and $t_j$ integrals run over time interval $T$, assumed to be a small fraction of the total observing time $T_{\rm obs}$. Note that the factor of $1/2$ arises because the two independent polarizations of photons do not correlate with each other.
Assuming that the noise is dominated by Poisson shot noise (i.e., signal covariance is negligible), then the variance in the cross-correlation between counts at telescopes $i$ and $j$ is
$\langle\delta N_i^2 \, \delta N_j^2\rangle - 
\langle\delta N_i \, \delta N_j\rangle^2 = (\Gamma \int dt)^2$, giving a simple expression for the signal to noise ratio of the detected photon correlations.

Photodetector timing jitter adds noise to the measured arrival times of the photons, which smears out the cross-correlation given in Eqn.\ \eqref{eq:countscovariance}.  If the probability distribution for each detector's timing jitter $\delta t$ is given by $W(\delta t)$, then we simply convolve $|{\cal V}|^2$ with $W$ in both the $t_i$ and $t_j$ directions before integrating, 
\begin{equation}
\langle\delta N_i \, \delta N_j\rangle =
\frac{\Gamma^2}2 
\int dt_i dt_j\, {\cal C}(\nu_0, \Delta\nu, \bm{B}, t_i-t_j),
\end{equation}
where
\begin{equation} \label{eq:convolved_vis}
{\cal C}(t_i-t_j) = \int d\tau_i d\tau_j
|{\cal V}(t_i-t_j + \tau_j - \tau_i)|^2\,
W(\tau_i)\,W(\tau_j).
\end{equation}
If $W$ is a Gaussian with rms dispersion given by $\sigma_t$, then  convolution with two jitters is equivalent to a single convolution with a Gaussian in $\Delta t$, with rms dispersion of $\sqrt{2} \sigma_t$.
Since the Poisson shot noise covariance is unchanged by the timing jitter, the signal to noise ratio in the detected intensity cross-correlations for total observing time $T_{\rm obs}$ becomes
\begin{equation}\label{eq:corrsnr}
{\rm SNR}^2 = \frac{\Gamma^2 T_{\rm obs}}{4} 
\int d\Delta t \,{\cal C}^2(\nu_0, \Delta\nu, \bm{B}, \Delta t)
\end{equation}

As a concrete example, let us consider 
a broadband source with a smooth emission spectrum, and suppose that the observed bandwidth is sufficiently narrow ($\Delta\nu \ll \nu_0$) that we can treat the spectrum as being almost constant, $I_\nu(\nu, \hat n) \approx I_\nu(\nu_0, \hat n)$ across the observed bandwidth.  Let us also suppose that the source is centred at location $\hat n_0$ on the sky, and then measure sky locations $\hat n$ relative to this reference location as $\hat n = \hat n_0 + \delta\hat n$, so that $\delta \hat n\cdot \hat n_0 \approx 0$ for $|{\delta \hat n}|\ll 1$.  We can decompose the baseline separation vector $\bm{B}$ into components parallel and perpendicular to $\hat n_0$, writing $\bm{B} = \bm{B}_\perp + (\bm{B}\cdot\hat n_0) \hat n_0$.  Then $\bm{k}\cdot\bm{B} = \bm{k}\cdot\bm{B}_\perp + \omega \bm{B}\cdot\hat n_0 /c$, and so $\bm{k}\cdot\bm{B} - \omega \Delta t = \bm{k}\cdot\bm{B}_\perp - \omega (\Delta t - \bm{B}\cdot\hat n_0 /c)$. Let us redefine $\Delta t$ to absorb the last term, i.e.\ $\Delta t \to \Delta t- \bm{B}\cdot\hat n_0 /c$.
In this limit, the time dependence of $\cal V$ in Eqn.\ \eqref{eq:normvis} factorizes out, as 
\begin{equation}
\label{eq:normvis_approx}
{\cal V} \approx {\rm sinc}\left(\frac{\Delta\omega\,\Delta t}2\right) e^{-i\omega_0 \Delta t} \,
{\cal V}(\nu_0, \bm{B}) 
\end{equation}
where  
\begin{eqnarray} \label{eq:vis}
{\cal V}(\nu_0, \bm{B}) &=& \frac{V(\nu_0, \bm{B})}{I_{\rm tot}} \\
&=& \frac{\int d^2\hat n\, I(\nu_0,\hat n) e^{2\pi i\bm{B}_\perp\cdot\hat n/\lambda_0}}
{\int d^2 \hat n\, I(\nu_0,\hat n)} \nonumber,
\end{eqnarray}
and $\lambda_0 = c/\nu_0$.  Note that in the aperture synthesis literature, $\bm{B}_\perp/\lambda_0$ is conventionally written as the two-component vector $(u,v)$, but we will not use that notation in the remainder of the paper. Similarly, for simplicity in the remainder of the paper, we will drop the $\perp$ subscript when discussing baseline lengths, so each time we write $B$, it is understood to be the magnitude of the perpendicular component, $|\bm{B}_\perp|$. 

For Gaussian time jitter $\sigma_t$ at each detector, we have
\begin{equation} \label{eq:FintermsofV}
{\cal C}(\nu_0, \Delta\nu, \bm{B}, \Delta t) = 
|{\cal V}(\nu_0, \bm{B})|^2 
f(\Delta\nu,\sigma_t, \Delta t),
\end{equation}
where
\begin{equation}
f = \int d\tau \,
{\rm sinc}^2\left(\frac{\Delta\omega\,(\Delta t-\tau)}2\right)
\frac{e^{-\tau^2/4\sigma_t^2}}{\sqrt{4\pi}\sigma_t}.
\end{equation}
In the limit $\sigma_t\Delta\omega\gg 1$, the Gaussian is much wider than the sinc function, giving
\begin{equation}
f\approx  \frac{e^{-\Delta t^2/4\sigma_t^2}}{
\sqrt{4\pi}\sigma_t \Delta\nu} .   
\end{equation}
Then Eqn.\ \eqref{eq:corrsnr} becomes
\begin{equation}
{\rm SNR}^2 = 
\frac{|{\cal V}(\nu_0, \bm{B})|^4}{\sqrt{128\pi}}
\left(\frac{\Gamma}{\Delta\nu} \right)^2
\frac{T_{\rm obs}}{\sigma_t}.
\end{equation}
In terms of the familiar visibility ${\cal V}(\nu_0, \bm{B})$, we can therefore write the signal to noise ratio of the intensity correlations as
\begin{equation}\label{eqn:snrvis}
{\rm SNR}= \frac{|{\cal V}|^2}{\sigma_{|{\cal V}|^2}}
\end{equation}
where
\begin{equation}\label{eqn:sigV}
\sigma^{-1}_{|{\cal V}|^2} = \frac{d\Gamma}{d\nu} 
\left(\frac{T_{\rm obs}}{\sigma_t}\right)^{1/2} (128\pi)^{-1/4}.
\end{equation}
Recall that $d\Gamma/d\nu = A\,F_v/(h \nu_0)$.

The quantity $\sigma_{|{\cal V}|^2}$ is a convenient way to characterize the noise in an intensity interferometry observation, in the limit when the time dependence of ${\cal V}$ factorizes as in Eqn.\ \eqref{eq:normvis_approx}.  
This expression encapsulates the various ways in which we can increase the SNR of intensity interferometry observations.  
For example, $\sigma_{|{\cal V}|^2}$ depends on the ratio $d\Gamma/d\nu$, the rate at which photons are detected, per unit frequency.  One way to increase  $d\Gamma/d\nu$ is to observe bright emission lines, in which many photons are emitted over a narrow frequency range.  In section \ref{sec:BLR}, we will discuss this case in some detail.  Another way to increase $d\Gamma/d\nu$ is simply to increase the rate at which photons are detected, i.e.\ we can achieve higher SNR by observing brighter sources using telescopes with larger collecting area.  We cannot observe arbitrarily bright sources, however, since single-photon detectors can have significant dead time following the detection of photons.  For SPADs, this dead time can be of order 5 nanoseconds \cite{deadtime}, meaning that for sources with photons arriving at rates $\Gamma \gtrsim 10^8\, {\rm s}^{-1}$, the detector can essentially saturate.  For extragalactic sources, this is usually not a concern: the examples given below typically have $\Gamma \sim 10^6\, {\rm s}^{-1}$, well below the saturation limit.

Besides collecting more photons, another way to increase the SNR is to use a spectroscopic element. The advantage of spectroscopy was already understood in the earliest days of intensity interferometry by Hanbury Brown ~\cite{HanburyBrown1974}, and, more recently, there have been proposals along this direction~\cite{2021sf2a.conf..335L,chen2023astrometry,Stankus_2022,trippe2014optical,vantilburg2023,galanis2023extendedpath}. To understand the enhancement, notice that if we subdivide the observed frequency band into $n_c$ smaller channels,  both $\Gamma$ and $\Delta\nu$ decrease by the approximately the same factor of $n_c$, leaving $d\Gamma/d\nu$ and hence $\sigma_{|{\cal V}|^2}$ essentially unchanged, for every one of the smaller channels. As long as the channels are independent, this increases the overall SNR by $n_c^{1/2}$.  This argument assumes that the visibility $\cal V$ is nearly independent of frequency $\nu$, which is reasonable for many sources, such as blackbodies whose temperature is nearly uniform across the surface of the source.  
This assumption can break down for emission lines, as we discuss in section \ref{sec:BLR}, but even in that case the basic result holds that the overall SNR increases significantly if we can divide the signal into many frequency channels.

The SNR in detecting photon correlations grows with decreasing $\Delta \nu$ until saturating at  $\Delta\nu\ \sim \Delta t ^{-1}$. Since modern detectors can achieve timing resolution of picoseconds, we therefore would like to achieve bandwidths $\Delta\nu \sim 10^{12}\,$Hz.  Since visible light has frequencies $\nu \sim 10^{15}\,$Hz, in order to avoid a significant loss in SNR we therefore require $\Delta\nu \ll \nu$, i.e.\ extremely narrow bandwidths.  This can be achieved by using a spectroscopic element, like a prism or a diffraction grating, to disperse the incoming light into spectral bands with much smaller bandwidths than that of the incoming light. Traditional intensity interferometry uses filters to reduce the bandwidth,  but this results in a significant loss of photons. A dispersive element can achieve the same effect \emph{per bandwidth}, but allow for a significant enhancement of the signal by the combination of many channels. The required spectral resolution for our application is very mild, ranging from hundreds to a few thousands of spectral channels, which is achievable even with off-the-shelf instruments. The trade-off is the need for a photodetector \emph{array} with at least the same number of pixels, and potentially some modification of the detection scheme at extremely fast time-resolutions due to a recently shown loss of coherence \cite{galanis2023extendedpath}, which we describe in Appendix \ref{sec:suppression}.

One takeaway message from this review is that intensity interferometry requires some combination of bright compact sources, large collecting areas, high precision timing, long observation times, and narrow bandwidths.  Eqn.\ \eqref{eqn:sigV} illustrates how the noise in the measured visibility depends on these observational parameters.
The simple theoretical expression for the signal to noise ratio given in Eqns.\ \eqref{eqn:snrvis} and \eqref{eqn:sigV} matches well with the actual SNR measured in intensity interferometry observations.  For example, recent work has observed three bright stars using 1m telescopes and SPADs with timing jitter with FWHM of 500 ps \cite{Guerin_time,Guerin_space}.  Using their measured photon fluxes at the detectors, the observed SNR in intensity correlations for their 3 sources agrees with Eqns.\ \eqref{eqn:snrvis} and \eqref{eqn:sigV} to $17-50\%$, varying between the sources.  

This agreement motivates us to consider whether intensity correlations could be measured not only for bright stars visible to the naked eye, but also for extragalactic sources, if we use telescope arrays similar in size to existing Cherenkov arrays and if we use detectors with picosecond timing precision. 
As our fiducial example of a telescope array, we will consider a hypothetical array similar in collecting area to the CTA South MST array, consisting of 14 telescopes, each of which has an effective collecting area of $88\, {\rm m}^2$ \footnote{\url{https://www.cta-observatory.org/project/technology/mst}}, but allowing for different baseline separations than the actual CTA array, and also with mirror surface errors at the millimeter level, somewhat better  than the actual CTA mirrors (but nowhere near the sub-micron requirements of diffraction limited mirrors).  For an example extragalactic source, we will consider an apparent magnitude of $g=12$.  For this source brightness and telescope size, each telescope receives photons at a rate $d\Gamma/d\nu \approx 1.4\cdot 10^{-7}$ \footnote{We convert from magnitudes to photon flux assuming a source with apparent magnitude $g=0$ has flux at the top of the atmosphere of 3730 Jy \cite{Schneider1983}.}.  
For a total observing time $T_{\rm obs}=10^5\,$s, with a timing jitter FWHM of 30 ps (corresponding to Gaussian rms $\sigma_t = 13$ ps), Eqn.\ \eqref{eq:visfisher} gives $\sigma_{|{\cal V}|^2}\approx 0.04$ for $n_c=1$ single channel.  If we instead use $n_c=5000$ independent channels, then $n_c^{-1/2}\sigma_{|{\cal V}|^2}\approx 5.6\cdot 10^{-4}$, or  
${\rm SNR}\approx 1800$ for an unresolved source.  This demonstrates that we can not only detect intensity correlations for extragalactic sources, but we can achieve a sufficiently high SNR that we can probe the physics of those distant sources.

This leads us to the question of how well intensity interferometry observations can be used to characterize  observed sources.  The way we will quantify this is to consider parametric models for our sources, and then to estimate the uncertainties on the derived model parameters using the Fisher matrix.  
If we observe random fields (like the fluctuating counts of photons measured at different telescopes and different times), we can express the Fisher matrix simply in terms of the covariance of those fluctuations \cite{Tegmark1997},
\begin{equation} \label{eq:Fisherdefn}
F_{\alpha\beta} = \frac12 {\rm Tr} 
\left[{\bf C}^{-1} {\bf C}_{,\alpha} {\bf C}^{-1} {\bf C}_{,\beta} \right] ,  
\end{equation}
where {\bf C} is the covariance matrix of the observables, and ${\bf C}_{,\alpha}\equiv \partial{\bf C}/\partial p_\alpha$ is the derivative of the covariance matrix with respect to parameter $p_\alpha$.  In the regime of interest, signal covariance is negligible compared to Poisson shot noise, and so Eqn.\ \eqref{eq:Fisherdefn} takes on a simple form,
\begin{equation} \label{eq:Fisherconvolveddefn}
F_{\alpha\beta} = 
\frac{\Gamma^2 T_{\rm obs}}4 
\int d\Delta t 
\frac{\partial{\cal C}}{\partial p_\alpha}
\frac{\partial{\cal C}}{\partial p_\beta},
\end{equation}
where ${\cal C}$ was defined in Eqn.\ \eqref{eq:convolved_vis}.  When we can express ${\cal C}$ in terms of the visibility ${\cal V}$ using Eqn.\ \eqref{eq:FintermsofV}, this further simplifies to
\begin{equation} \label{eq:Fishervisdefn}
F_{\alpha\beta} = 
\frac{1}{\sigma_{|{\cal V}|^2}^2}
\frac{\partial |{\cal V}|^2}{\partial p_\alpha} 
\frac{\partial |{\cal V}|^2}{\partial p_\beta}, 
\end{equation}
where $\sigma_{|{\cal V}|^2}$ was defined in Eqn.\ \eqref{eqn:sigV}.
When we observe multiple baselines and multiple independent frequency channels, then we simply sum their Fisher matrices to obtain joint constraints.  Below we will use these expressions for the Fisher matrix to forecast how well intensity interferometers could be used to study various sources.

\section{AGN accretion disks} \label{sec:AGN}

Intensity interferometry is best suited for sources that are bright and compact.  Traditionally, nearby stars have been the main targets of intensity interferometric observations (e.g., \cite{Dravins2012}), however extragalactic sources such as supernovae and active galactic nuclei (AGN) can also be sufficiently bright and compact to warrant interferometric observations.  

AGN are powered by supermassive black holes that are accreting gas \cite{Peterson1997,Krolik1999,netzer_2013}. The observed emission from luminous AGN is widely believed to originate from a geometrically thin, optically thick accretion disk, see Ref.\ \cite{Davis2020} for a recent review, and see \cite{Hopkins2023a,Hopkins2023b} for alternative models involving thick disks.  Basic aspects of disk physics may be understood using simple models (e.g., \cite{Shakura1973}) but numerous puzzles remain \cite{Davis2020} and measurement of the emission profile of these disks would greatly help to elucidate the physics governing disk structure.  

Long baseline interferometers can resolve the structure of AGN accretion disks.  For example, the Schwarzschild radius for a SMBH of mass $M=10^9 M_\odot$ at a distance of 20 Mpc corresponds to an angle of $\approx 1 \mu$as, which may be resolved with baselines $B \sim 100\,$km at visible wavelengths.  The visible emission from AGN accretion disks is expected to arise from radii of order $R \sim 100 R_S$ (or possibly larger, see \cite{2012ApJ...751..106J,2014ApJ...783...47J,Morgan2018}), meaning that telescope arrays with baselines spanning 1-100 km should cover the relevant range of scales for AGN observations.

\subsection{Fringe visibility for disk}

\begin{figure}
    \centering
    \includegraphics[width=0.48\textwidth]{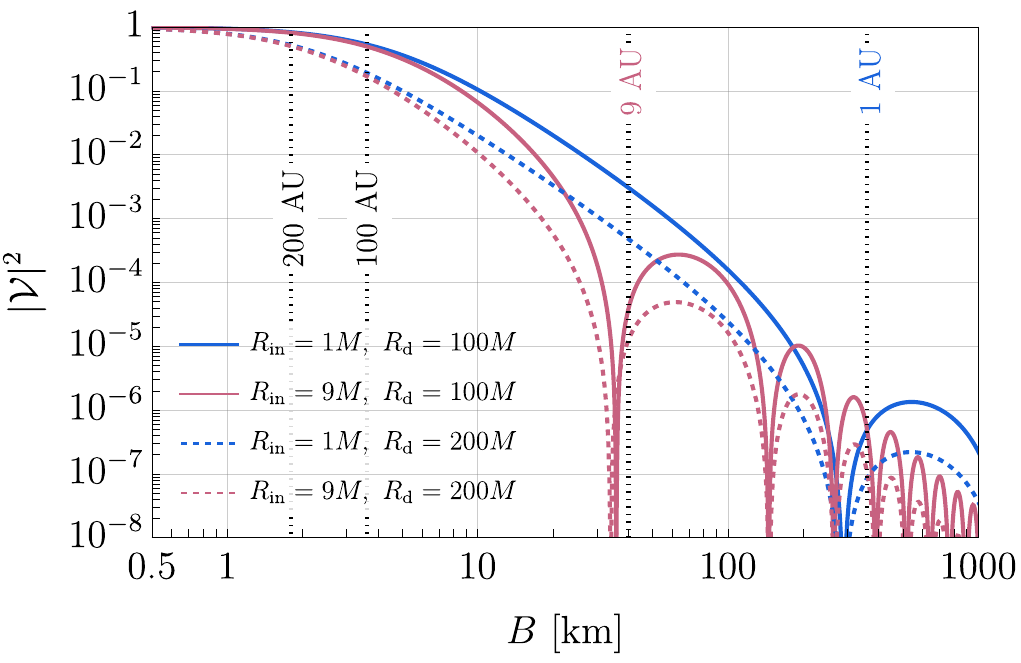}
    \caption{Squared visibility $|{\cal V}|^2$ for Shakura-Sunyaev \cite{Shakura1973} disk, with $GM/c^2 = 1\,$AU, at distance $D=20\,$Mpc, observed at wavelength $\lambda=5500$\AA.  The two choices of $R_{\rm in}$ correspond to the ISCO for maximal spin, for prograde and retrograde orbits respectively.  The vertical dotted lines indicate the baselines  $B=\lambda\,D/(2\pi\,R_{\rm in})$, where we expect the interferometer to start to resolve a uniform disk of radius $R_{\rm in}$.}
    \label{fig:AGNvis}
\end{figure}

Let us consider a thin accretion disk, using cylindrical coordinates $R$ and $\phi$ in the disk plane.  If the emergent intensity from the disk depends only on the local radius of a point in the disk plane, $I = I(R)$, then Eqn.\ \eqref{eq:vis} gives
\begin{equation}\label{eq:diskvis}
{\cal V}(\nu, \bm{B}) =
\frac{\int dR\, R\, I(R)\, J_0
\left(\frac{2\pi q\nu B R}{cD}\right)}
{\int dR\, R\, I(R)} ,
\end{equation}
where $q = \sqrt{\cos^2 i \, \cos^2 \phi_B + \sin^2 \phi_B}$, for disk inclination $i$ and baseline of length $B$ and position angle $\phi_B$ measured relative to the position angle of the apparent disk minor axis.  Also, 
$J_0$ is a Bessel function, and $D$ is the angular diameter distance from the observer to the AGN.  Fig.~\ref{fig:AGNvis} shows examples of the visibility for various disk profiles.

By measuring $|{\cal V}|^2$ at many different baselines, we can study the properties of the accretion disk.  Since we cannot directly image disks using intensity interferometry, we instead model the observed visibilities using a parametric model, and then constrain the model parameters. We note, apropos, that the need for modeling is not unique to intensity interferometry, as the sparse sampling of the $u$-$v$ plane in long baseline radio interferometry demands similar procedures. The parameter constraints derived from intensity interferometry observations may be estimated using the Fisher matrix defined in Eqn.\ \eqref{eq:Fishervisdefn},
\begin{eqnarray} \label{eq:visfisher}
F_{\alpha\beta}&=&\binom{n_t}{2} \frac{T_{\rm obs}}{\sigma_t} 
\frac{1}{\sqrt{128\pi}}\left(\frac{d\Gamma}{d\nu}\right)^2 \\
&& \times \sum_{\nu,\bm{B}} f(\bm{B})
\frac{\partial |{\cal V}(\nu,\bm{B})|^2}{\partial p_\alpha} 
\frac{\partial |{\cal V}(\nu,\bm{B})|^2}{\partial p_\beta} .
\nonumber \\
&=& \frac{1}{\sigma_{|{\cal V}|^2}^2}
\sum_{\nu,\bm{B}} f(\bm{B})
\frac{\partial |{\cal V}(\nu,\bm{B})|^2}{\partial p_\alpha} 
\frac{\partial |{\cal V}(\nu,\bm{B})|^2}{\partial p_\beta} \nonumber
\end{eqnarray}
Here, $T_{\rm obs}$ is the total observing time, 
$n_t$ is the number of telescopes, so that the number of pairs is $\binom{n_t}{2}$, and $f(\bm{B})$ is the fraction of time $T_{\rm obs}$ observing on baseline $\bm{B}$, with $\sum f = 1$.  We also assume that every telescope has the same area $A$ and the same mean rate of photon counts $\Gamma$, and that each channel's bandwidth $\Delta\omega=2\pi\Delta\nu$ satisfies $\sigma_t\,\Delta\omega \gg 1$, where $\sigma_t$ is the uncertainty in the photon time of arrival due to detector timing jitter.  Note that $\sigma_{|\mathcal{V}|^2}$ in  Eqn.~\eqref{eq:visfisher} generalizes  $\sigma_{|\mathcal{V}^2|}$ in Eqn.~\eqref{eqn:sigV} to account for $n_t$ telescopes.

\begin{figure}
\centering
\includegraphics[width=0.48\textwidth]{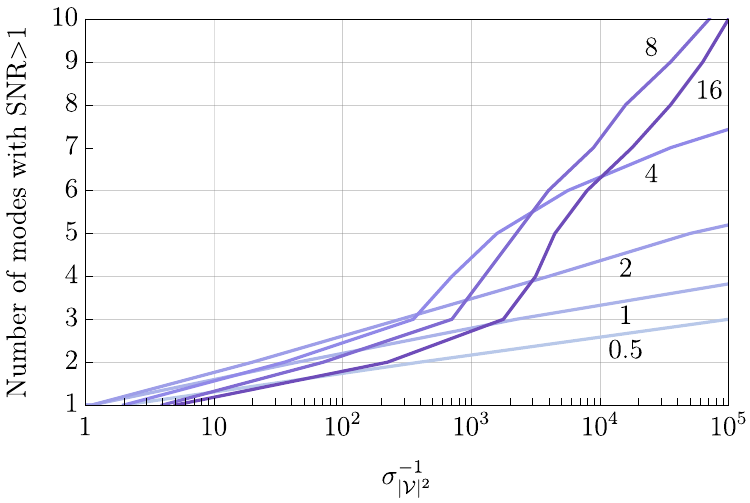}
\caption{Principal component analysis. We use 2D Bessel functions as our basis set, and take as the fiducial profile 
a lognormal profile $\beta(R)=\frac{1}{R\sqrt{2\pi} R_0\sigma}\exp\parea{-\frac{\log^2(R/R_0)}{2\sigma^2}-\frac{\sigma^2}{2}}$, with $\sigma=0.5$. The darker the color, the longer the maximum baseline: $kR_0B_\text{max}/D={0.5,1,2,4,8,16}$.}
    \label{fig:pca}
\end{figure}

\subsection{Principal component analysis}

Using Eqn.\ \eqref{eq:visfisher}, we can estimate how well the parameters of an assumed disk model for $I(R)$ may be constrained using a given set of observations.
To understand most generally what information may be gleaned from intensity interferometry observations of nearby AGN accretion disks, we perform a principal component analysis \cite{Huterer2003}.  We write an arbitrary disk profile as a sum over basis functions,
\begin{equation}
I(R) = \sum_i a_i \, u_i(R),    
\end{equation}
treating the coefficients $a_i$ as parameters.  Using Eqns.\ \eqref{eq:diskvis} and \eqref{eq:visfisher}, we can compute the Fisher matrix for this choice of parameters.  If we instead treat $\log(a_i)$ as the parameters, then the Fisher matrix tells us the signal to noise ratio on the measured coefficients.  The eigenvectors of the Fisher matrix tell us the parameter combinations that are measured independently, and the eigenvalues of those eigenvectors tell us the signal to noise ratio with which those parameter combinations are determined.  This PCA therefore can be used to determine the number of independent quantities that a given observation can measure above some SNR threshold.  Note that the choice of basis function set does not affect the recovered principal components, as long as the basis functions are complete.

Figure \ref{fig:pca} shows an example.  In this example, we use Bessel functions $J_1(R/(R_{\rm max} x_i))$ as the basis functions, where $R_{\rm max}$ is the maximum radius over which we reconstruct the disk profile $I(R)$ and $x_i$ is the $i^{\rm th}$ root of $J_1$.  As a fiducial profile, we arbitrarily adopt a lognormal profile $I \propto \exp[-(\log R/\sigma)^2/2]/R$; the basic conclusions below do not depend strongly on this choice of fiducial profile.   For ${\rm SNR} < 10$, we can only measure one number, e.g. the fraction of light that is unresolved $kB_{\rm max} R/D \lesssim 1$.  As we increase the SNR, it becomes possible to measure additional independent numbers, but doing so requires longer baselines that resolve the disk structure better.  The number of well-measured parameters grows slowly for ${\rm SNR}<1000$, but once we hit  ${\rm SNR}>1000$ the number of measurable parameters grows more quickly.

\subsection{Numerical example: Shakura-Sunyaev disk}\label{sec:SS}

The above PCA estimate suggests that if we can achieve SNR $>10^3$ it becomes possible to study disk physics by measuring multiple independent parameters describing the disks.  Achieving this SNR for nearby AGN appears feasible, as discussed in section \ref{sec:review}. There, we used Eqn.\ \eqref{eqn:sigV} to show that a hypothetical array of telescopes similar to the CTA South MST array, observing a source with apparent magnitude $g=12$, could achieve SNR of 25 per channel, giving ${\rm SNR}\approx 1800$ for $n_c=5000$ channels.  Thus, using realistic arrays with reasonable observing times, we can achieve a sufficiently high SNR that probing disk physics becomes feasible.  Note that apparent magnitude $g=12$ is quite bright for AGN, but several Type 1 Seyferts this bright may be found in public catalogs \cite{AGNcatalog,TurinSyCAT}.  As we can see from Fig.\ \ref{fig:pca}, in this high SNR regime, we would ideally like baseline coverage in the range $k\,B\,R_{\rm disk}/D \sim 5-10$ to measure many independent parameters describing the disk.  

For a concrete example, we use Eqn. \eqref{eq:visfisher} to estimate how well the parameters of the Shakura-Sunyaev \cite{Shakura1973} profile may be measured.  This profile is given by:
\begin{equation}\label{eq:SSprofile}
I(R) = I_0 \left[e^{f(R)} - 1\right]^{-1},
\end{equation}
where
\begin{equation}\label{eq:fSS}
f(R) = \frac{\nu}{\nu_0(R)} = 
\left[\left(\frac{R_0}{R}\right)^n\,
\left(1-
\sqrt{\frac{R_{\rm in}}{R}}
\right)
\right]^{-1/4}.
\end{equation}
The parameters are: an effective radius $R_0$ for the light at frequency $\nu$, the power-law index $n$, and the inner edge of the disk $R_{\rm in}$.  The Shakura-Sunyaev disk has $n=3$, but other models can allow different behaviour \cite{2014ApJ...783...47J}. Note that the normalization $I_0$ does not affect ${\cal V}$, and its value is fixed by the observed AGN flux once we have chosen the other parameters.  Also note that this neglects relativistic effects like Doppler beaming, gravitational lensing, and gravitational redshift, for simplicity.  Below we repeat this analysis including these relativistic effects (see \S\ref{sec:edge}).

\begin{figure}
\centering
\includegraphics[width=0.48\textwidth]{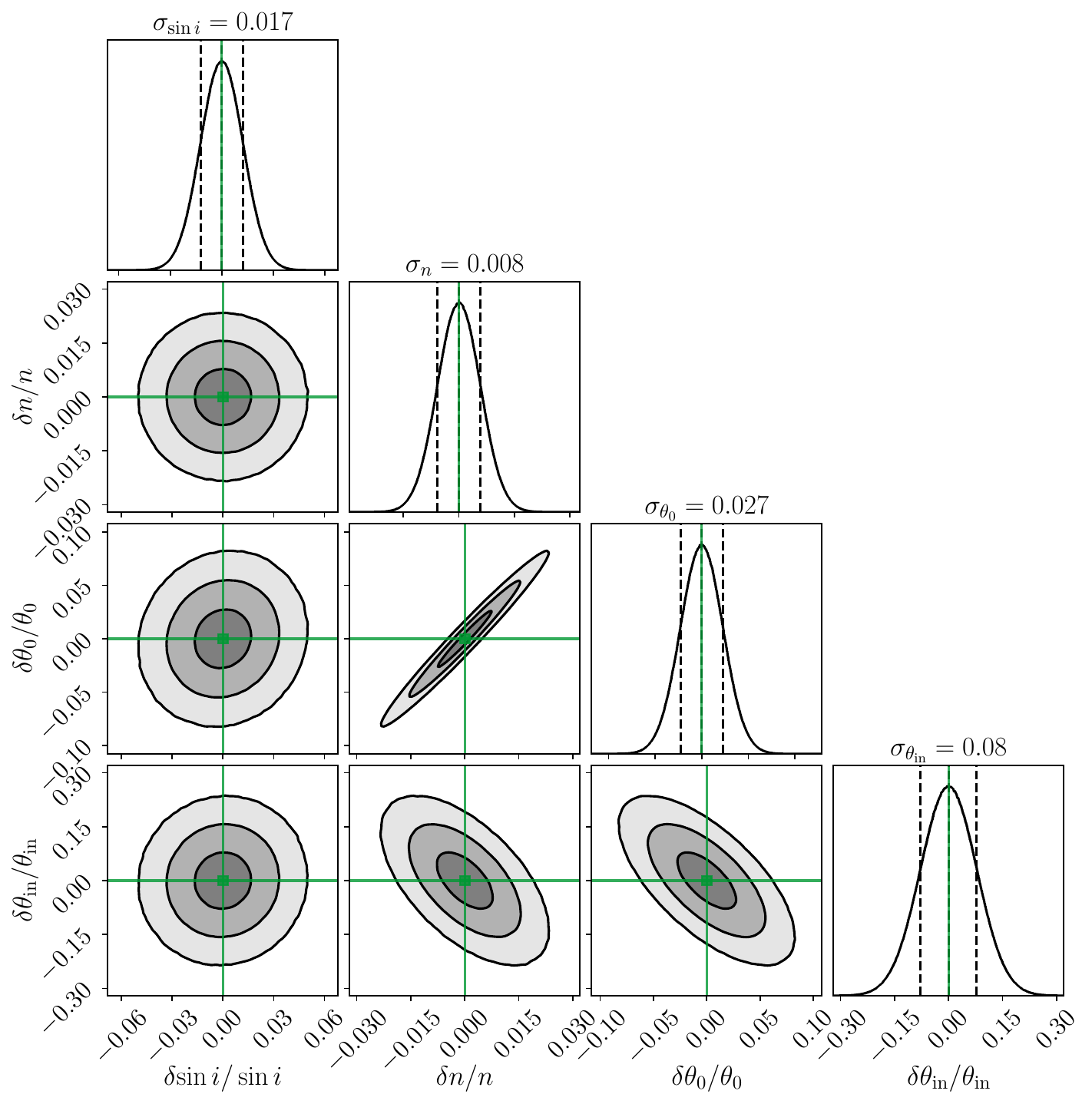}
\caption{Error covariance for intensity interferometry on an AGN accretion disk with visibility error $n_c^{-1/2}\sigma_{|\mathcal{V}|^2}\approx5\cdot 10^{-4}$. We assume a Shakura-Sunyaev profile with parameters $\sin i=1/2$, $n=3$, $\theta_\text{0}=42.8 ~GMc^{-2}/D$ at 550nm and $\theta_\text{in}=6~ GMc^{-2}/D$. The observation here is broadband in the range 350-900nm.}
    \label{fig:disk_corner}
\end{figure}

Following the discussion above, we set $n_c^{1/2}\sigma_{|{\cal V}|^2}^{-1} = 2000$.  We assume fiducial parameter values $n=3$, $R_{\rm in}=6\, GM/c^2$ (i.e., the Schwarzschild ISCO), and $R_0=43\, GM/c^2$ at $\lambda=500\,$nm, so that the mean emission-weighted radius is $\bar R \equiv \int R^2 I(R)dR / \int R\,I(R)dR \approx 150\, GM/c^2$ at this wavelength.  For $M=10^8 M_\odot$, note that the gravitational radius is $GM/c^2 \approx 1\,$AU.  We assume $f(\bm{B}) = (2\pi B_{\rm max}\,B)^{-1}$ for $|\bm{B}|<B_{\rm max}$, so that $\int f(\bm{B}) B\,dB d\phi = 1$, and take $k\,B_{\rm max}/D = 0.045\, (GM/c^2)^{-1}$.  For $\lambda\approx 550\,$nm and $D=20\,$Mpc, this would correspond to $B_{\rm max} \approx 16\,$km.  
For these parameters, we find that the parameters of the disk model are well constrained, and show the predicted error covariance in Fig.\ \ref{fig:disk_corner}.

Note that broadband observations are advantageous not only because of the $n_c^{1/2}$ enhancement to the SNR, but also because the apparent disk size is expected to vary with wavelength, since the local disk temperature is a function of radius.  If we observe an $\mathcal{O}(1)$ range of wavelengths, we can detect this wavelength dependence of the effective size, measuring the radial slope $n$ and thereby helping to break degeneracies that may be present in narrowband observations.

This illustrates that the optimal observing strategy will  depend on the science objective of the observations. Because emission of shorter wavelengths peaks at smaller radii, if the observational target is to measure the innermost part of the disk, it is always advantageous to restrict to narrowband observations at the shortest wavelengths possible. In contrast, for applications like the overall disk profile, it is more useful to use shorter baselines and do broadband observations. For instance, in the error covariance of Fig.~\ref{fig:disk_corner}, 
the fractional error on the power-law index is $\sim 0.8\%$ for broadband observations in the 350-900nm range, while for narrowband observations we find the fractional uncertainty on $n$ to be $\sim 4\%$ at all wavelengths. 

We leave a detailed analysis of the optimal observing strategy to future work. Next, we discuss the physical interest of a few of the parameters we can constrain.

\subsubsection{Inclination angle}

Our simple Fisher analysis finds that the disk inclination angle $i$ is measured well, at the percent level for the parameters described above.  As Figure \ref{fig:disk_corner} illustrates, the uncertainty on the inclination parameter is not correlated with uncertainties on other parameters.  This is because the inclination is the only parameter that controls the non-circularity of the appearance of the disk (e.g., Eqn.\ \eqref{eq:diskvis}), and the apparent shape of the disk is determined well, as long as baselines with many different orientations are sampled.  

One caveat to this argument is that relativistic effects like Doppler beaming can modify the appearance of this disk, generating non-circularity that could be mistaken for inclination.  We estimate that the systematic error on the inferred inclination arising from beaming should be $\sim {\cal O}(1\%)$.  One way to see this is to consider the multipole expansion of the light emitted from the disk.  Using coordinates centered on the black hole, we can expand the intensity profile into angular multipoles,
\begin{equation}\label{eq:multipole}
I(\omega,\hat n) = I(\omega,\theta,\phi) = \sum_l a_l(\omega,\theta) \cos\left(l (\phi-\phi_l)\right). 
\end{equation}
Inserting this multipole expansion into Eqn.\ \eqref{eq:vis} gives
\begin{equation}
{\cal V}(\omega, \bm{B})=\sum_l i^l A_l(kB) \cos(l(\phi_l-\phi_B)),
\end{equation}
where $A_l(kB) = 2\pi\, F^{-1}\int a_l(\omega,\theta)\, J_l(kB\theta)\, \theta \, d\theta$, and $F=\int I\, d^2 \hat n$ is the total flux. Note that even multipoles make a purely real contribution to the visibility, while odd multipoles make a purely imaginary contribution.  

A circular disk viewed at some inclination angle has only even nonzero multipoles.  Beaming, to lowest order, generates a dipole ($l=1$), which is therefore purely imaginary.  When we square the visibility to compute $|{\cal V}|^2$, the real and imaginary parts add in quadrature.  Using simulations of disk appearances that include relativistic effects like beaming (see \S\ref{sec:edge} below), we estimate that the dipole/monopole ratio has amplitude $\sim 10-15\%$ on the scales of interest for typical disk parameters, and would therefore make a $\sim 1-2\%$ effect on the inferred inclination angle.  We thus conclude that relativistic beaming will not significantly contaminate measurements of disk inclinations.

Measurement of the inclination angle for a large sample of bright AGN (e.g., $\sim100$ objects) would be useful for probing AGN physics. One example would be testing the `unification model' of AGN \cite{Antonucci1993}, which postulates that different types of AGN such as Seyfert 1, Seyfert 2, etc., correspond to similar physical objects that are merely viewed at different orientation angles.  A significant correlation between AGN type and disk inclination angle $i$  would represent a stunning success for AGN theory.

\subsubsection{Radial slope}

We parameterized the radial slope of the temperature profile as $T^4 \propto R^{-n}$ in Eqn. \eqref{eq:fSS}.  Our simple Fisher estimate finds that the radial slope of the disk is measured precisely, as seen in Fig.~\ref{fig:disk_corner}. A measurement of this slope would dramatically constrain models of the central accretion disk.  A generic prediction of steady, thin, viscous disk models is that $n \approx 3$.  This prediction is not satisfied in some accreting white dwarf in binary systems (``cataclysmic variables'') where the profile can be determined using eclipse mapping \cite{Baptista1995,Linnell2008,Linnell2010}.  Analyses of quasar microlensing observations have also been used to constrain the disk radial profile \cite{2014ApJ...783...47J} and find $n\approx 5.3$, significantly different than $n=3$.
Departures from $n = 3$ signal departures from classical thin disk theory and could indicate the presence of a warped disk, driving of accretion by a magnetized wind, or disk flaring.  Long-baseline interferometric observations of AGN disks could test whether this behavior is common in near-Eddington  supermassive black holes.

\subsubsection{Inner disk structure} \label{sec:edge}

Another parameter measured well is the inner disk edge, with our Fisher matrix estimate giving $R_{\rm in}/\sigma_R \approx 6$.  In our assumed Shakura-Sunyaev model profile, the disk sharply truncates at this inner edge, inside of which is a completely dark hole in the emission profile.  In appendix \ref{sec:hole} we explore in a more model-independent context whether intensity interferometry observations can establish the existence of a sharply edged hole in the disk.

More generally, the small uncertainty on the inferred $R_{\rm in}$ parameter suggests that long-baseline observations are sensitive to the central structure of AGN accretion disks.  In this inner region, relativistic effects such as beaming and lensing become significant, invalidating our simplistic Newtonian analysis.  To account for these effects, we repeat our Fisher matrix analysis for a spin-aligned, infinitely thin Novikov-Thorne \cite{NovikovThorne} disk profile, with vanishing emission inside the ISCO. We observe this disk using the {\tt IPOLE} code \cite{ipole}, which accounts for relativistic effects such as frame dragging, beaming, etc.  To quantify how well we can probe the innermost disk structure, we estimate how well the spin parameter $a$ of the central black hole may be determined.  We compute the Fisher matrix by varying the parameters of the disk, like the accretion rate and BH mass, as well as observer parameters like the inclination angle and the distance to the AGN.  We treat these as nuisance parameters and marginalize over them to estimate the error on the inferred spin.  For fiducial choices of $M_{\rm BH}=10^8 M_\odot$, $D=10\,$Mpc, Eddington ratio of 0.1, inclination $i=60^\circ$, and spin $a=0.5$, we find that $\sigma_a\approx 0.3$ for observations with $\sigma_{|{\cal V}|^2}^{-1} = 2000$.  As noted above, for bright AGN this SNR can be achieved in about 24 hours of observations with modestly sized interferometric arrays.  

We stress that this result does not mean that we can measure BH spins precisely, since this calculation is highly simplistic, ignoring important effects such as disk warping, misalignment, self-illumination, or the existence of a disk atmosphere.  Instead, we interpret this result as a demonstration that long-baseline interferometric observations can indeed probe the innermost emission from AGN disks.  A detailed comparison of such observations to realistic GRMHD simulations of AGN accretion disks could provide a wealth of insight into the physics of the brightest black holes in the universe.

\section{Broad line region} \label{sec:BLR}

The visible radiation from bright AGN is believed to originate from geometrically thin, optically thick accretion disks.  There are two key aspects of this visible emission that will be important for the discussion below.  First, the spectra of AGN light exhibit not only broadband continuum emission, but also extremely bright emission lines, which in some cases have broad linewidths corresponding to velocity dispersions of thousands of km/s \cite{Peterson2006}.  Secondly, the emission from AGN is typically variable over time, both for the broadband continuum and also for the line emission \cite{SDSSRM,2023arXiv230501014S,OzDES}. 
Crucially, the line variability typically lags the continuum variability in time, with delays ranging from days to months, depending on the AGN and depending on the line.  These time lags are understood to arise from the large spatial separation between the line emitting regions and the continuum emitting region.   The physical picture is that the continuum radiation from the central AGN illuminates and heats the broad line region, which then cools via line radiation.  The time lag between the continuum variability and the line variability can be used to infer the size of the broad line region, using a method called reverberation mapping \cite{Bahcall1972,Blandford1982,Peterson1993}.

Interferometric observations of broad line emission from AGN would be useful in several ways.  Resolved images of the broad line region would help constrain the physics of the outer accretion disk and radiative transfer near the AGN.  Additionally, interferometric images would also allow the determination of the angular size of the broad line region.  As noted above, reverberation mapping measures the physical size (in cm) of the broad line region.  The combination of reverberation mapping and resolved interferometry would therefore measure both the physical size and the angular size of the regions emitting the same line photons, which then measures the angular diameter distance to the AGN. Since the redshifts of these AGN are already known from their spectra, a distance determination would measure the Hubble constant $H_0$, one of the most important and most controversial quantities in modern cosmology \cite{Kamionkowski2023}.  We discuss this in more detail below.

\subsection{Reverberation mapping}

Let us write the continuum lightcurve as $c(t)$, and the broad line (BL) lightcurves as $\bm{b}(t)$, related to the continuum lightcurve via transfer functions $\bm{\psi}$, given by $\bm{b} = c*\bm{\psi}$ \cite{Blandford1982}.  Here, we consider $\bm{b}$ as an array of lightcurves measured in each independent channel covering the emission line. 
The transfer functions $\bm{\psi}(\Delta t)$ may be determined from the two-point correlations of $c$ and $\bm{b}$.  As long as we do not observe the system at a special time, i.e.\ as long as the 2-point correlations are stationary, then 
it is convenient to work in Fourier space, where the two-point correlations of $c(\omega)$ and $\bm{b}(\omega)$ become block diagonal, in the sense that different $\omega$'s are uncorrelated.  Note that in this context, $\omega$ refers to the Fourier conjugate of $\Delta t$, and is not the observed frequency of the light $2\pi\nu$.

Suppose we observe $\bm{b}$ and $c$ with some measurement noise $\epsilon_b$ and $\epsilon_c$, each uncorrelated with anything else.  At each frequency, let us write the data as a vector $\bm{d}(\omega) = \{c_{\rm obs}(\omega), \bm{b}_{\rm obs}(\omega)\}$, with covariance
\begin{equation}
\langle\bm{d}(\omega_1) \bm{d}^\dagger(\omega_2)\rangle = 2\pi \delta(\omega_1-\omega_2) \left[P_c\, \bm{G}\, \bm{G}^\dagger + {\bf N}\right]
\end{equation}
where $P_c$ is the emitted continuum power spectrum, 
$\bm{G}=\{1,\bm{\psi}\}$, and ${\bf N}$ is the noise matrix, whose nonzero elements are along the diagonal, corresponding to the measurement noise for continuum ($N_c$) and each of the BL channels ($N_b$).  Note that $\bm{G}$ and $\bf N$ are functions of $\omega$, and their indices label different channels (and the continuum).  

For simplicity, let us suppose that $P_c$ and ${\bf N}$ are known, so that the only quantities to be determined from the data are the transfer functions $\bm{\psi}(\omega)$.  Let us describe $\bm{\psi}$ using a parametric model with parameters $p_\alpha$.  We can again use the Fisher matrix to estimate the uncertainties on the parameters.  Writing the data covariance as ${\bf C} = P_c\, \bm{G}\, \bm{G}^\dagger + {\bf N}$, the Fisher matrix takes the form \cite{Tegmark1997}
\begin{equation}\label{eq:RMfisher}
F_{\alpha\beta} = \sum_\omega \frac{1}{2}{\rm Tr}
\left[
{\bf C}^{-1}{\bf C}_{,\alpha}\,{\bf C}^{-1}{\bf C}_{,\beta}
\right]
\end{equation}
where ${\bf C}_{,\alpha}\equiv \partial{\bf C}/\partial p_\alpha$, and the sum in Eqn.\ \eqref{eq:RMfisher} runs over all independent Fourier modes in the lightcurve.
Making use of the Sherman-Morrison formula, we have 
\begin{equation}
{\bf C}^{-1} = {\bf N}^{-1}-P_c
\frac{{\bf N}^{-1} \bm{G}\, \bm{G}^\dagger{\bf N}^{-1}}
{1 + P_c \bm{G}^\dagger {\bf N}^{-1} \bm{G}},
\end{equation}
and the partial derivatives are 
\begin{equation}
{\bf C}_{,\alpha} = P_c\,\left[\bm{G}_{,\alpha} \bm{G}^\dagger
+ \bm{G}\,\bm{G}_{,\alpha}^\dagger\right].
\end{equation}
If we observe the emission line with only one channel, then the Fisher matrix simplifies to 
\begin{equation}\label{eq:RMsinglefisher}
    F_{\alpha\beta} = \sum_\omega
\frac12 \frac{P_c^2}{d^2} \left(d\, g_1 + N_c^2 g_2\right),
\end{equation}
where 
\begin{eqnarray}
d &=& P_c N_b + N_c N_b + |\psi|^2 P_c N_c \\
g_1 &=& \partial_\alpha \psi^*\, \partial_\beta \psi + \partial_\beta \psi^* \, \partial_\alpha \psi \\
g_2 &=& \partial_\alpha(|\psi|^2) \, \partial_\beta(|\psi|^2).
\end{eqnarray}

\subsection{Transfer function}

Next let us discuss the transfer functions relating continuum fluctuations and BL fluctuations.  We can obtain a parametric form for $\bm{\psi}(\Delta t$) by assuming a model.  Here, we focus on the simple example where the continuum emission may be treated as unresolved (point-like), and the BL emission arises from a Keplerian disk inclined relative to the line of sight at angle $i$, as suggested by recent interferometric observations of AGN broad line regions by the GRAVITY collaboration \cite{2018Natur.563..657G,2020A&A...643A.154G,2021A&A...648A.117G,2024arXiv240107676G,2024arXiv240114567A}.  In the case of an inclined Keplerian disk, the transfer function takes a simple form \cite{Blandford1982}, which we summarize below.  

Let us describe locations in the plane of the disk using coordinates $R$ and $\phi$, where the azimuthal angle $\phi$ is measured relative to the apparent minor axis on the sky.  In the AGN rest frame, the time delay $\Delta t$ for a path that starts at the origin, reaches point $(R,\phi)$ in the disk, and then proceeds to us (at infinity) is 
\begin{equation}
c\,\Delta t = R\,(1+\sin i\,\cos\phi),
\end{equation}
where $i$ is the disk inclination angle relative to the line of sight.  
If the AGN is at redshift $z$, then the observed time delay is larger by a factor of $1+z$.  From context, it is hopefully clear that the time delay $\Delta t$ appearing here is unrelated to the time difference $\Delta t$ between two telescopes that appeared in Eqns.\ \eqref{eq:visibility} and \eqref{eq:normvis}.  

Suppose that BL emission from the disk may be written as 
\begin{equation} \label{eq:profile}
I_{\rm BL}(t,R,\phi) = \beta(R,\phi) F_c(t-\Delta t(R,\phi)),
\end{equation}
where $F_c$ is the continuum flux from the unresolved central source, and $\beta$ is some response function that depends on the distribution of broad line emitters around the central source.  For simplicity, we will assume that $\beta$ is a purely radial function, $\beta=\beta(R)$, independent of $\phi$.  In reality, this is not the case: radiative transfer effects cause line photons to escape the disk preferentially along certain directions that leads to a predictable $\phi$ dependence that we will ignore below.  
With these assumptions, the integrated flux from the BLR is
\begin{eqnarray}
F(t) &=& |\cos i| \int  I_{\rm BL}(t,R,\phi) R\, dR\, d\phi \nonumber \\  
&=& |\cos i| \int \beta(R) F_c(t-\Delta t)  R\, dR\, d\phi\nonumber \\
&=& \int  F_c(t^\prime)\, \psi(t-t^\prime) dt^\prime
\end{eqnarray}
where the transfer function is 
\begin{equation}
\psi(\delta t) = |\cos i|\int \beta(R)\, \delta(\Delta t(R,\phi) -\delta t) R\,dR\,d\phi.
\end{equation}
This is the transfer function for the BL emission, integrated over the entire line.  We can similarly compute the transfer function at any frequency $\nu$ that is redshifted relative to the line center $\nu_c$, by using the line-of-sight component of the Keplerian velocity, 
\begin{equation}
v_{\rm LOS}(R,\phi) = \left(\frac{GM}{R}\right)^{1/2} \sin i\,\sin \phi,    
\end{equation}
which gives Doppler shift  $\nu \approx \nu_c (1-v_{\rm LOS}/c)$.

The velocity-dependent transfer function then becomes
\begin{widetext}
\begin{equation}
\psi(\delta t, v) = |\cos i|
\int\beta(R)\, \delta(\Delta t(R,\phi) -\delta t)  
\delta(v_{\rm LOS}(R,\phi) -v) R\,dR\,d\phi.
\end{equation}
In Fourier space, this is
\begin{eqnarray} \label{eqn:psivel}
\psi(\omega, v) &=& \int \psi(\delta t, v) e^{i\,\omega\, \delta t}
d\delta t \\
&=& |\cos i| \int\beta(R)\delta(v_{\rm LOS}(R,\phi) -v) e^{i\,\omega\, \Delta t(R,\phi)} R\,dR\,d\phi. \nonumber\\
&=& 2 |\cos i| \int_0^{R_{\rm max}(v)}
\beta(R)\, R \, W_{\rm RM}(R,v,\omega)\, dR, \nonumber
\end{eqnarray}
\end{widetext}
where 
\begin{equation}\label{eq:RMweight}
W_{\rm RM}(R,v,\omega) = \frac{e^{ikR} 
\cos\left(kR\sqrt{\sin^2 i - \frac{v^2 R}{GM}}\right)}
{\sqrt{\frac{GM}R \sin^2 i - v^2}}
\end{equation}
for $k=\omega/c$, and $R_{\rm max}(v) = (GM/v^2)\sin^2 i$.

Equation \eqref{eqn:psivel} gives the transfer function at a specific velocity.  For a channel of some finite bandwidth $\nu_{\rm min} < \nu < \nu_{\rm max}$, the transfer function is the integral of this expression,
\begin{equation}
\psi_{\rm channel}(\omega) = \int_{v_{\rm min}}^{v_{\rm max}} \psi(\omega, v) dv
\end{equation}
for $v_{\rm min}=c\,(1-\nu_{\rm max}/\nu_c)$ and $v_{\rm max}=c\,(1-\nu_{\rm min}/\nu_c)$.

This completes the expression for the transfer functions appearing in the previous subsection.  If we sum over all channels to obtain an integrated transfer function for the total BL flux, then the expression simplifies, giving
\begin{equation}
 \psi(\omega) = 2\pi |\cos i| \int dR \, R
\,\beta(R) \, e^{ikR} J_0(k R \sin i),
\end{equation}
where again $k=\omega/c$.

\subsection{Interferometry for emission line}

Along with the Fisher matrix for reverberation mapping Eqn.\ \eqref{eq:RMfisher}, we also require the Fisher matrix for intensity interferometry observations, in order to forecast the joint constraints.  Eqn.\ \eqref{eq:visfisher} gives an expression for the Fisher matrix, under the assumption that the sky emission $I(\nu, \hat n)$ factorizes into $I(\nu) f(\hat n)$, and that for a narrow channel, the spectrum $I(\nu)$ could be approximated as a constant, top-hat spectrum, i.e.\ $I(\nu) = I(\nu_0)$ for $|\nu-\nu_0| < \Delta\nu/2$.  Both of these assumptions become invalid for line emission from a rotating source like an accretion disk.  For example, opposite sides of the disk can have oppositely signed line-of-sight velocities, giving blueshifts on one side of the disk and redshifts on the other side.  That is, different regions of the source can have systematically different emission spectra.

\begin{figure*}
\centering
\includegraphics[width=0.48\textwidth]{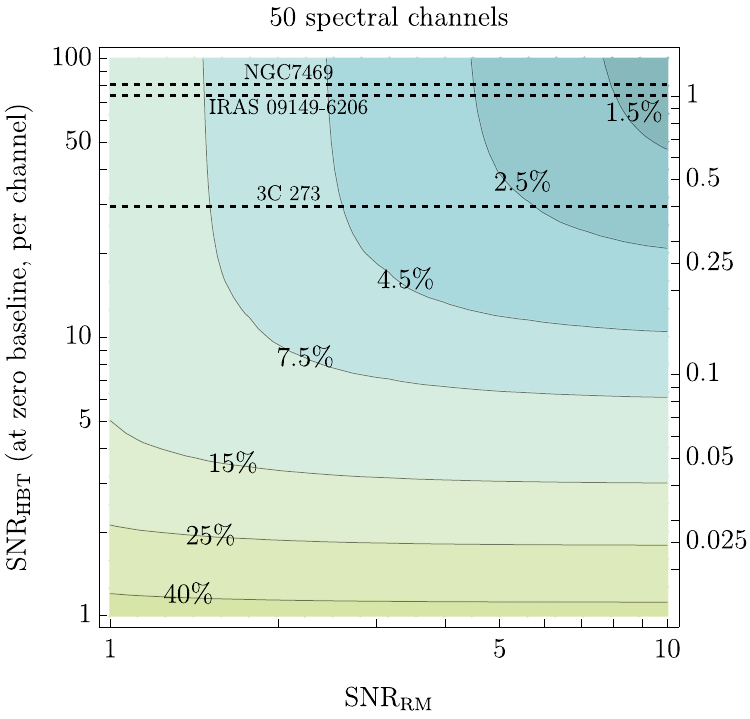}
    \quad
\includegraphics[width=0.48\textwidth]{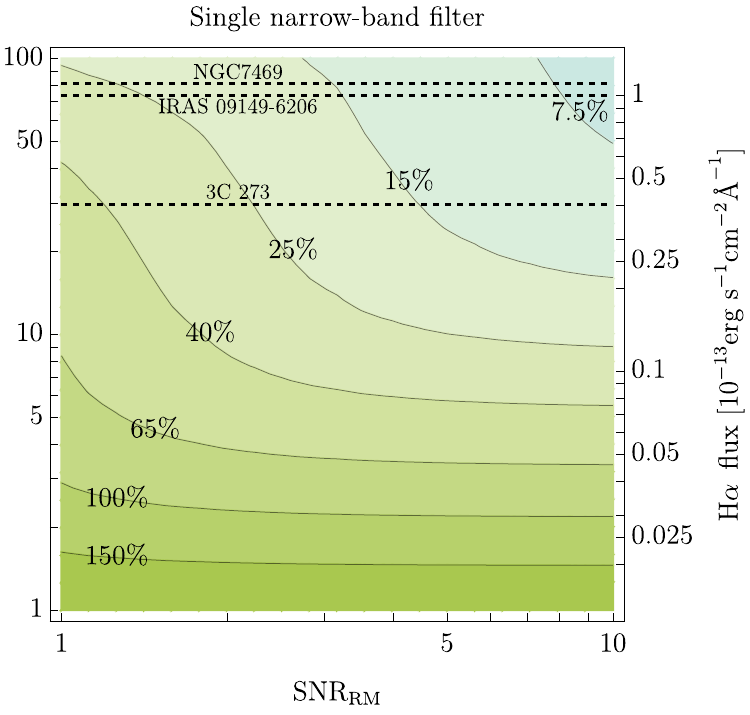}
\caption{
Distance error for joint reverberation mapping and intensity interferometry (RM-HBT) observations, as a function of BLR SNR, $(P_b/N_b)^{1/2}$, and intensity correlation SNR (Eqn.\ \eqref{eq:corrsnr}).  The right-axis shows the H$\alpha$ line flux that would give the corresponding SNR for intensity interferometry assuming total observing time $T_{\rm obs}=24\,$hr, timing jitter FWHM of 30 ps,  and total collecting area $n_t\,A = 1232\,{\rm m}^2$, equal to the CTA MST South array as discussed in the text.  Colours correspond to fractional error in the derived distance to the AGN, ${\sigma_D}/{D}$. The horizontal dashed lines indicate the H$\alpha$ fluxes reported for 3 example AGN \cite{BASS_DR2,Dietrich1999}. 
}
    \label{fig:distance}
\end{figure*}

This somewhat changes the expression for the visibility ${\cal V}$.  Rather than using the approximation used in Eqn.\ \eqref{eq:normvis_approx}, we instead must evaluate the exact expression in Eqns.\ \eqref{eq:visibility} and \eqref{eq:normvis} to define the time-dependent $V$ and ${\cal V}$.  If we repeat the reasoning used to derive Eqn.\ \eqref{eqn:psivel}, we obtain
\begin{widetext}
\begin{equation}\label{eq:Vis1}
V(\nu,\bm{B},\Delta t) = 2 F_c |\cos i|\, e^{-i\omega\Delta t}
 \int_0^{R_{\rm max}(v)}
\beta(R)\, R \, W_{\rm I}(R,v,\bm{B})\, dR  
\end{equation}
where $\omega=2\pi\nu$, $v=c(1-\nu/\nu_c)$, $\nu_c$ is the line's central frequency, $R_{\rm max}(v) = (GM/v^2)\sin^2 i$, and 
\begin{equation}    
\label{eq:visweight}
W_{\rm I}(R,v,\bm{B}) = 
\frac{
e^{i\frac{kBR}{D}\sqrt{\frac{v^2R}{GM}}\frac{\sin\phi_B}{\sin i}} 
\cos\left[\frac{kBR}{D}\sqrt{1-\frac{v^2R}{GM\sin^2 i}} \cos \phi_B\right]}
{\sqrt{\frac{GM}R \sin^2 i - v^2}}
\end{equation}
We integrate Eqn.\ \eqref{eq:Vis1} to obtain the visibility for a channel with $\nu_{\rm min}<\nu<\nu_{\rm max}$, giving
\begin{equation}\label{eq:BLvis}
{\cal V}(\bm{B},\Delta t) = 
\frac{\int_{\nu_{\rm min}}^{\nu_{\rm max}} d\nu\,
e^{-i\omega\Delta t} \int_0^{R_{\rm max}(v)}
\beta(R)\, R \, W_{\rm I}(R,v,\bm{B})\, dR } 
{
\int_{\nu_{\rm min}}^{\nu_{\rm max}} d\nu\,
\int_0^{R_{\rm max}(v)}
\frac{\beta(R)\, R}{\sqrt{\frac{GM}R \sin^2 i - v^2}} 
dR 
}.
\end{equation}
\end{widetext}
An efficient way to evaluate Eqn.\ \eqref{eq:BLvis} simultaneously for many different values of $\Delta t$ is by fast Fourier transform (FFT) over the frequency range $\nu_{\rm min}<\nu<\nu_{\rm max}$.  Given ${\cal V}(\bm{B},\Delta t)$, we then convolve $|{\cal V}|^2$ with the timing jitter for each telescope to compute the observed intensity correlations using Eqn.\ \eqref{eq:convolved_vis}, and then compute the Fisher matrix using Eqn.\ \eqref{eq:Fisherconvolveddefn}, generalized to an array of $n_t$ telescopes observing many independent frequency channels $\nu$ and baselines $\bm{B}$,  
\begin{equation} \label{eq:Fisherconvolvedarray}
F_{\alpha\beta} = \binom{n_t}{2} \sum_{\nu,\bm{B}}
\frac{\Gamma^2 T_{\rm obs}}4 
\int d\Delta t 
\frac{\partial{\cal C}}{\partial p_\alpha}
\frac{\partial{\cal C}}{\partial p_\beta}.
\end{equation}

\subsection{Numerical example}
\label{sec:RM_example}

The transfer function and the visibility both depend on the form of $\beta(R)$ in Eqn.\ \eqref{eq:profile}.  We'll consider a simple illustrative example, a power law $\beta\propto R^n$ between some inner and outer radii $R_{\rm in}$ and $R_{\rm out}$ \cite{Murray1997}.  For concreteness, we assume fiducial parameter values $n=2$ and $R_{\rm out}/R_{\rm in}=100$.

Let us consider IRAS 09149-6206 as a concrete example. Its redshift of $z=0.057$ corresponds to a comoving distance $d\approx 170\,h^{-1}$Mpc.  In the H$\alpha$ line at $\lambda=6560\,$\AA, its flux density reported in the BASS DR2 catalog \cite{BASS_DR2} is $F_\lambda \sim 10^{-13}\,{\rm erg}\,{\rm s}^{-1}{\rm cm}^{-2}$\AA$^{-1}$.  This is a bright source, but not uniquely exceptional, since other sources like NGC 7469 or 3C 273 exhibit similar flux density \cite{BASS_DR2,Dietrich1999}.  We will again consider a hypothetical array similar to the CTA South MST, with $n_t=14$ telescopes of area $A=88\,$m$^2$ each.  For this area and line brightness, each telescope observes
$d\Gamma/d\nu = 4\cdot 10^{-7}$.  Assuming timing jitter with FWHM of 30 ps gives $\sigma_t \approx 13\,$ps, and if $T_{\rm obs}=24\,$hr, then $T_{\rm obs}/\sigma_t \approx 7\cdot10^{15}$.  If we do not use spectroscopy, i.e.\ setting $n_c=1$, then Eqn.\ \eqref{eq:visfisher} gives $\sigma_{|V|^2}^{-1}=73$.  For moderate spectral resolution of ${\cal R}=5000$, a linewidth $\Delta v \approx 0.01\,c$ would give $n_c=50$ channels across the line, improving the expected SNR.

Next, let us evaluate the Fisher matrix from 
reverberation mapping, equation \eqref{eq:RMfisher}.
Let us assume the AGN lightcurve is monitored with a cadence of 1 week, observed for 200 times over 4 years.  Note that this lightcurve monitoring may be done at small telescopes, completely different than the array used for intensity interferometry.  Indeed, numerous RM monitoring campaigns for AGN are already ongoing \cite{SDSSRM,2023arXiv230501014S,OzDES}.  With 200 observations, we measure $n_\omega=100$ independent frequencies.  
If we set $N_c = 0.01 P_c$, and $N_b=0.1 P_c$, corresponding to noise of 10\% and 30\% for continuum and broad line, respectively, then we get $\sigma_D/D = 0.036$ for $n_c=50$, and $\sigma_D/D = 0.17$ for $n_c=1$, for each AGN.  Examining Fig.\ \ref{fig:distance}, we can see that for these parameters, the error on $H_0$ is limited by the reverberation mapping uncertainties, rather than the intensity interferometry observations.  Our assumed errors on the AGN variability appear reasonable for shallow SDSS observations of AGN lightcurves at Gpc distances \cite{SDSSRM}, and may be significantly smaller for deep observations of more nearby, apparently brighter AGN.  
With 5 such AGN observed with 50 channels, then adding these errors in quadrature to the sample variance from peculiar velocities shown in Fig.\ \ref{fig:Hubble},  we measure $H_0$ to about 3\%, which would distinguish the $\sim 10\%$ discrepancy in $H_0$ values by about $3\sigma$.  These errors could be further reduced by more intensive observing campaigns that would give higher SNR in both reverberation mapping and intensity interferometry, or simply by observing a larger sample of AGN.

\begin{figure}[t]
\centering
\includegraphics[width=0.48\textwidth]{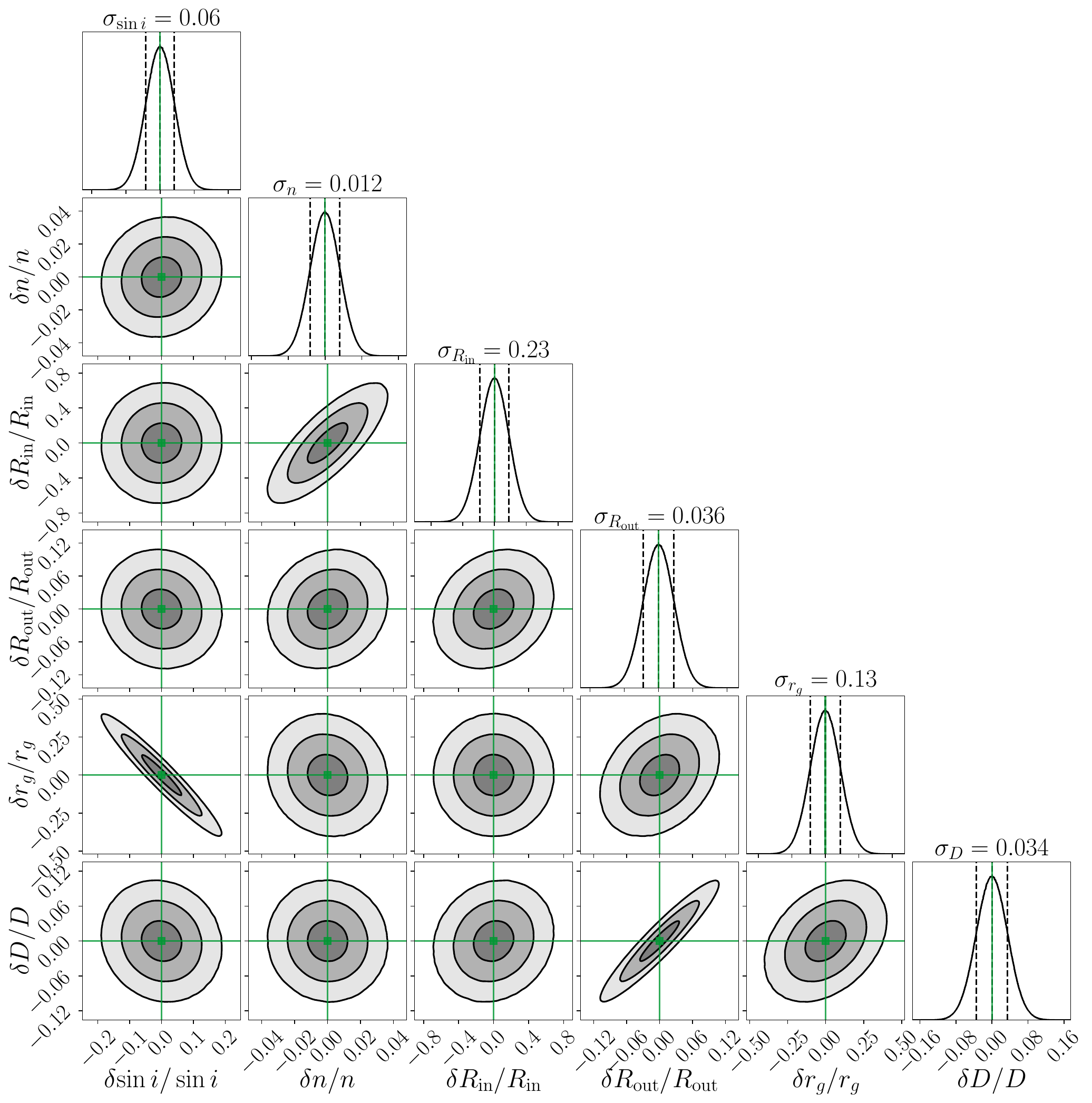}
\caption{
Error covariance for combined constraints from reverberation mapping and intensity interferometry, for observations with RM noise $N_b = 0.1\, P_b$, HBT SNR=73 per channel, and $n_c=50$ frequency channels as discussed in \S\ref{sec:RM_example}. The parameters of the BLR region chosen here are $\beta(R)\propto R^n$ and $\{\sin i, n, R_\text{in}, R_\text{out}\}=\{1/2, 2, 10^3\,GM, 10^5 \,GM\}.$
}
\label{fig:corner}
\end{figure}

\begin{figure}
\includegraphics[width=0.48\textwidth]{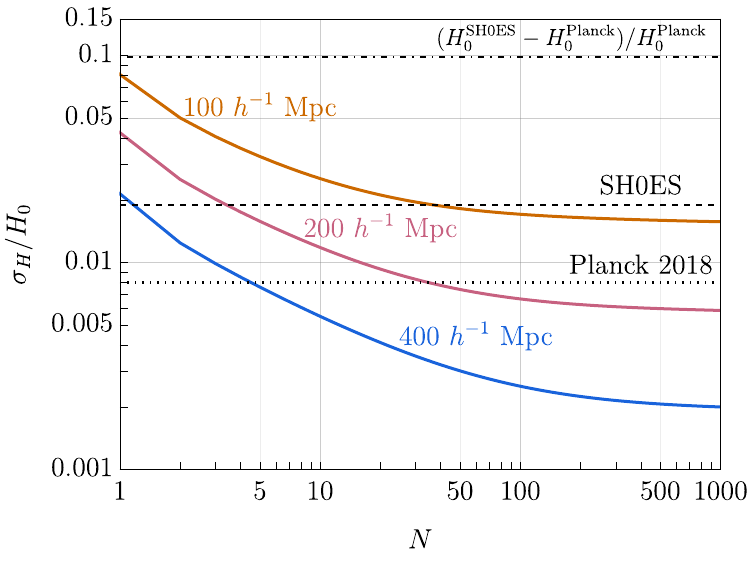}
\caption{Sample variance on $H_0$ from a sample of $N$ objects  with perfectly measured redshifts and distances.  We used halos with $M_{\rm vir}\geq 10^{13} h^{-1} M_\odot$ from the publicly available {\tt HugeMDPL} N-body simulation \cite{MultiDark}, and measured pairwise separations and relative velocities for $N$ randomly selected pairs found within separation $R<R_{\rm max}$, for different values of $R_{\rm max}$, measured in units of $h^{-1}$Mpc.  The behaviour is well described by $\sigma_H^2(R_{\rm max}) = \sigma_{\rm CV}^2(R_{\rm max}) + v^2 R_{\rm max}^{-2} N^{-1}$, where $v\approx 630\,$km/s and $\sigma_{\rm CV}$ is the cosmic variance on $H_0$ that arises from the fluctuations in the matter overdensity in specific regions of size $R_{\rm max}$. The dashed line corresponds to the fractional errors on $H_0$ from direct measurement using Cepheids and type Ia supernovae~\cite{Riess_2022}, while the dotted line shows the $H_0$ uncertainty from Planck observations of CMB anisotropies ~\cite{Planck2018}.} 
\label{fig:Hubble}
\end{figure}

In summary, the combination of intensity interferometry and reverberation mapping appears to be a promising method to measure absolute distances and the Hubble constant.  Our simple Fisher estimates of infinitely thin Keplerian disks provide motivation for more detailed future studies of whether Hubble forecasts remain promising for more realistic disks with turbulent velocities, nonzero thickness, proper treatment of line radiative transfer effects, etc.

\section{Discussion}

\subsection{Other AGN applications}

The previous sections describe two applications of intensity interferometry observations of AGN, for measuring the Hubble expansion rate and for studying the physics of AGN accretion disks.  In both examples, interesting results may be achieved using reasonable observation times (e.g., days-weeks) on feasible arrays, similar in size to existing atmospheric Cherenkov arrays.  Using larger arrays, or longer observing campaigns, other science applications also become feasible.

\subsubsection{Changes in the BLR}

The examples sketched above focus on measuring the time-independent profiles of AGN emission, but the time dependence of the signal also allows new probes of AGN physics.  Previous reverberation mapping studies have established that brighter AGN have larger BLRs \cite{Peterson2006}, so we might expect that when an individual AGN brightens, the size of its BLR will grow in size.  Intensity interferometry should be able to measure this effect.  One simple way to perform this measurement would be to split the BL visibility measurements, based on the brightness of the continuum emission at earlier times.  Since we measure $|{\cal V}|^2$ using two-point intensity correlations $\langle I(t) I(t+\delta t)\rangle$, we can think of splitting the signal as a kind of 3-point correlation, $\langle F[I_c(t-\Delta t)] I_{\rm BL}(t) I_{\rm BL}(t+\delta t)\rangle$, where $\delta t$ is of order picoseconds and $\Delta t$ is of order the RM lag, e.g.\ weeks or months, and $F$ is some function that selects or upweights times when the continuum intensity $I_c$ is in some specified range.  Using appropriately chosen weighting functions $F$, we can measure the size of the BL region as a function of the continuum brightness, and by varying $\Delta t$, we can observe how the BLR responds in time, e.g. detecting outgoing waves in the BLR as the central source varies in brightness.

\subsubsection{Continuum reverberation mapping}

Similar reasoning can also be applied to the continuum emission itself.  In the same way that line emission exhibits reverberations in response to continuum variability, the continuum emission also exhibits reverberations \cite{Sergeev2005,Cackett2007}.  Continuum reverberation mapping has recently been used to reveal surprising patterns of slow ingoing waves of temperature fluctuations in AGN accretion disks \cite{Neustadt2022MNRAS}.  Intensity interferometry can resolve these waves, revealing their structure and helping to elucidate their physics.  In \S\ref{sec:AGN} we discussed how 2-point intensity correlations $\langle I(t) I(t+\delta t)\rangle$ measure the mean profile of the accretion disk, and similarly to the above discussion, 3-point correlations $\langle I(t-\Delta t)] I(t) I(t+\delta t)\rangle$ can pick out the fluctuating part of the emission that correlates with variability at earlier times. This can be measured separately for different bands, e.g.\ to measure how the bluer and redder parts of the disk respond differently to UV or X-ray fluctuations \cite{Shappee2014,Fausnaugh2016} that can be contemporaneously observed with upcoming satellites \cite{QUVIK}.

\subsubsection{Tidal disruption events}

As the total collecting area of the telescope array grows in size, intensity interferometry observations may be performed for fainter sources, enabling additional applications.  For instance, when individual stars are tidally disrupted by supermassive black holes, they can generate outbursts called tidal disruption events (TDEs) that can last for weeks to months \cite{Stone2019,Gezari2021}.   The physics of these TDEs remains poorly understood at present, in particular the relative importance of different processes expected to be responsible for the visible light emitted by these events, like shocks from self-intersecting gas streams or the formation of a central accretion disk \cite{Piran2015,Metzger2016,Steinberg2022}.  Some TDEs exhibit late-time plateaus in their lightcurves for reasons that are not understood.  Long-baseline intensity interferometry can spatially resolve TDE emission at every stage of the event, helping to illuminate a phenomenon that is expected to be observed in large numbers when wide-area, high-cadence imaging surveys like LSST begin.

\subsubsection{Photon rings}

Another interesting science target would be the photon ring(s) that arise from strong-field gravitational lensing \cite{luminet1979,Gralla2019}.  The shape of the photon ring can be used to test modifications from GR in the strong-field regime \cite{Johnson2020,gralla2020b,Staelens:2023jgr,Cardenas-Avendano:2023obg,Cardenas-Avendano:2023dzo,Lupsasca:2024wkp}.  The rings are expected to occur in the vicinity of the horizon and the ISCO, with a fractional width of order 1\%  for the $n>1$ rings depending on the geometry of the accretion flow.  In principle, an intensity interferometer observing at $\lambda\approx 500\,$nm with baselines of order $B\sim 10^4\,$km could resolve angular scales $\lambda/B \approx 0.01\mu$as, adequate to resolve photon rings around nearby supermassive black holes.

In practice, this observation would be incredibly challenging, since 
the visibility on these baselines is expected to be extremely small, meaning that an enormous SNR would be required to unambiguously detect features of these rings.  Additionally, theoretically it is unclear if rings will be detectable from bright AGN.  If the accretion disk does not sharply truncate at some inner edge, but instead gradually diminishes towards the horizon, then it is possible that the gas flow will remain optically thick inside the ISCO, potentially blocking the photon rings \cite{Gammie1999,Agol2000}).  Future intensity interferometry observations of the innermost disk structure should help illuminate whether disks truncate at sufficiently large radii to avoid obscuring photon rings.  Similar observations may also be useful for detecting intensity correlations arising from multiple imaging in photon rings \cite{Hadar2021}.

\subsubsection{Closure phases}

At the very highest rates of photon arrivals, it becomes possible to detect not only the two-point correlations of intensity, but also higher-order correlations like the 3-point function.  
Here, we do not mean simply correlating 2-point intensity interferometry (on picosecond timescales) with lightcurve variations on much longer timescales as discussed earlier in this section, but instead we mean detecting picosecond correlations of intensities at 3 different telescopes.  Similar to the use of baseline triplets in amplitude interferometry, 3-point correlations allow the measurement of closure phases \cite{ThompsonBook,Lawson2000}, which provide phase information.

Repeating the arguments given in section \ref{sec:review}, the 3-point correlation observed over a window of time $\Delta t$ is
\begin{equation}
\langle \delta N_1\,\delta N_2\,\delta N_3\rangle
=  \frac{\Gamma^3\, \Delta t}{2 \Delta\nu^2} {\rm Re}({\cal V}_{12} {\cal V}_{23} {\cal V}_{31}).
\end{equation}
The Poisson  noise is $\langle\delta N^2\rangle^{3/2} = (\Gamma\,\Delta t)^{3/2}$.  Allowing for timing jitter that is Gaussian distributed with rms dispersion $\sigma_t$,  the signal to noise ratio in detecting the 3-point correlation becomes
\begin{eqnarray}
{\rm SNR} &=& \frac{1}{(768\pi^2)^{1/4}}
\left(\frac{d\Gamma}{d\nu}\right)^{3/2}
\left(\frac{T}{\sigma_t}\right)^{1/2}
\frac{1}{(\sigma_t \Delta\nu)^{1/2} }\nonumber \\
&&{\times\,\rm Re}({\cal V}_{12} {\cal V}_{23} {\cal V}_{31}).
\end{eqnarray}
Due to the extra 1/2-power of $d\Gamma/d\nu$, the SNR in the 3-point function is usually orders of magnitude smaller than the SNR in the 2-point function, meaning that detecting closure phases is only possible for either the brightest sources (like naked-eye stars) or for huge collecting areas.
For a numerical example, consider a source with apparent magnitude $V=12$, observed by an array with total area $n_t\,A=10^6 {\rm m}^2$, with spectral resolution $\Delta\nu=10^{12}\,$Hz, and timing jitter rms $\sigma_t=10\,$ps.   For these parameters, in $10^5\,$s observing time we reach ${\rm SNR}\approx 94$ for the 3-point correlation.  This means that detecting closure phases for extragalactic sources would require an observatory on the scale of the Square Kilometre Array \cite{SKA}, but composed of optical rather than radio telescopes.  This would obviously be a prodigious undertaking, but it is worth noting that enormous optical arrays have been proposed for studying exoplanet atmospheres \cite{LFAST}, and so large interferometric arrays may be able to use similar economies of scale.
The measurement of closure phases on multiple baselines would be transformative, since it allows for true imaging with interferometers \cite{Chael2018}.

\subsection{The future}

We conclude by speculating on possible strategies for implementing intensity interferometry arrays capable of studying extragalactic sources like AGN.  As mentioned in the introduction, high-precision timing technology for single photon detection is rapidly advancing, with significant recent breakthroughs in timing jitter, multiplexing, and cost.  Since we can expect the expense of these detectors to continue to decrease in the near future, for early generations of intensity interferometers, one possible strategy might be to initially neglect multiplexing, and instead instrument each telescope with a small number of detectors.  This would mean initially forgoing spectroscopy as a way of increasing the SNR, and instead focusing on bright line-emitters like AGN broad lines.  One advantage of neglecting spectroscopy is that the spectroscopic suppression effect discussed in appendix \ref{sec:suppression} will not be a concern, even for large-aperture mirrors.  

A related question is whether it is better to build many small telescopes, or a small number of large-area telescopes.  As we can see from Eqn.\ \eqref{eq:visfisher}, the SNR in detecting intensity correlations scales like $n_t\,A$, which is the total collecting area.  We would like to maximize the total SNR for a given total expenditure in cost.  Suppose that the total expenditure is $C$, and that the cost of each telescope may be written as $c_0 + c(A)$, where $A$ is the collecting area of the telescope, and $c_0$ is the fixed cost associated with each telescope, e.g.\ due to the detector, atomic clock, electronics, etc.  The cost of constructing a telescope of area $A$ may be some nonlinear function of $A$, so let us write $c(A)= c_0\times (A/A_0)^p$, where $A_0$ is the area that costs $c_0$, and $p$ is some exponent to allow for nonlinear dependence of the expense.  For a fixed total $C$, then the optimal number of telescopes $n_t$ and optimal area $A$ per telescope becomes $A = A_0\times (p-1)^{-1/p}$, and $n_t = (C/c_0)\times (1-p^{-1})$.  Unless $p\approx 1$, this says that generically the optimal telescope area is that whose cost is of order the other fixed costs.  In practice, it appears that $p\approx 1.3$ \cite{bely2003} 
for diffraction limited mirrors, suggesting that it may be preferable to construct a small number of large-area telescopes, i.e.\ to devote most of the expenditure to telescope area rather than other expenses.  On the other hand, diffraction-limited mirrors are not required to measure intensity interferometry, as demonstrated by Cherenkov arrays \cite{kieda2021veritasstellar,Zmija2023,MAGIC2024}.  To achieve picosecond-level timing, mirror blanks with $\sim$mm surface errors may be adequate; see Appendix \ref{sec:aberrations} for a discussion of this as well as other technical requirements to implement intensity interferometers.
This analysis is obviously extremely crude, and its conclusions could change as the costs of the constituent components of the interferometric array evolve over time.   It may also be worth considering even more  speculative ideas, such as using refractors like Fresnel lenses instead of reflectors to achieve large collecting areas \cite{Milster2020,Kim:20}.

Independent of the exact choice of design, we stress that the array sizes needed to achieve the science objectives discussed above are not wildly futuristic.  Intensity interferometry has already been demonstrated on Cherenkov arrays \cite{kieda2021veritasstellar,Zmija2023,MAGIC2024}, and is expected to be demonstrated on the CTA array, whose size we have adopted as our fiducial example.  Similarly, intensity interferometry has been detected using SPAD detectors with timing jitter similar to what we assumed \cite{horch2022observations}.  As we discussed above, construction of an intensity interferometer of this scale would provide myriad benefits to AGN science, cosmology, and strong-field gravity, and potentially other targets like supernovae ~\cite{Chen2024} or microarcsecond astrometry \cite{Stankus_2022,chen2023astrometry,vantilburg2023,galanis2023extendedpath}. 


\begin{acknowledgments}
We thank Masha Baryakhtar, Chad Bender, Daniel Ega\~na-Ugrinovic, Jacob Isbell, Yue Shen, Jon Sievers, Aaron Tohuvavohu, Ken Van Tilburg, and Tanya Zelevinsky for helpful discussions.
Research at Perimeter Institute is supported in part by the Government of Canada through the Department of Innovation, Science and Economic Development Canada and by the Province of Ontario through the Ministry of Colleges and Universities. This research was enabled in part by resources provided by Compute Ontario and Compute Canada. This work was supported by grants from the Simons Foundation (MP-SCMPS-00001470) and the National Science Foundation (PHY-230919).  The CosmoSim database used in this paper is a service by the Leibniz-Institute for Astrophysics Potsdam (AIP). The MultiDark database was developed in cooperation with the Spanish MultiDark Consolider Project CSD2009-00064.
This work has made use of the {\tt GSL} \cite{GSL}, {\tt FFTW} \cite{FFTW}, and {\tt CUBA} \cite{CUBA} numerical libraries, and we thank the respective authors for making their software publicly available. 
We have also used the VizieR catalogue access tool \cite{vizier}, the Astrophysics Data Service (\href{http://adsabs.harvard.edu/abstract_service.html}{\tt ADS}), and \href{https://arxiv.org}{\tt arXiv} preprint repository extensively during this project and the writing of the paper.

\end{acknowledgments}

\appendix

\section{Signature of a hole} \label{sec:hole}

In section \ref{sec:edge} we found that the radius of the dark area (``hole'') in the Shakura-Sunyaev profile is precisely encoded in the observed visibility amplitudes.  This result is potentially intriguing for accretion disk physics, since it is somewhat controversial whether physically self-consistent disks should have sharp inner edges bounding interior holes, or if instead the disk profile more gradually decreases inside the ISCO (e.g., \cite{Gammie1999,Agol2000,Afshordi2003,Shafee2008,Beckwith2008}).
Since our Fisher matrix estimate is model dependent, in this section we explore in a more model-independent context whether visibility amplitude measurements may be capable of measuring the presence of a hole.  The basic idea is that a profile with a hole can be regarded as the sum of a wider, hole-free profile (``main component'') and a narrower profile of negative brightness (``hole component'').  The main component has more total flux but varies more slowly, so its complex visibility  starts larger and falls off more rapidly.  The hole component has less total flux but varies more rapidly, so its complex visibility starts smaller, but also falls off more slowly.  The two components have opposite sign, so one finds a  ``null'' in the visibility (place where it vanishes) where the two components are first of comparable importance.  At larger baselines the hole begins to dominate, and the visibility should here encode its properties.

\begin{figure}
\includegraphics[width=\linewidth]{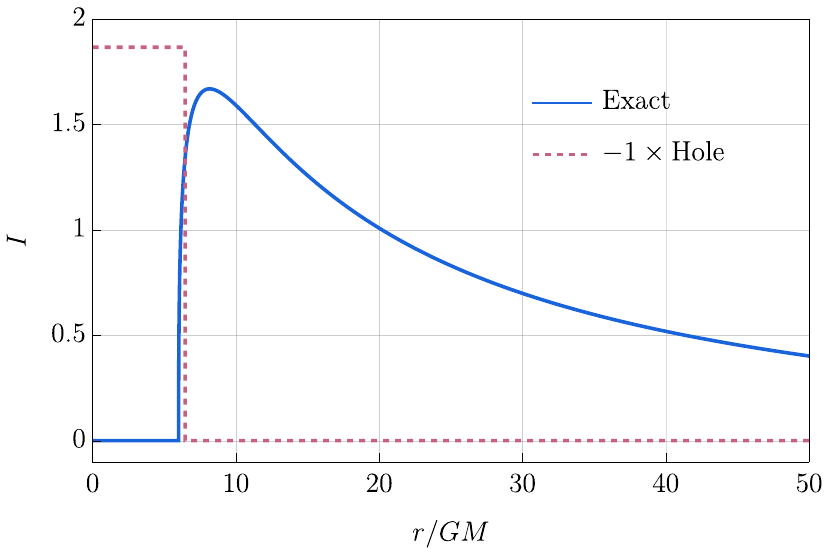}
\includegraphics[width=\linewidth]{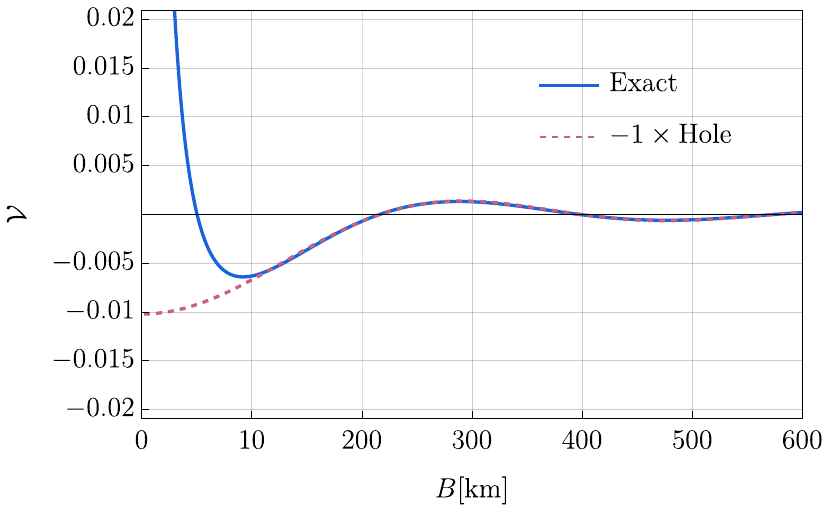}
\caption{A canonical Shakura-Sunyaev profile (defined in Eqn.\ \eqref{eq:SSprofile} with parameters given in Sec.~\ref{sec:SS}) together with the best fitting hole profile Eqn.\ \eqref{eq:holeProfile} using baselines 70km-200km for the fit.}\label{fig:holeWorks}
\end{figure}

We can test this idea by modeling the hole component as a sharp disk of negative brightness,
\begin{align}\label{eq:holeProfile}
I(r) = -I_{\rm hole} \Theta(r-R_{\rm hole}),
\end{align}
whose un-normalized visibility $V$ (as defined in Eqn.\ \eqref{eq:vis}) is 
\begin{align}
    V_{\rm hole} = -2 \pi I_{\rm hole} R_{\rm hole}^2 \frac{J_1(2 \pi u R_{\rm hole})}{2\pi u R_{\rm hole}},
\end{align}
where $J_n$ is the Bessel function of order $n$.  For the SS profile, we expect the hole to have radius 
$R_{\rm hole} \approx R_{\rm in}$ and brightness near  the maximum value of the SS profile. 
 Fig.~\ref{fig:holeWorks} shows the full visibility and best-fitting hole visibility for our canonical parameters, confirming the validity of the ``disk plus hole'' heuristic in this case.

\begin{figure}
\includegraphics[width=\linewidth]{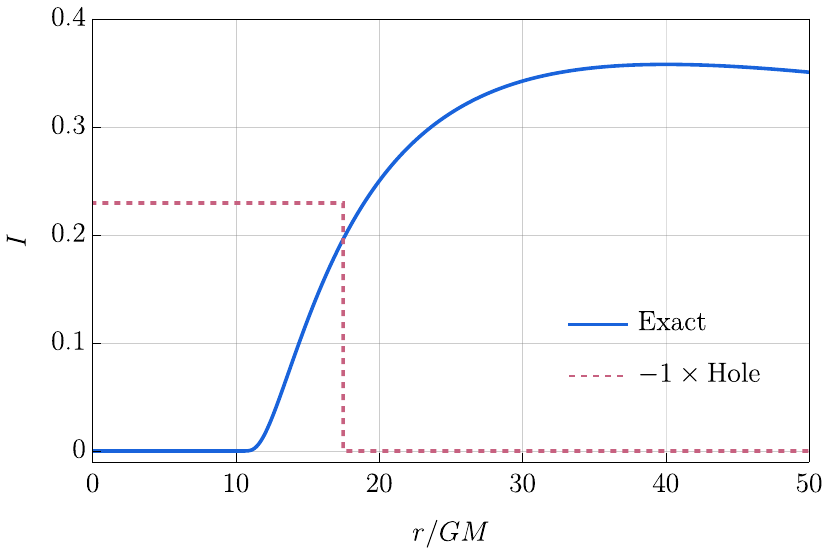}
\includegraphics[width=\linewidth]{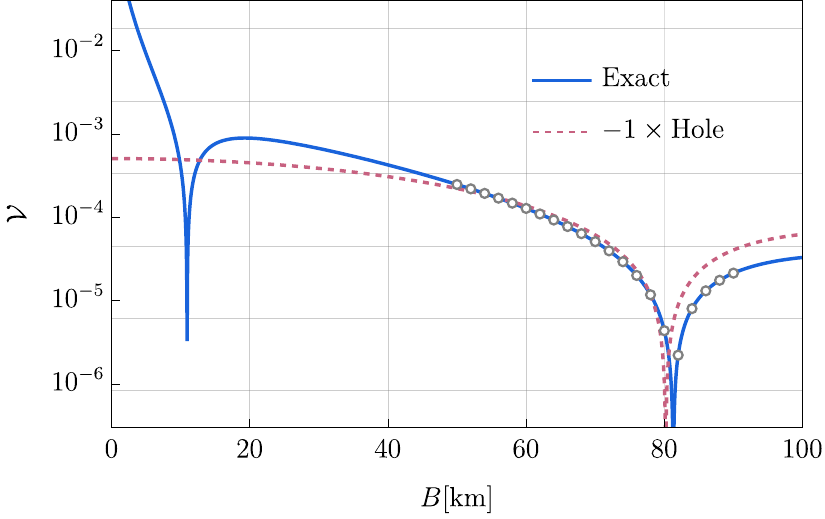}
\caption{Hole fit for a profile with a smoother drop to zero brightness.  The parameters are $(n,m,R_{\rm in},R_0)=(2,4,10,90)$.  The points used for the fit are shown as gray circles.
 }\label{fig:fitPlots}
\end{figure}

The SS profile has a rather sharp dropoff to zero brightness, making the hole signature especially clear.  To test whether this heuristic could be useful in a broader context, we generalize the SS profile \eqref{eq:SSprofile} to allow a smoother rise from zero.  This is easily done by adding another parameter, $m$, to \eqref{eq:fSS},
\begin{equation}\label{eq:fSS2}
f(R) = 
\left[\left(\frac{R_0}{R}\right)^n\,
\left(1-
\sqrt{\frac{R_{\rm in}}{R}}
\right)^m
\right]^{-1/4}.
\end{equation}
We perform a parameter survey in the ranges $n \in [2,6]$, $m\in [2,6]$, $R_{\rm in}\in [2,10]\, GM$, and $R_0\in [30,90]\, GM$.  The procedure for each parameter choice is the following.  We compute the visibility amplitude and find the first two nulls. We find the best-fitting hole visibility amplitude using baselines between the first and second nulls, including some points around the second null.  We record the results and make the plots shown in Fig.~\ref{fig:fitPlots}.  We visually inspect the plots and confirm that in all cases, the visibility fit is qualitatively acceptable and the inferred hole radius and brightness are qualitatively consistent with the profile.  The figure shows one of the worst-fitting examples in the parameter survey.  Even in this case, the presence of a hole is inferred with reasonable parameters.

We also explored this fitting procedure with profiles that do not feature a hole; in these cases the second null (if any) has very low visibility amplitude and any inferred hole flux is correspondingly negligible.  We conclude that the general approach of comparing the second null in the visibility amplitude to that of a hole is a robust way to detect and constrain holes in smooth axisymmetric profiles.  

Real profiles will depart from these assumptions in a few important ways.  Even a perfectly axisymmetric disk will produce a nonaxisymmetric profile when viewed at non-zero inclination. 
 The observational appearance in this case was  explored in Refs.~ \cite{luminet1979,hollywood1997,beckwith2005}.  There are two important inclination-dependent effects: Doppler boosting from orbiting matter introduces a brightness variation around the disk, and line-of-sight projection effects give the hole (and in general the level sets of intensity---see e.g. Fig.~6 of Ref.~\cite{gralla2020}) an an oblong shape.  For small observer inclinations, the hole remains roughly circular and the Doppler boosting adds a subdominant dipolar component ($\propto \cos \theta$, if $\theta$ is angle on the image) to the dominant axisymmetric component.  The main effect of the dipole is to partially ``fill in'' the null so it is not perfectly deep---the amplitude does not dip all the way to zero.  However, the location of the null is determined by the dominant axisymmetric component, and we expect that a hole can be robustly detected for small inclinations.  The situation is less obvious for moderate to extreme inclinations, but we note that significant additional information can be gained by observing the visibility amplitude at different angles (baseline orientations) in the visibility plane.

Real profiles will also not be as perfectly smooth as ours; there will be spatial (and temporal) variation on smaller scales due to stochastic and/or turbulent fluctuations in the disk.  Such features will introduce power on larger baselines, with the potential to contaminate the disk signature.  Again, the main effect would be to partially ``fill in'' the null.  Absent a fundamental understanding of small-scale plasma processes, it is difficult to predict the size of these effects.  However, we note that particularly large fluctuations would still be time-variable, and information from multiple observations could help remove them.

Finally, real profiles may also contain a ``photon ring,'' which is the highly-demagnified image of the back side of the disk \cite{luminet1979,gralla2020}.  (Higher-order images will have negligible flux density.)  This feature is a generic consequence of strong gravitational lensing by black holes, and may be visible through the hole in the disk for profiles.  The photon ring will also contribute power on long baselines.  Modeling the photon ring as a delta function, its visibility will fall off as $B^{-1/2}$, as compared to the $B^{-3/2}$ falloff of a hole, so there is some potential for distinguishing the two features.  However, in numerical experiments for axisymmetric profiles in the Kerr spacetime we have found that, for the parameter ranges where the photon ring is visible, there is no range of baselines where it dominates.  In practice, the regime where the theoretical falloffs ($B^{-1/2}$ and $B^{-1/3}$) occur is too narrow for the two to be distinguished.  Instead, we find that the hole component dominates near the second null (as expected), whereas both components contribute at longer baselines.  In principle one could glean the presence of two components by observing the beating between them, but this would take a truly exquisite observation.

\section{Spectroscopic suppression of intensity correlations} \label{sec:suppression}

Any spectroscopic element produces a series of copies of the image on the sky at different wavelengths. These wavelengths are spatially separated on the photodetector array, with a spacing that depends on the resolution. Crucially, these images can be spatially separated even within a \emph{single} pixel.

For a point-like source, what is produced is a series of Airy patterns, whose width is set by the diffraction limit of the telescope. If a pixel is large enough to contain spatially non-overlapping Airy patterns of different wavelengths, the correlation function does not approach unity in the limit of infinitely good time-resolution. Intuitively, this is because two frequencies that are $\lesssim\sigma_t^{-1}$ apart are phase-coherent on timescales $\sim \sigma_t$ on which the photodetector clicks, and so contribute to the correlation function by surviving the time-smearing. However, if the Airy patterns of these frequencies become spatially separated, even on the same pixel, this contribution is lost, as the product of their electric fields vanishes. In contrast, spatially separating frequencies that are $\gtrsim \sigma^{-1}_t$ apart has no such effect, because these frequencies are incoherent and their product would not survive the time-smearing even if they were spatially overlapping. 

The simple solution to this would be to make pixels subtending an angle as large as the diffraction limit of the telescope. However, due to atmospheric fluctuations, which occur on ms timescales, any pixel (or collection of pixels) devoted to a single spectral channel has to be at least as large as the seeing angle. In the limit where the spectroscopic resolution spatially separates frequencies $\lesssim\sigma_t^{-1}$, the mismatch between the seeing angle and the diffraction limit of the telescope is the resulting suppression. The correlation function at zero baseline and zero time-delay becomes
\begin{equation}
    \mathcal{C}\approx\min\left\{1,\frac{1}{\sqrt{4\pi}\Delta\nu\sigma_t},\frac{\theta_\text{diffraction}}{\theta_\text{seeing}}\right\}.
\end{equation}
We refer the reader to section 3.4 of \cite{galanis2023extendedpath} for a detailed proof of this effect and for a discussion of potential mitigating solutions.

\section{Atmospheric and instrumental aberrations}
\label{sec:aberrations}

One key feature of intensity interferometry, which makes it attractive in terms of cost and implementation, is its broad insensitivity to atmospheric and instrumental aberrations. 

First, light paths from different angles are affected by \emph{spatial} fluctuations of the atmospheric index of refraction. Intensity interferometry is only sensitive to the light-path \emph{difference} of the two simultaneous wavefronts at \emph{each} telescope. Thus, within the isoplanatic patch angle (several arcseconds in the optical) where these fluctuations are correlated, atmospheric distortions are irrelevant~\cite{bely2003,galanis2023extendedpath,vantilburg2023}.

Second, temporal fluctuations of the index of refraction affect light paths in different telescopes in an uncorrelated way, introducing an effective time-jitter. Path delays due to the atmosphere are calculated~\cite{mendes2004high,hulley2007ray,stuhl2021atmospheric} to be less than $\sim$mm (3ps) at zenith and $\sim$10mm (30ps) at low elevation angles, thus unimportant for the photodetectors considered here. Indeed, the intensity interferometer at the Southern Connecticut State University uses $50$ps SPADs without issue~\cite{horch2022observations}.

Third, for single channel intensity interferometry, instrumental aberrations can similarly be tolerated if they do not introduce differential path-delays longer than mm-10mm (3ps-30ps), similar to the surface errors of mirror blanks.  This could potentially reduce the production costs of the required mirrors by a significant factor.  If spectroscopy is used, tighter tolerances for the mirror surface may be needed, as discussed at the end of \S\ref{sec:review}. Even then, these are much milder than the need of diffraction limited mirrors~\cite{galanis2023extendedpath}.

The fundamental path-length uncertainty of an intensity interferometer results from the photodetectors' time-jitter and, in this work, is on the order of 10mm. However, other experimental errors, such as drifting of the telescope baseline and of the local clocks, may result in a larger \emph{effective} time-jitter.

Precise geodesy is set to improve with the next-generation VLBI Global Observing System (VGOS)~\cite{hase2012emerging,behrend2019roll}, which has already demonstrated a root-mean-square deviation of 1.6mm~\cite{niell2018demonstration} and is planning to reach a precision of 1mm~\cite{rioja2020precise}. This is already better than the required $9$mm precision for a 30ps photodetector.  Baseline drift errors can, therefore, be eliminated.

Next, maintaining a time-resolution $\sigma_t \sim 10\,$ps during a 3hr observing night requires stability of the local clocks to about 1 part in $10^{15}$. This is achievable with commercial hydrogen masers~\cite{mhm,CH1-76A}, already employed by the Event Horizon Telescope~\cite{EHT_instrumentation}. If correlations can be found (i.e. SNR $>1$) faster than 3hr, then the local clocks can be re-calibrated over shorter periods. This can be achieved with much cheaper Rb or Cs clocks \cite{8040C,3352A,5071A}, or possibly iodine clocks \cite{Roslund2023},
and thus, the required synchronization equipment can constitute only a fraction of the overall cost of an intensity interferometry array. 
Another potential solution is to transmit precise timing over fiber optic, which has been demonstrated over distances of tens of km \cite{fiber_transmit_time}, potentially even hundreds of km \cite{Koke_2021}.
The choice of synchronization equipment will depend not only on the cost per clock, but also on the collecting area per observing site, the use of spectroscopy, the spacing of the baselines and the science targets, among other parameters. We thus leave a more detailed study to future work. 

Finally, it is worthwhile to compare the measured photon flux of real intensity interferometry observations to the expected photon flux for the sources considered, as instrumental imperfections may hamper the total number of recorded photons and, hence, the SNR. To this end, we compare to the results of the group in Universit\'{e} C\^{o}te d' Azur~\cite{Guerin_time,Guerin_space}. The expected photon flux in their measurement, given by the apparent magnitude of the observed stars, differs from the measured one by a factor of $\sim 7$~\footnote{They also use a polarizer, which cuts the expected photon flux in half and which may account for some flux loss too. A polarizer does not change the SNR for intensity interferometry, so we ignore it in this estimate}. However, all sources of flux loss are accounted for. For instance, for $\alpha$ CMi (I band magnitude of $-0.28$) they suffer the following flux losses: $40\%$ from the use of two narrow-band filters, $44\%$ due to optics imperfections, $34\%$ due to the use of fibers, while their SPAD has a quantum efficiency of $0.64$. Accounting for these factors reconciles the average recorded photon rate to the expected one. We expect that optimization of optical configurations will minimize flux loss. For instance, filters may be unnecessary, especially if spectroscopy is employed, while single photon detectors with higher quantum efficiencies have been demonstrated~\cite{high_eta}, with ongoing technological improvements towards this goal~\cite{NISTQuantumDetectors}. We thus deem the forecasts of this work sufficient.

\bibliographystyle{apsrev4-2}

\newcommand{\aap}{Astron.\ Astrophys.}
\newcommand{\aaps}{Astron.\ Astrophys.\ Supp.\ Ser.}
\newcommand{\apjl}{\apj\ Lett.}
\newcommand{\apjs}{\apj\ Supp.\ Ser.}
\newcommand{\aj}{Astron.\ J.}
\newcommand{\araa}{Ann.\ Rev.\ Astron.\ Astrophys.}
\newcommand{\grl}{Geophysical Research Letters}
\newcommand{\jcap}{J.\ Cosm.\ Astropart.\ Phys.}
\newcommand{\mnras}{Mon.\ Not.\ Royal Astron.\ Soc.}
\newcommand{\pasp}{Proc.\  Astron.\ Soc.\ Pacific}
\bibliography{hbt}

\begin{thebibliography}{150}%
\makeatletter
\providecommand \@ifxundefined [1]{%
 \@ifx{#1\undefined}
}%
\providecommand \@ifnum [1]{%
 \ifnum #1\expandafter \@firstoftwo
 \else \expandafter \@secondoftwo
 \fi
}%
\providecommand \@ifx [1]{%
 \ifx #1\expandafter \@firstoftwo
 \else \expandafter \@secondoftwo
 \fi
}%
\providecommand \natexlab [1]{#1}%
\providecommand \enquote  [1]{``#1''}%
\providecommand \bibnamefont  [1]{#1}%
\providecommand \bibfnamefont [1]{#1}%
\providecommand \citenamefont [1]{#1}%
\providecommand \href@noop [0]{\@secondoftwo}%
\providecommand \href [0]{\begingroup \@sanitize@url \@href}%
\providecommand \@href[1]{\@@startlink{#1}\@@href}%
\providecommand \@@href[1]{\endgroup#1\@@endlink}%
\providecommand \@sanitize@url [0]{\catcode `\\12\catcode `\$12\catcode
  `\&12\catcode `\#12\catcode `\^12\catcode `\_12\catcode `\%12\relax}%
\providecommand \@@startlink[1]{}%
\providecommand \@@endlink[0]{}%
\providecommand \url  [0]{\begingroup\@sanitize@url \@url }%
\providecommand \@url [1]{\endgroup\@href {#1}{\urlprefix }}%
\providecommand \urlprefix  [0]{URL }%
\providecommand \Eprint [0]{\href }%
\providecommand \doibase [0]{https://doi.org/}%
\providecommand \selectlanguage [0]{\@gobble}%
\providecommand \bibinfo  [0]{\@secondoftwo}%
\providecommand \bibfield  [0]{\@secondoftwo}%
\providecommand \translation [1]{[#1]}%
\providecommand \BibitemOpen [0]{}%
\providecommand \bibitemStop [0]{}%
\providecommand \bibitemNoStop [0]{.\EOS\space}%
\providecommand \EOS [0]{\spacefactor3000\relax}%
\providecommand \BibitemShut  [1]{\csname bibitem#1\endcsname}%
\let\auto@bib@innerbib\@empty
\bibitem [{\citenamefont {{Born}}\ and\ \citenamefont
  {{Wolf}}(2019)}]{BornWolf}%
  \BibitemOpen
  \bibfield  {author} {\bibinfo {author} {\bibfnamefont {M.}~\bibnamefont
  {{Born}}}\ and\ \bibinfo {author} {\bibfnamefont {E.}~\bibnamefont
  {{Wolf}}},\ }\href {https://doi.org/10.1017/9781108769914} {\emph {\bibinfo
  {title} {{Principles of Optics}}}}\ (\bibinfo  {publisher} {Cambridge Univ.
  Pr.},\ \bibinfo {year} {2019})\BibitemShut {NoStop}%
\bibitem [{\citenamefont {{Thompson}}\ \emph {et~al.}(2017)\citenamefont
  {{Thompson}}, \citenamefont {{Moran}},\ and\ \citenamefont
  {{Swenson}}}]{ThompsonBook}%
  \BibitemOpen
  \bibfield  {author} {\bibinfo {author} {\bibfnamefont {A.~R.}\ \bibnamefont
  {{Thompson}}}, \bibinfo {author} {\bibfnamefont {J.~M.}\ \bibnamefont
  {{Moran}}},\ and\ \bibinfo {author} {\bibfnamefont {J.}~\bibnamefont
  {{Swenson}}, \bibfnamefont {George~W.}},\ }\href
  {https://doi.org/10.1007/978-3-319-44431-4} {\emph {\bibinfo {title}
  {{Interferometry and Synthesis in Radio Astronomy, 3rd Edition}}}}\ (\bibinfo
   {publisher} {{Springer}},\ \bibinfo {year} {2017})\BibitemShut {NoStop}%
\bibitem [{\citenamefont {{Lawson}}(2000)}]{Lawson2000}%
  \BibitemOpen
  \bibfield  {author} {\bibinfo {author} {\bibfnamefont {P.~R.}\ \bibnamefont
  {{Lawson}}},\ }\href@noop {} {\emph {\bibinfo {title} {Principles of Long
  Baseline Stellar Interferometry}}}\ (\bibinfo  {publisher} {NASA JPL},\
  \bibinfo {year} {2000})\BibitemShut {NoStop}%
\bibitem [{\citenamefont {{Labeyrie}}\ \emph {et~al.}(2014)\citenamefont
  {{Labeyrie}}, \citenamefont {{Lipson}},\ and\ \citenamefont
  {{Nisenson}}}]{Labeyrie.book}%
  \BibitemOpen
  \bibfield  {author} {\bibinfo {author} {\bibfnamefont {A.}~\bibnamefont
  {{Labeyrie}}}, \bibinfo {author} {\bibfnamefont {S.~G.}\ \bibnamefont
  {{Lipson}}},\ and\ \bibinfo {author} {\bibfnamefont {P.}~\bibnamefont
  {{Nisenson}}},\ }\href@noop {} {\emph {\bibinfo {title} {{An Introduction to
  Optical Stellar Interferometry}}}}\ (\bibinfo  {publisher} {Cambridge
  University Press},\ \bibinfo {year} {2014})\BibitemShut {NoStop}%
\bibitem [{\citenamefont {{CHIME/FRB Collaboration}}\ \emph
  {et~al.}(2018)\citenamefont {{CHIME/FRB Collaboration}}, \citenamefont
  {{Amiri}}, \citenamefont {{Bandura}}, \citenamefont {{Berger}} \emph
  {et~al.}}]{CHIMEFRB}%
  \BibitemOpen
  \bibfield  {author} {\bibinfo {author} {\bibnamefont {{CHIME/FRB
  Collaboration}}}, \bibinfo {author} {\bibfnamefont {M.}~\bibnamefont
  {{Amiri}}}, \bibinfo {author} {\bibfnamefont {K.}~\bibnamefont {{Bandura}}},
  \bibinfo {author} {\bibfnamefont {P.}~\bibnamefont {{Berger}}}, \emph
  {et~al.},\ }\href {https://doi.org/10.3847/1538-4357/aad188} {\bibfield
  {journal} {\bibinfo  {journal} {\apj}\ }\textbf {\bibinfo {volume} {863}},\
  \bibinfo {eid} {48} (\bibinfo {year} {2018})},\ \Eprint
  {https://arxiv.org/abs/1803.11235} {arXiv:1803.11235 [astro-ph.IM]}
  \BibitemShut {NoStop}%
\bibitem [{\citenamefont {{Event Horizon Telescope Collaboration}}\ \emph
  {et~al.}(2019)\citenamefont {{Event Horizon Telescope Collaboration}},
  \citenamefont {{Akiyama}}, \citenamefont {{Alberdi}}, \citenamefont {{Alef}}
  \emph {et~al.}}]{EHT}%
  \BibitemOpen
  \bibfield  {author} {\bibinfo {author} {\bibnamefont {{Event Horizon
  Telescope Collaboration}}}, \bibinfo {author} {\bibfnamefont
  {K.}~\bibnamefont {{Akiyama}}}, \bibinfo {author} {\bibfnamefont
  {A.}~\bibnamefont {{Alberdi}}}, \bibinfo {author} {\bibfnamefont
  {W.}~\bibnamefont {{Alef}}}, \emph {et~al.},\ }\href
  {https://doi.org/10.3847/2041-8213/ab0e85} {\bibfield  {journal} {\bibinfo
  {journal} {\apjl}\ }\textbf {\bibinfo {volume} {875}},\ \bibinfo {eid} {L4}
  (\bibinfo {year} {2019})},\ \Eprint {https://arxiv.org/abs/1906.11241}
  {arXiv:1906.11241 [astro-ph.GA]} \BibitemShut {NoStop}%
\bibitem [{\citenamefont {{ALMA Partnership}}\ \emph
  {et~al.}(2015)\citenamefont {{ALMA Partnership}}, \citenamefont {{Brogan}},
  \citenamefont {{P{\'e}rez}}, \citenamefont {{Hunter}} \emph {et~al.}}]{ALMA}%
  \BibitemOpen
  \bibfield  {author} {\bibinfo {author} {\bibnamefont {{ALMA Partnership}}},
  \bibinfo {author} {\bibfnamefont {C.~L.}\ \bibnamefont {{Brogan}}}, \bibinfo
  {author} {\bibfnamefont {L.~M.}\ \bibnamefont {{P{\'e}rez}}}, \bibinfo
  {author} {\bibfnamefont {T.~R.}\ \bibnamefont {{Hunter}}}, \emph {et~al.},\
  }\href {https://doi.org/10.1088/2041-8205/808/1/L3} {\bibfield  {journal}
  {\bibinfo  {journal} {\apjl}\ }\textbf {\bibinfo {volume} {808}},\ \bibinfo
  {eid} {L3} (\bibinfo {year} {2015})},\ \Eprint
  {https://arxiv.org/abs/1503.02649} {arXiv:1503.02649 [astro-ph.SR]}
  \BibitemShut {NoStop}%
\bibitem [{\citenamefont {{Hezaveh}}\ \emph {et~al.}(2016)\citenamefont
  {{Hezaveh}}, \citenamefont {{Dalal}}, \citenamefont {{Marrone}} \emph
  {et~al.}}]{Hezaveh2016}%
  \BibitemOpen
  \bibfield  {author} {\bibinfo {author} {\bibfnamefont {Y.~D.}\ \bibnamefont
  {{Hezaveh}}}, \bibinfo {author} {\bibfnamefont {N.}~\bibnamefont {{Dalal}}},
  \bibinfo {author} {\bibfnamefont {D.~P.}\ \bibnamefont {{Marrone}}}, \emph
  {et~al.},\ }\href {https://doi.org/10.3847/0004-637X/823/1/37} {\bibfield
  {journal} {\bibinfo  {journal} {\apj}\ }\textbf {\bibinfo {volume} {823}},\
  \bibinfo {eid} {37} (\bibinfo {year} {2016})},\ \Eprint
  {https://arxiv.org/abs/1601.01388} {arXiv:1601.01388 [astro-ph.CO]}
  \BibitemShut {NoStop}%
\bibitem [{\citenamefont {{Hanbury Brown}}\ and\ \citenamefont
  {{Twiss}}(1954)}]{HanburyBrown1954}%
  \BibitemOpen
  \bibfield  {author} {\bibinfo {author} {\bibfnamefont {R.}~\bibnamefont
  {{Hanbury Brown}}}\ and\ \bibinfo {author} {\bibfnamefont {R.~G.}\
  \bibnamefont {{Twiss}}},\ }\href {https://doi.org/10.1080/14786440708520475}
  {\bibfield  {journal} {\bibinfo  {journal} {Philosophical Magazine}\ }\textbf
  {\bibinfo {volume} {45}},\ \bibinfo {pages} {663} (\bibinfo {year}
  {1954})}\BibitemShut {NoStop}%
\bibitem [{\citenamefont {{Hanbury Brown}}(1956)}]{HanburyBrown1956}%
  \BibitemOpen
  \bibfield  {author} {\bibinfo {author} {\bibfnamefont {R.}~\bibnamefont
  {{Hanbury Brown}}},\ }\href {https://doi.org/10.1038/1781046a0} {\bibfield
  {journal} {\bibinfo  {journal} {\nat}\ }\textbf {\bibinfo {volume} {178}},\
  \bibinfo {pages} {1046} (\bibinfo {year} {1956})}\BibitemShut {NoStop}%
\bibitem [{\citenamefont {{Purcell}}(1956)}]{Purcell1956}%
  \BibitemOpen
  \bibfield  {author} {\bibinfo {author} {\bibfnamefont {E.~M.}\ \bibnamefont
  {{Purcell}}},\ }\href {https://doi.org/10.1038/1781449a0} {\bibfield
  {journal} {\bibinfo  {journal} {\nat}\ }\textbf {\bibinfo {volume} {178}},\
  \bibinfo {pages} {1449} (\bibinfo {year} {1956})}\BibitemShut {NoStop}%
\bibitem [{\citenamefont {{Hanbury Brown}}\ and\ \citenamefont
  {{Twiss}}(1957)}]{1957RSPSA.242..300B}%
  \BibitemOpen
  \bibfield  {author} {\bibinfo {author} {\bibfnamefont {R.}~\bibnamefont
  {{Hanbury Brown}}}\ and\ \bibinfo {author} {\bibfnamefont {R.~Q.}\
  \bibnamefont {{Twiss}}},\ }\href {https://doi.org/10.1098/rspa.1957.0177}
  {\bibfield  {journal} {\bibinfo  {journal} {Proceedings of the Royal Society
  of London Series A}\ }\textbf {\bibinfo {volume} {242}},\ \bibinfo {pages}
  {300} (\bibinfo {year} {1957})}\BibitemShut {NoStop}%
\bibitem [{\citenamefont {{Twiss}}\ and\ \citenamefont
  {{Little}}(1957)}]{Twiss1957}%
  \BibitemOpen
  \bibfield  {author} {\bibinfo {author} {\bibfnamefont {R.~Q.}\ \bibnamefont
  {{Twiss}}}\ and\ \bibinfo {author} {\bibfnamefont {A.~G.}\ \bibnamefont
  {{Little}}},\ }\href {https://doi.org/10.1038/180324a0} {\bibfield  {journal}
  {\bibinfo  {journal} {\nat}\ }\textbf {\bibinfo {volume} {180}},\ \bibinfo
  {pages} {324} (\bibinfo {year} {1957})}\BibitemShut {NoStop}%
\bibitem [{\citenamefont {{Hanbury Brown}}\ and\ \citenamefont
  {{Twiss}}(1958{\natexlab{a}})}]{1958RSPSA.243..291B}%
  \BibitemOpen
  \bibfield  {author} {\bibinfo {author} {\bibfnamefont {R.}~\bibnamefont
  {{Hanbury Brown}}}\ and\ \bibinfo {author} {\bibfnamefont {R.~Q.}\
  \bibnamefont {{Twiss}}},\ }\href {https://doi.org/10.1098/rspa.1958.0001}
  {\bibfield  {journal} {\bibinfo  {journal} {Proceedings of the Royal Society
  of London Series A}\ }\textbf {\bibinfo {volume} {243}},\ \bibinfo {pages}
  {291} (\bibinfo {year} {1958}{\natexlab{a}})}\BibitemShut {NoStop}%
\bibitem [{\citenamefont {{Hanbury Brown}}\ and\ \citenamefont
  {{Twiss}}(1958{\natexlab{b}})}]{1958RSPSA.248..199B}%
  \BibitemOpen
  \bibfield  {author} {\bibinfo {author} {\bibfnamefont {R.}~\bibnamefont
  {{Hanbury Brown}}}\ and\ \bibinfo {author} {\bibfnamefont {R.~Q.}\
  \bibnamefont {{Twiss}}},\ }\href {https://doi.org/10.1098/rspa.1958.0239}
  {\bibfield  {journal} {\bibinfo  {journal} {Proceedings of the Royal Society
  of London Series A}\ }\textbf {\bibinfo {volume} {248}},\ \bibinfo {pages}
  {199} (\bibinfo {year} {1958}{\natexlab{b}})}\BibitemShut {NoStop}%
\bibitem [{\citenamefont {{Hanbury Brown}}\ and\ \citenamefont
  {{Twiss}}(1958{\natexlab{c}})}]{1958RSPSA.248..222B}%
  \BibitemOpen
  \bibfield  {author} {\bibinfo {author} {\bibfnamefont {R.}~\bibnamefont
  {{Hanbury Brown}}}\ and\ \bibinfo {author} {\bibfnamefont {R.~Q.}\
  \bibnamefont {{Twiss}}},\ }\href {https://doi.org/10.1098/rspa.1958.0240}
  {\bibfield  {journal} {\bibinfo  {journal} {Proceedings of the Royal Society
  of London Series A}\ }\textbf {\bibinfo {volume} {248}},\ \bibinfo {pages}
  {222} (\bibinfo {year} {1958}{\natexlab{c}})}\BibitemShut {NoStop}%
\bibitem [{\citenamefont {{Hanbury Brown}}(1974)}]{HanburyBrown1974}%
  \BibitemOpen
  \bibfield  {author} {\bibinfo {author} {\bibfnamefont {R.}~\bibnamefont
  {{Hanbury Brown}}},\ }\href@noop {} {\emph {\bibinfo {title} {{The intensity
  interferometer. Its applications to astronomy}}}}\ (\bibinfo  {publisher}
  {{Halsted Press}},\ \bibinfo {year} {1974})\BibitemShut {NoStop}%
\bibitem [{\citenamefont {{Twiss}}(1969)}]{Twiss1969}%
  \BibitemOpen
  \bibfield  {author} {\bibinfo {author} {\bibfnamefont {R.~Q.}\ \bibnamefont
  {{Twiss}}},\ }\href {https://doi.org/10.1080/713818198} {\bibfield  {journal}
  {\bibinfo  {journal} {Optica Acta}\ }\textbf {\bibinfo {volume} {16}},\
  \bibinfo {pages} {423} (\bibinfo {year} {1969})}\BibitemShut {NoStop}%
\bibitem [{\citenamefont {{Baym}}(1998)}]{Baym1998}%
  \BibitemOpen
  \bibfield  {author} {\bibinfo {author} {\bibfnamefont {G.}~\bibnamefont
  {{Baym}}},\ }\href {https://doi.org/10.48550/arXiv.nucl-th/9804026}
  {\bibfield  {journal} {\bibinfo  {journal} {Acta Physica Polonica B}\
  }\textbf {\bibinfo {volume} {29}},\ \bibinfo {pages} {1839} (\bibinfo {year}
  {1998})},\ \Eprint {https://arxiv.org/abs/nucl-th/9804026}
  {arXiv:nucl-th/9804026 [nucl-th]} \BibitemShut {NoStop}%
\bibitem [{\citenamefont {{Loudon}}(2000)}]{Loudon2000}%
  \BibitemOpen
  \bibfield  {author} {\bibinfo {author} {\bibfnamefont {R.}~\bibnamefont
  {{Loudon}}},\ }\href@noop {} {\emph {\bibinfo {title} {{The Quantum Theory of
  Light}}}}\ (\bibinfo  {publisher} {{Oxford University Press}},\ \bibinfo
  {year} {2000})\BibitemShut {NoStop}%
\bibitem [{\citenamefont {{Fienup}}(1982)}]{Fienup1982}%
  \BibitemOpen
  \bibfield  {author} {\bibinfo {author} {\bibfnamefont {J.~R.}\ \bibnamefont
  {{Fienup}}},\ }\href {https://doi.org/10.1364/AO.21.002758} {\bibfield
  {journal} {\bibinfo  {journal} {\ao}\ }\textbf {\bibinfo {volume} {21}},\
  \bibinfo {pages} {2758} (\bibinfo {year} {1982})}\BibitemShut {NoStop}%
\bibitem [{\citenamefont {{Dong}}\ \emph {et~al.}(2023)\citenamefont {{Dong}},
  \citenamefont {{Valzania}}, \citenamefont {{Maillard}}, \citenamefont
  {{Pham}}, \citenamefont {{Gigan}},\ and\ \citenamefont {{Unser}}}]{Dong2023}%
  \BibitemOpen
  \bibfield  {author} {\bibinfo {author} {\bibfnamefont {J.}~\bibnamefont
  {{Dong}}}, \bibinfo {author} {\bibfnamefont {L.}~\bibnamefont {{Valzania}}},
  \bibinfo {author} {\bibfnamefont {A.}~\bibnamefont {{Maillard}}}, \bibinfo
  {author} {\bibfnamefont {T.-a.}\ \bibnamefont {{Pham}}}, \bibinfo {author}
  {\bibfnamefont {S.}~\bibnamefont {{Gigan}}},\ and\ \bibinfo {author}
  {\bibfnamefont {M.}~\bibnamefont {{Unser}}},\ }\href
  {https://doi.org/10.1109/MSP.2022.3219240} {\bibfield  {journal} {\bibinfo
  {journal} {IEEE Signal Processing Magazine}\ }\textbf {\bibinfo {volume}
  {40}},\ \bibinfo {pages} {45} (\bibinfo {year} {2023})},\ \Eprint
  {https://arxiv.org/abs/2204.03554} {arXiv:2204.03554 [physics.optics]}
  \BibitemShut {NoStop}%
\bibitem [{\citenamefont {{Hanbury Brown}}\ \emph {et~al.}(1967)\citenamefont
  {{Hanbury Brown}}, \citenamefont {{Davis}},\ and\ \citenamefont
  {{Allen}}}]{1967MNRAS.137..375H}%
  \BibitemOpen
  \bibfield  {author} {\bibinfo {author} {\bibfnamefont {R.}~\bibnamefont
  {{Hanbury Brown}}}, \bibinfo {author} {\bibfnamefont {J.}~\bibnamefont
  {{Davis}}},\ and\ \bibinfo {author} {\bibfnamefont {L.~R.}\ \bibnamefont
  {{Allen}}},\ }\href {https://doi.org/10.1093/mnras/137.4.375} {\bibfield
  {journal} {\bibinfo  {journal} {\mnras}\ }\textbf {\bibinfo {volume} {137}},\
  \bibinfo {pages} {375} (\bibinfo {year} {1967})}\BibitemShut {NoStop}%
\bibitem [{\citenamefont {Kieda}\ \emph {et~al.}(2021)\citenamefont {Kieda},
  \citenamefont {Davis}, \citenamefont {LeBohec}, \citenamefont {Lisa},\ and\
  \citenamefont {Matthews}}]{kieda2021veritasstellar}%
  \BibitemOpen
  \bibfield  {author} {\bibinfo {author} {\bibfnamefont {D.}~\bibnamefont
  {Kieda}}, \bibinfo {author} {\bibfnamefont {J.}~\bibnamefont {Davis}},
  \bibinfo {author} {\bibfnamefont {T.}~\bibnamefont {LeBohec}}, \bibinfo
  {author} {\bibfnamefont {M.}~\bibnamefont {Lisa}},\ and\ \bibinfo {author}
  {\bibfnamefont {N.~K.}\ \bibnamefont {Matthews}},\ }\href@noop {} {\bibinfo
  {title} {The veritas-stellar intensity interferometry (vsii) survey of
  stellar diameters}} (\bibinfo {year} {2021}),\ \Eprint
  {https://arxiv.org/abs/2108.09774} {arXiv:2108.09774 [astro-ph.SR]}
  \BibitemShut {NoStop}%
\bibitem [{\citenamefont {Zmija}\ \emph {et~al.}(2023)\citenamefont {Zmija},
  \citenamefont {Vogel}, \citenamefont {Wohlleben}, \citenamefont {Anton},
  \citenamefont {Zink},\ and\ \citenamefont {Funk}}]{Zmija2023}%
  \BibitemOpen
  \bibfield  {author} {\bibinfo {author} {\bibfnamefont {A.}~\bibnamefont
  {Zmija}}, \bibinfo {author} {\bibfnamefont {N.}~\bibnamefont {Vogel}},
  \bibinfo {author} {\bibfnamefont {F.}~\bibnamefont {Wohlleben}}, \bibinfo
  {author} {\bibfnamefont {G.}~\bibnamefont {Anton}}, \bibinfo {author}
  {\bibfnamefont {A.}~\bibnamefont {Zink}},\ and\ \bibinfo {author}
  {\bibfnamefont {S.}~\bibnamefont {Funk}},\ }\bibfield  {journal} {\bibinfo
  {journal} {Monthly Notices of the Royal Astronomical Society}\ }\href
  {https://doi.org/10.1093/mnras/stad3676} {10.1093/mnras/stad3676} (\bibinfo
  {year} {2023}),\ \Eprint {https://arxiv.org/abs/2312.08015} {arXiv:2312.08015
  [astro-ph.IM]} \BibitemShut {NoStop}%
\bibitem [{\citenamefont {{MAGIC Collaboration}}\ \emph
  {et~al.}(2024)\citenamefont {{MAGIC Collaboration}}, \citenamefont {{Abe}},
  \citenamefont {{Abhir}}, \citenamefont {{Acciari}} \emph
  {et~al.}}]{MAGIC2024}%
  \BibitemOpen
  \bibfield  {author} {\bibinfo {author} {\bibnamefont {{MAGIC
  Collaboration}}}, \bibinfo {author} {\bibfnamefont {S.}~\bibnamefont
  {{Abe}}}, \bibinfo {author} {\bibfnamefont {J.}~\bibnamefont {{Abhir}}},
  \bibinfo {author} {\bibfnamefont {V.~A.}\ \bibnamefont {{Acciari}}}, \emph
  {et~al.},\ }\href {https://doi.org/10.48550/arXiv.2402.04755} {\bibfield
  {journal} {\bibinfo  {journal} {arXiv e-prints}\ ,\ \bibinfo {eid}
  {arXiv:2402.04755}} (\bibinfo {year} {2024})},\ \Eprint
  {https://arxiv.org/abs/2402.04755} {arXiv:2402.04755 [astro-ph.IM]}
  \BibitemShut {NoStop}%
\bibitem [{\citenamefont {Steinhauer}\ \emph {et~al.}(2021)\citenamefont
  {Steinhauer}, \citenamefont {Gyger},\ and\ \citenamefont
  {Zwiller}}]{SteinhauerReview}%
  \BibitemOpen
  \bibfield  {author} {\bibinfo {author} {\bibfnamefont {S.}~\bibnamefont
  {Steinhauer}}, \bibinfo {author} {\bibfnamefont {S.}~\bibnamefont {Gyger}},\
  and\ \bibinfo {author} {\bibfnamefont {V.}~\bibnamefont {Zwiller}},\ }\href
  {https://doi.org/10.1063/5.0044057} {\bibfield  {journal} {\bibinfo
  {journal} {Applied Physics Letters}\ }\textbf {\bibinfo {volume} {118}},\
  \bibinfo {pages} {100501} (\bibinfo {year} {2021})},\ \Eprint
  {https://arxiv.org/abs/https://doi.org/10.1063/5.0044057}
  {https://doi.org/10.1063/5.0044057} \BibitemShut {NoStop}%
\bibitem [{\citenamefont {Hadfield}(2020)}]{Hadfield2020}%
  \BibitemOpen
  \bibfield  {author} {\bibinfo {author} {\bibfnamefont {R.~H.}\ \bibnamefont
  {Hadfield}},\ }\href {https://doi.org/10.1038/s41566-020-0614-0} {\bibfield
  {journal} {\bibinfo  {journal} {Nature Photonics}\ }\textbf {\bibinfo
  {volume} {14}},\ \bibinfo {pages} {201} (\bibinfo {year} {2020})}\BibitemShut
  {NoStop}%
\bibitem [{\citenamefont {Mueller}\ \emph {et~al.}(2021)\citenamefont
  {Mueller}, \citenamefont {Korzh}, \citenamefont {Runyan} \emph
  {et~al.}}]{SpiropuluShaw2021}%
  \BibitemOpen
  \bibfield  {author} {\bibinfo {author} {\bibfnamefont {A.~S.}\ \bibnamefont
  {Mueller}}, \bibinfo {author} {\bibfnamefont {B.}~\bibnamefont {Korzh}},
  \bibinfo {author} {\bibfnamefont {M.}~\bibnamefont {Runyan}}, \emph
  {et~al.},\ }\href {https://doi.org/10.1364/OPTICA.444108} {\bibfield
  {journal} {\bibinfo  {journal} {Optica}\ }\textbf {\bibinfo {volume} {8}},\
  \bibinfo {pages} {1586} (\bibinfo {year} {2021})}\BibitemShut {NoStop}%
\bibitem [{\citenamefont {Korzh}\ \emph {et~al.}(2020)\citenamefont {Korzh},
  \citenamefont {Zhao}, \citenamefont {Allmaras} \emph
  {et~al.}}]{ShawBerggren2020}%
  \BibitemOpen
  \bibfield  {author} {\bibinfo {author} {\bibfnamefont {B.}~\bibnamefont
  {Korzh}}, \bibinfo {author} {\bibfnamefont {Q.-Y.}\ \bibnamefont {Zhao}},
  \bibinfo {author} {\bibfnamefont {J.~P.}\ \bibnamefont {Allmaras}}, \emph
  {et~al.},\ }\href@noop {} {\bibfield  {journal} {\bibinfo  {journal} {Nature
  Photonics}\ }\textbf {\bibinfo {volume} {14}},\ \bibinfo {pages} {250}
  (\bibinfo {year} {2020})}\BibitemShut {NoStop}%
\bibitem [{\citenamefont {Wollman}\ \emph {et~al.}(2017)\citenamefont
  {Wollman}, \citenamefont {Verma}, \citenamefont {Beyer}, \citenamefont
  {Briggs}, \citenamefont {Korzh}, \citenamefont {Allmaras}, \citenamefont
  {Marsili}, \citenamefont {Lita}, \citenamefont {Mirin}, \citenamefont {Nam},\
  and\ \citenamefont {Shaw}}]{WollmanUV}%
  \BibitemOpen
  \bibfield  {author} {\bibinfo {author} {\bibfnamefont {E.~E.}\ \bibnamefont
  {Wollman}}, \bibinfo {author} {\bibfnamefont {V.~B.}\ \bibnamefont {Verma}},
  \bibinfo {author} {\bibfnamefont {A.~D.}\ \bibnamefont {Beyer}}, \bibinfo
  {author} {\bibfnamefont {R.~M.}\ \bibnamefont {Briggs}}, \bibinfo {author}
  {\bibfnamefont {B.}~\bibnamefont {Korzh}}, \bibinfo {author} {\bibfnamefont
  {J.~P.}\ \bibnamefont {Allmaras}}, \bibinfo {author} {\bibfnamefont
  {F.}~\bibnamefont {Marsili}}, \bibinfo {author} {\bibfnamefont {A.~E.}\
  \bibnamefont {Lita}}, \bibinfo {author} {\bibfnamefont {R.~P.}\ \bibnamefont
  {Mirin}}, \bibinfo {author} {\bibfnamefont {S.~W.}\ \bibnamefont {Nam}},\
  and\ \bibinfo {author} {\bibfnamefont {M.~D.}\ \bibnamefont {Shaw}},\ }\href
  {https://doi.org/10.1364/OE.25.026792} {\bibfield  {journal} {\bibinfo
  {journal} {Opt. Express}\ }\textbf {\bibinfo {volume} {25}},\ \bibinfo
  {pages} {26792} (\bibinfo {year} {2017})}\BibitemShut {NoStop}%
\bibitem [{\citenamefont {Sanzaro}\ \emph {et~al.}(2018)\citenamefont
  {Sanzaro}, \citenamefont {Gattari}, \citenamefont {Villa}, \citenamefont
  {Tosi}, \citenamefont {Croce},\ and\ \citenamefont {Zappa}}]{Zappa2018}%
  \BibitemOpen
  \bibfield  {author} {\bibinfo {author} {\bibfnamefont {M.}~\bibnamefont
  {Sanzaro}}, \bibinfo {author} {\bibfnamefont {P.}~\bibnamefont {Gattari}},
  \bibinfo {author} {\bibfnamefont {F.}~\bibnamefont {Villa}}, \bibinfo
  {author} {\bibfnamefont {A.}~\bibnamefont {Tosi}}, \bibinfo {author}
  {\bibfnamefont {G.}~\bibnamefont {Croce}},\ and\ \bibinfo {author}
  {\bibfnamefont {F.}~\bibnamefont {Zappa}},\ }\href
  {https://doi.org/10.1109/JSTQE.2017.2762464} {\bibfield  {journal} {\bibinfo
  {journal} {IEEE Journal of Selected Topics in Quantum Electronics}\ }\textbf
  {\bibinfo {volume} {24}},\ \bibinfo {pages} {1} (\bibinfo {year}
  {2018})}\BibitemShut {NoStop}%
\bibitem [{\citenamefont {Becker}\ \emph {et~al.}(2022)\citenamefont {Becker},
  \citenamefont {Gramuglia}, \citenamefont {Wu}, \citenamefont {Charbon},\ and\
  \citenamefont {Bruschini}}]{becker2022}%
  \BibitemOpen
  \bibfield  {author} {\bibinfo {author} {\bibfnamefont {W.}~\bibnamefont
  {Becker}}, \bibinfo {author} {\bibfnamefont {F.}~\bibnamefont {Gramuglia}},
  \bibinfo {author} {\bibfnamefont {M.-L.}\ \bibnamefont {Wu}}, \bibinfo
  {author} {\bibfnamefont {E.}~\bibnamefont {Charbon}},\ and\ \bibinfo {author}
  {\bibfnamefont {C.}~\bibnamefont {Bruschini}},\ }\href@noop {} {\bibinfo
  {title} {8.7 ps fwhm irf width from ultrafast spad}},\ \bibinfo
  {howpublished}
  {\url{https://www.becker-hickl.com/literature/application-notes/8-7-ps-fwhm-irf-width-from-ultrafast-spad/}}
  (\bibinfo {year} {2022}),\ \bibinfo {note} {[Online; accessed
  06-Apr-2023]}\BibitemShut {NoStop}%
\bibitem [{\citenamefont {{Oripov}}\ \emph {et~al.}(2023)\citenamefont
  {{Oripov}}, \citenamefont {{Rampini}}, \citenamefont {{Allmaras}} \emph
  {et~al.}}]{Oripov2023}%
  \BibitemOpen
  \bibfield  {author} {\bibinfo {author} {\bibfnamefont {B.~G.}\ \bibnamefont
  {{Oripov}}}, \bibinfo {author} {\bibfnamefont {D.~S.}\ \bibnamefont
  {{Rampini}}}, \bibinfo {author} {\bibfnamefont {J.}~\bibnamefont
  {{Allmaras}}}, \emph {et~al.},\ }\href
  {https://doi.org/10.1038/s41586-023-06550-2} {\bibfield  {journal} {\bibinfo
  {journal} {Nature}\ }\textbf {\bibinfo {volume} {622}},\ \bibinfo {eid} {730}
  (\bibinfo {year} {2023})},\ \Eprint {https://arxiv.org/abs/2306.09473}
  {arXiv:2306.09473 [quant-ph]} \BibitemShut {NoStop}%
\bibitem [{\citenamefont {Aull}\ \emph {et~al.}(2018)\citenamefont {Aull},
  \citenamefont {Duerr}, \citenamefont {Frechette} \emph
  {et~al.}}]{LargeFormatAull}%
  \BibitemOpen
  \bibfield  {author} {\bibinfo {author} {\bibfnamefont {B.~F.}\ \bibnamefont
  {Aull}}, \bibinfo {author} {\bibfnamefont {E.~K.}\ \bibnamefont {Duerr}},
  \bibinfo {author} {\bibfnamefont {J.~P.}\ \bibnamefont {Frechette}}, \emph
  {et~al.},\ }\href {https://doi.org/10.1109/JSTQE.2017.2736440} {\bibfield
  {journal} {\bibinfo  {journal} {IEEE Journal of Selected Topics in Quantum
  Electronics}\ }\textbf {\bibinfo {volume} {24}},\ \bibinfo {pages} {1}
  (\bibinfo {year} {2018})}\BibitemShut {NoStop}%
\bibitem [{\citenamefont {Punjiya}\ \emph {et~al.}(2021)\citenamefont
  {Punjiya}, \citenamefont {Ryu}, \citenamefont {Gregory}, \citenamefont
  {McIntosh}, \citenamefont {Duerr},\ and\ \citenamefont
  {Katake}}]{EuropaSlides}%
  \BibitemOpen
  \bibfield  {author} {\bibinfo {author} {\bibfnamefont {M.}~\bibnamefont
  {Punjiya}}, \bibinfo {author} {\bibfnamefont {K.}~\bibnamefont {Ryu}},
  \bibinfo {author} {\bibfnamefont {M.}~\bibnamefont {Gregory}}, \bibinfo
  {author} {\bibfnamefont {A.}~\bibnamefont {McIntosh}}, \bibinfo {author}
  {\bibfnamefont {E.}~\bibnamefont {Duerr}},\ and\ \bibinfo {author}
  {\bibfnamefont {A.}~\bibnamefont {Katake}},\ }in\ \href
  {https://trs.jpl.nasa.gov/handle/2014/54810} {\emph {\bibinfo {booktitle}
  {Advanced Technology for National Security Workshop (ATNS) 2021, Pasadena,
  California, April 6-8, 2021}}}\ (\bibinfo  {publisher} {Pasadena, CA: Jet
  Propulsion Laboratory, National Aeronautics and Space Administration},\
  \bibinfo {year} {2021})\BibitemShut {NoStop}%
\bibitem [{\citenamefont {Punjiya}\ \emph {et~al.}(2022)\citenamefont
  {Punjiya}, \citenamefont {Ryu}, \citenamefont {Aull} \emph
  {et~al.}}]{EuropaLidar}%
  \BibitemOpen
  \bibfield  {author} {\bibinfo {author} {\bibfnamefont {M.}~\bibnamefont
  {Punjiya}}, \bibinfo {author} {\bibfnamefont {K.}~\bibnamefont {Ryu}},
  \bibinfo {author} {\bibfnamefont {B.}~\bibnamefont {Aull}}, \emph {et~al.},\
  }in\ \href {https://doi.org/10.1117/12.2618800} {\emph {\bibinfo {booktitle}
  {Advanced Photon Counting Techniques XVI}}},\ Vol.\ \bibinfo {volume}
  {12089},\ \bibinfo {editor} {edited by\ \bibinfo {editor} {\bibfnamefont
  {M.~A.}\ \bibnamefont {Itzler}}, \bibinfo {editor} {\bibfnamefont {J.~C.}\
  \bibnamefont {Bienfang}},\ and\ \bibinfo {editor} {\bibfnamefont {K.~A.}\
  \bibnamefont {McIntosh}}},\ \bibinfo {organization} {International Society
  for Optics and Photonics}\ (\bibinfo  {publisher} {SPIE},\ \bibinfo {year}
  {2022})\ p.\ \bibinfo {pages} {1208903}\BibitemShut {NoStop}%
\bibitem [{\citenamefont {Ryu}\ \emph {et~al.}(2022)\citenamefont {Ryu},
  \citenamefont {Aull}, \citenamefont {Collins} \emph {et~al.}}]{SOIspad}%
  \BibitemOpen
  \bibfield  {author} {\bibinfo {author} {\bibfnamefont {K.~K.}\ \bibnamefont
  {Ryu}}, \bibinfo {author} {\bibfnamefont {B.~F.}\ \bibnamefont {Aull}},
  \bibinfo {author} {\bibfnamefont {M.}~\bibnamefont {Collins}}, \emph
  {et~al.},\ }in\ \href {https://doi.org/10.1117/12.2618610} {\emph {\bibinfo
  {booktitle} {Advanced Photon Counting Techniques XVI}}},\ Vol.\ \bibinfo
  {volume} {12089},\ \bibinfo {editor} {edited by\ \bibinfo {editor}
  {\bibfnamefont {M.~A.}\ \bibnamefont {Itzler}}, \bibinfo {editor}
  {\bibfnamefont {J.~C.}\ \bibnamefont {Bienfang}},\ and\ \bibinfo {editor}
  {\bibfnamefont {K.~A.}\ \bibnamefont {McIntosh}}},\ \bibinfo {organization}
  {International Society for Optics and Photonics}\ (\bibinfo  {publisher}
  {SPIE},\ \bibinfo {year} {2022})\ p.\ \bibinfo {pages} {120890C}\BibitemShut
  {NoStop}%
\bibitem [{\citenamefont {{Horch}}\ \emph {et~al.}(2022)\citenamefont
  {{Horch}}, \citenamefont {{Weiss}}, \citenamefont {{Klaucke}}, \citenamefont
  {{Pellegrino}},\ and\ \citenamefont {{Rupert}}}]{horch2022observations}%
  \BibitemOpen
  \bibfield  {author} {\bibinfo {author} {\bibfnamefont {E.~P.}\ \bibnamefont
  {{Horch}}}, \bibinfo {author} {\bibfnamefont {S.~A.}\ \bibnamefont
  {{Weiss}}}, \bibinfo {author} {\bibfnamefont {P.~M.}\ \bibnamefont
  {{Klaucke}}}, \bibinfo {author} {\bibfnamefont {R.~A.}\ \bibnamefont
  {{Pellegrino}}},\ and\ \bibinfo {author} {\bibfnamefont {J.~D.}\ \bibnamefont
  {{Rupert}}},\ }\href {https://doi.org/10.3847/1538-3881/ac43bb} {\bibfield
  {journal} {\bibinfo  {journal} {\aj}\ }\textbf {\bibinfo {volume} {163}},\
  \bibinfo {eid} {92} (\bibinfo {year} {2022})},\ \Eprint
  {https://arxiv.org/abs/2112.07758} {arXiv:2112.07758 [astro-ph.IM]}
  \BibitemShut {NoStop}%
\bibitem [{\citenamefont {{Guerin}}\ \emph {et~al.}(2018)\citenamefont
  {{Guerin}}, \citenamefont {{Rivet}}, \citenamefont {{Fouch{\'e}}},
  \citenamefont {{Labeyrie}}, \citenamefont {{Vernet}}, \citenamefont
  {{Vakili}},\ and\ \citenamefont {{Kaiser}}}]{Guerin_space}%
  \BibitemOpen
  \bibfield  {author} {\bibinfo {author} {\bibfnamefont {W.}~\bibnamefont
  {{Guerin}}}, \bibinfo {author} {\bibfnamefont {J.~P.}\ \bibnamefont
  {{Rivet}}}, \bibinfo {author} {\bibfnamefont {M.}~\bibnamefont
  {{Fouch{\'e}}}}, \bibinfo {author} {\bibfnamefont {G.}~\bibnamefont
  {{Labeyrie}}}, \bibinfo {author} {\bibfnamefont {D.}~\bibnamefont
  {{Vernet}}}, \bibinfo {author} {\bibfnamefont {F.}~\bibnamefont {{Vakili}}},\
  and\ \bibinfo {author} {\bibfnamefont {R.}~\bibnamefont {{Kaiser}}},\ }\href
  {https://doi.org/10.1093/mnras/sty1792} {\bibfield  {journal} {\bibinfo
  {journal} {\mnras}\ }\textbf {\bibinfo {volume} {480}},\ \bibinfo {pages}
  {245} (\bibinfo {year} {2018})},\ \Eprint {https://arxiv.org/abs/1805.06653}
  {arXiv:1805.06653 [astro-ph.IM]} \BibitemShut {NoStop}%
\bibitem [{\citenamefont {{Rybicki}}\ and\ \citenamefont
  {{Lightman}}(1986)}]{RybickiLightman}%
  \BibitemOpen
  \bibfield  {author} {\bibinfo {author} {\bibfnamefont {G.~B.}\ \bibnamefont
  {{Rybicki}}}\ and\ \bibinfo {author} {\bibfnamefont {A.~P.}\ \bibnamefont
  {{Lightman}}},\ }\href@noop {} {\emph {\bibinfo {title} {{Radiative Processes
  in Astrophysics}}}}\ (\bibinfo  {publisher} {{Wiley}},\ \bibinfo {year}
  {1986})\BibitemShut {NoStop}%
\bibitem [{\citenamefont {Severini}\ \emph {et~al.}(2022)\citenamefont
  {Severini}, \citenamefont {Cusini}, \citenamefont {Berretta}, \citenamefont
  {Pasquinelli}, \citenamefont {Incoronato},\ and\ \citenamefont
  {Villa}}]{deadtime}%
  \BibitemOpen
  \bibfield  {author} {\bibinfo {author} {\bibfnamefont {F.}~\bibnamefont
  {Severini}}, \bibinfo {author} {\bibfnamefont {I.}~\bibnamefont {Cusini}},
  \bibinfo {author} {\bibfnamefont {D.}~\bibnamefont {Berretta}}, \bibinfo
  {author} {\bibfnamefont {K.}~\bibnamefont {Pasquinelli}}, \bibinfo {author}
  {\bibfnamefont {A.}~\bibnamefont {Incoronato}},\ and\ \bibinfo {author}
  {\bibfnamefont {F.}~\bibnamefont {Villa}},\ }\href
  {https://doi.org/10.1109/JSTQE.2021.3124825} {\bibfield  {journal} {\bibinfo
  {journal} {IEEE Journal of Selected Topics in Quantum Electronics}\ }\textbf
  {\bibinfo {volume} {28}},\ \bibinfo {pages} {1} (\bibinfo {year}
  {2022})}\BibitemShut {NoStop}%
\bibitem [{\citenamefont {{Lai}}\ \emph {et~al.}(2021)\citenamefont {{Lai}},
  \citenamefont {{Labeyrie}}, \citenamefont {{Guerin}}, \citenamefont
  {{Vakili}}, \citenamefont {{Kaiser}}, \citenamefont {{Rivet}}, \citenamefont
  {{Hugbart}}, \citenamefont {{Matthews}}, \citenamefont {{Chab{\'e}}},
  \citenamefont {{Courde}}, \citenamefont {{Samain}},\ and\ \citenamefont
  {{Vernet}}}]{2021sf2a.conf..335L}%
  \BibitemOpen
  \bibfield  {author} {\bibinfo {author} {\bibfnamefont {O.}~\bibnamefont
  {{Lai}}}, \bibinfo {author} {\bibfnamefont {G.}~\bibnamefont {{Labeyrie}}},
  \bibinfo {author} {\bibfnamefont {W.}~\bibnamefont {{Guerin}}}, \bibinfo
  {author} {\bibfnamefont {F.}~\bibnamefont {{Vakili}}}, \bibinfo {author}
  {\bibfnamefont {R.}~\bibnamefont {{Kaiser}}}, \bibinfo {author}
  {\bibfnamefont {J.~P.}\ \bibnamefont {{Rivet}}}, \bibinfo {author}
  {\bibfnamefont {M.}~\bibnamefont {{Hugbart}}}, \bibinfo {author}
  {\bibfnamefont {N.}~\bibnamefont {{Matthews}}}, \bibinfo {author}
  {\bibfnamefont {J.}~\bibnamefont {{Chab{\'e}}}}, \bibinfo {author}
  {\bibfnamefont {C.}~\bibnamefont {{Courde}}}, \bibinfo {author}
  {\bibfnamefont {E.}~\bibnamefont {{Samain}}},\ and\ \bibinfo {author}
  {\bibfnamefont {D.}~\bibnamefont {{Vernet}}},\ }in\ \href@noop {} {\emph
  {\bibinfo {booktitle} {SF2A-2021: Proceedings of the Annual meeting of the
  French Society of Astronomy and Astrophysics}}},\ \bibinfo {editor} {edited
  by\ \bibinfo {editor} {\bibfnamefont {A.}~\bibnamefont {{Siebert}}}, \bibinfo
  {editor} {\bibfnamefont {K.}~\bibnamefont {{Bailli{\'e}}}}, \bibinfo {editor}
  {\bibfnamefont {E.}~\bibnamefont {{Lagadec}}}, \bibinfo {editor}
  {\bibfnamefont {N.}~\bibnamefont {{Lagarde}}}, \bibinfo {editor}
  {\bibfnamefont {J.}~\bibnamefont {{Malzac}}}, \bibinfo {editor}
  {\bibfnamefont {J.~B.}\ \bibnamefont {{Marquette}}}, \bibinfo {editor}
  {\bibfnamefont {M.}~\bibnamefont {{N'Diaye}}}, \bibinfo {editor}
  {\bibfnamefont {J.}~\bibnamefont {{Richard}}},\ and\ \bibinfo {editor}
  {\bibfnamefont {O.}~\bibnamefont {{Venot}}}}\ (\bibinfo {year} {2021})\ pp.\
  \bibinfo {pages} {335--338}\BibitemShut {NoStop}%
\bibitem [{\citenamefont {Chen}\ \emph {et~al.}(2023)\citenamefont {Chen},
  \citenamefont {Nomerotski}, \citenamefont {Slosar}, \citenamefont {Stankus},\
  and\ \citenamefont {Vintskevich}}]{chen2023astrometry}%
  \BibitemOpen
  \bibfield  {author} {\bibinfo {author} {\bibfnamefont {Z.}~\bibnamefont
  {Chen}}, \bibinfo {author} {\bibfnamefont {A.}~\bibnamefont {Nomerotski}},
  \bibinfo {author} {\bibfnamefont {A.}~\bibnamefont {Slosar}}, \bibinfo
  {author} {\bibfnamefont {P.}~\bibnamefont {Stankus}},\ and\ \bibinfo {author}
  {\bibfnamefont {S.}~\bibnamefont {Vintskevich}},\ }\href@noop {} {\bibfield
  {journal} {\bibinfo  {journal} {Phys. Rev. D}\ }\textbf {\bibinfo {volume}
  {107}},\ \bibinfo {pages} {023015} (\bibinfo {year} {2023})}\BibitemShut
  {NoStop}%
\bibitem [{\citenamefont {Stankus}\ \emph {et~al.}(2022)\citenamefont
  {Stankus}, \citenamefont {Nomerotski}, \citenamefont {Slosar},\ and\
  \citenamefont {Vintskevich}}]{Stankus_2022}%
  \BibitemOpen
  \bibfield  {author} {\bibinfo {author} {\bibfnamefont {P.}~\bibnamefont
  {Stankus}}, \bibinfo {author} {\bibfnamefont {A.}~\bibnamefont {Nomerotski}},
  \bibinfo {author} {\bibfnamefont {A.}~\bibnamefont {Slosar}},\ and\ \bibinfo
  {author} {\bibfnamefont {S.}~\bibnamefont {Vintskevich}},\ }\bibfield
  {journal} {\bibinfo  {journal} {The Open Journal of Astrophysics}\ }\textbf
  {\bibinfo {volume} {5}},\ \href {https://doi.org/10.21105/astro.2010.09100}
  {10.21105/astro.2010.09100} (\bibinfo {year} {2022})\BibitemShut {NoStop}%
\bibitem [{\citenamefont {Trippe}\ \emph {et~al.}(2014)\citenamefont {Trippe},
  \citenamefont {Kim}, \citenamefont {Lee}, \citenamefont {Choi}, \citenamefont
  {Oh}, \citenamefont {Lee}, \citenamefont {Yoon}, \citenamefont {Im},\ and\
  \citenamefont {Park}}]{trippe2014optical}%
  \BibitemOpen
  \bibfield  {author} {\bibinfo {author} {\bibfnamefont {S.}~\bibnamefont
  {Trippe}}, \bibinfo {author} {\bibfnamefont {J.-Y.}\ \bibnamefont {Kim}},
  \bibinfo {author} {\bibfnamefont {B.}~\bibnamefont {Lee}}, \bibinfo {author}
  {\bibfnamefont {C.}~\bibnamefont {Choi}}, \bibinfo {author} {\bibfnamefont
  {J.}~\bibnamefont {Oh}}, \bibinfo {author} {\bibfnamefont {T.}~\bibnamefont
  {Lee}}, \bibinfo {author} {\bibfnamefont {S.-C.}\ \bibnamefont {Yoon}},
  \bibinfo {author} {\bibfnamefont {M.}~\bibnamefont {Im}},\ and\ \bibinfo
  {author} {\bibfnamefont {Y.-S.}\ \bibnamefont {Park}},\ }\href@noop {}
  {\bibfield  {journal} {\bibinfo  {journal} {Journal of Korean Astronomical
  Society}\ }\textbf {\bibinfo {volume} {47}},\ \bibinfo {pages} {235}
  (\bibinfo {year} {2014})}\BibitemShut {NoStop}%
\bibitem [{\citenamefont {Tilburg}\ \emph {et~al.}(2023)\citenamefont
  {Tilburg}, \citenamefont {Baryakhtar}, \citenamefont {Galanis},\ and\
  \citenamefont {Weiner}}]{vantilburg2023}%
  \BibitemOpen
  \bibfield  {author} {\bibinfo {author} {\bibfnamefont {K.~V.}\ \bibnamefont
  {Tilburg}}, \bibinfo {author} {\bibfnamefont {M.}~\bibnamefont {Baryakhtar}},
  \bibinfo {author} {\bibfnamefont {M.}~\bibnamefont {Galanis}},\ and\ \bibinfo
  {author} {\bibfnamefont {N.}~\bibnamefont {Weiner}},\ }\href@noop {}
  {\bibinfo {title} {Astrometry with extended-path intensity correlation}}
  (\bibinfo {year} {2023}),\ \Eprint {https://arxiv.org/abs/2307.03221}
  {arXiv:2307.03221 [astro-ph.IM]} \BibitemShut {NoStop}%
\bibitem [{\citenamefont {Galanis}\ \emph {et~al.}(2023)\citenamefont
  {Galanis}, \citenamefont {Tilburg}, \citenamefont {Baryakhtar},\ and\
  \citenamefont {Weiner}}]{galanis2023extendedpath}%
  \BibitemOpen
  \bibfield  {author} {\bibinfo {author} {\bibfnamefont {M.}~\bibnamefont
  {Galanis}}, \bibinfo {author} {\bibfnamefont {K.~V.}\ \bibnamefont
  {Tilburg}}, \bibinfo {author} {\bibfnamefont {M.}~\bibnamefont
  {Baryakhtar}},\ and\ \bibinfo {author} {\bibfnamefont {N.}~\bibnamefont
  {Weiner}},\ }\href@noop {} {\bibinfo {title} {Extended-path intensity
  correlation: Microarcsecond astrometry with an arcsecond field of view}}
  (\bibinfo {year} {2023}),\ \Eprint {https://arxiv.org/abs/2307.06989}
  {arXiv:2307.06989 [astro-ph.IM]} \BibitemShut {NoStop}%
\bibitem [{\citenamefont {{Guerin}}\ \emph {et~al.}(2017)\citenamefont
  {{Guerin}}, \citenamefont {{Dussaux}}, \citenamefont {{Fouch{\'e}}},
  \citenamefont {{Labeyrie}}, \citenamefont {{Rivet}}, \citenamefont
  {{Vernet}}, \citenamefont {{Vakili}},\ and\ \citenamefont
  {{Kaiser}}}]{Guerin_time}%
  \BibitemOpen
  \bibfield  {author} {\bibinfo {author} {\bibfnamefont {W.}~\bibnamefont
  {{Guerin}}}, \bibinfo {author} {\bibfnamefont {A.}~\bibnamefont {{Dussaux}}},
  \bibinfo {author} {\bibfnamefont {M.}~\bibnamefont {{Fouch{\'e}}}}, \bibinfo
  {author} {\bibfnamefont {G.}~\bibnamefont {{Labeyrie}}}, \bibinfo {author}
  {\bibfnamefont {J.~P.}\ \bibnamefont {{Rivet}}}, \bibinfo {author}
  {\bibfnamefont {D.}~\bibnamefont {{Vernet}}}, \bibinfo {author}
  {\bibfnamefont {F.}~\bibnamefont {{Vakili}}},\ and\ \bibinfo {author}
  {\bibfnamefont {R.}~\bibnamefont {{Kaiser}}},\ }\href
  {https://doi.org/10.1093/mnras/stx2143} {\bibfield  {journal} {\bibinfo
  {journal} {\mnras}\ }\textbf {\bibinfo {volume} {472}},\ \bibinfo {pages}
  {4126} (\bibinfo {year} {2017})},\ \Eprint {https://arxiv.org/abs/1708.06119}
  {arXiv:1708.06119 [astro-ph.IM]} \BibitemShut {NoStop}%
\bibitem [{Note1()}]{Note1}%
  \BibitemOpen
  \bibinfo {note} {\protect \url
  {https://www.cta-observatory.org/project/technology/mst}}\BibitemShut
  {NoStop}%
\bibitem [{Note2()}]{Note2}%
  \BibitemOpen
  \bibinfo {note} {We convert from magnitudes to photon flux assuming a source
  with apparent magnitude $g=0$ has flux at the top of the atmosphere of 3730
  Jy \cite {Schneider1983}.}\BibitemShut {Stop}%
\bibitem [{\citenamefont {{Tegmark}}\ \emph {et~al.}(1997)\citenamefont
  {{Tegmark}}, \citenamefont {{Taylor}},\ and\ \citenamefont
  {{Heavens}}}]{Tegmark1997}%
  \BibitemOpen
  \bibfield  {author} {\bibinfo {author} {\bibfnamefont {M.}~\bibnamefont
  {{Tegmark}}}, \bibinfo {author} {\bibfnamefont {A.~N.}\ \bibnamefont
  {{Taylor}}},\ and\ \bibinfo {author} {\bibfnamefont {A.~F.}\ \bibnamefont
  {{Heavens}}},\ }\href {https://doi.org/10.1086/303939} {\bibfield  {journal}
  {\bibinfo  {journal} {\apj}\ }\textbf {\bibinfo {volume} {480}},\ \bibinfo
  {pages} {22} (\bibinfo {year} {1997})},\ \Eprint
  {https://arxiv.org/abs/astro-ph/9603021} {arXiv:astro-ph/9603021 [astro-ph]}
  \BibitemShut {NoStop}%
\bibitem [{\citenamefont {{Dravins}}\ \emph {et~al.}(2012)\citenamefont
  {{Dravins}}, \citenamefont {{LeBohec}}, \citenamefont {{Jensen}},\ and\
  \citenamefont {{Nu{\~n}ez}}}]{Dravins2012}%
  \BibitemOpen
  \bibfield  {author} {\bibinfo {author} {\bibfnamefont {D.}~\bibnamefont
  {{Dravins}}}, \bibinfo {author} {\bibfnamefont {S.}~\bibnamefont
  {{LeBohec}}}, \bibinfo {author} {\bibfnamefont {H.}~\bibnamefont
  {{Jensen}}},\ and\ \bibinfo {author} {\bibfnamefont {P.~D.}\ \bibnamefont
  {{Nu{\~n}ez}}},\ }\href {https://doi.org/10.1016/j.newar.2012.06.001}
  {\bibfield  {journal} {\bibinfo  {journal} {New Astronomy Reviews}\ }\textbf
  {\bibinfo {volume} {56}},\ \bibinfo {pages} {143} (\bibinfo {year} {2012})},\
  \Eprint {https://arxiv.org/abs/1207.0808} {arXiv:1207.0808 [astro-ph.IM]}
  \BibitemShut {NoStop}%
\bibitem [{\citenamefont {{Peterson}}(1997)}]{Peterson1997}%
  \BibitemOpen
  \bibfield  {author} {\bibinfo {author} {\bibfnamefont {B.~M.}\ \bibnamefont
  {{Peterson}}},\ }\href@noop {} {\emph {\bibinfo {title} {{An Introduction to
  Active Galactic Nuclei}}}}\ (\bibinfo  {publisher} {{Cambridge University
  Press}},\ \bibinfo {year} {1997})\BibitemShut {NoStop}%
\bibitem [{\citenamefont {{Krolik}}(1999)}]{Krolik1999}%
  \BibitemOpen
  \bibfield  {author} {\bibinfo {author} {\bibfnamefont {J.~H.}\ \bibnamefont
  {{Krolik}}},\ }\href@noop {} {\emph {\bibinfo {title} {{Active galactic
  nuclei : from the central black hole to the galactic environment}}}}\
  (\bibinfo  {publisher} {{Princeton University Press}},\ \bibinfo {year}
  {1999})\BibitemShut {NoStop}%
\bibitem [{\citenamefont {Netzer}(2013)}]{netzer_2013}%
  \BibitemOpen
  \bibfield  {author} {\bibinfo {author} {\bibfnamefont {H.}~\bibnamefont
  {Netzer}},\ }\href {https://doi.org/10.1017/CBO9781139109291} {\emph
  {\bibinfo {title} {The Physics and Evolution of Active Galactic Nuclei}}}\
  (\bibinfo  {publisher} {Cambridge University Press},\ \bibinfo {year}
  {2013})\BibitemShut {NoStop}%
\bibitem [{\citenamefont {{Davis}}\ and\ \citenamefont
  {{Tchekhovskoy}}(2020)}]{Davis2020}%
  \BibitemOpen
  \bibfield  {author} {\bibinfo {author} {\bibfnamefont {S.~W.}\ \bibnamefont
  {{Davis}}}\ and\ \bibinfo {author} {\bibfnamefont {A.}~\bibnamefont
  {{Tchekhovskoy}}},\ }\href
  {https://doi.org/10.1146/annurev-astro-081817-051905} {\bibfield  {journal}
  {\bibinfo  {journal} {\araa}\ }\textbf {\bibinfo {volume} {58}},\ \bibinfo
  {pages} {407} (\bibinfo {year} {2020})},\ \Eprint
  {https://arxiv.org/abs/2101.08839} {arXiv:2101.08839 [astro-ph.HE]}
  \BibitemShut {NoStop}%
\bibitem [{\citenamefont {{Hopkins}}\ \emph
  {et~al.}(2023{\natexlab{a}})\citenamefont {{Hopkins}}, \citenamefont
  {{Grudic}}, \citenamefont {{Su}} \emph {et~al.}}]{Hopkins2023a}%
  \BibitemOpen
  \bibfield  {author} {\bibinfo {author} {\bibfnamefont {P.~F.}\ \bibnamefont
  {{Hopkins}}}, \bibinfo {author} {\bibfnamefont {M.~Y.}\ \bibnamefont
  {{Grudic}}}, \bibinfo {author} {\bibfnamefont {K.-Y.}\ \bibnamefont {{Su}}},
  \emph {et~al.},\ }\href {https://doi.org/10.48550/arXiv.2309.13115}
  {\bibfield  {journal} {\bibinfo  {journal} {arXiv e-prints}\ ,\ \bibinfo
  {eid} {arXiv:2309.13115}} (\bibinfo {year} {2023}{\natexlab{a}})},\ \Eprint
  {https://arxiv.org/abs/2309.13115} {arXiv:2309.13115 [astro-ph.GA]}
  \BibitemShut {NoStop}%
\bibitem [{\citenamefont {{Hopkins}}\ \emph
  {et~al.}(2023{\natexlab{b}})\citenamefont {{Hopkins}}, \citenamefont
  {{Squire}}, \citenamefont {{Su}} \emph {et~al.}}]{Hopkins2023b}%
  \BibitemOpen
  \bibfield  {author} {\bibinfo {author} {\bibfnamefont {P.~F.}\ \bibnamefont
  {{Hopkins}}}, \bibinfo {author} {\bibfnamefont {J.}~\bibnamefont {{Squire}}},
  \bibinfo {author} {\bibfnamefont {K.-Y.}\ \bibnamefont {{Su}}}, \emph
  {et~al.},\ }\href {https://doi.org/10.48550/arXiv.2310.04506} {\bibfield
  {journal} {\bibinfo  {journal} {arXiv e-prints}\ ,\ \bibinfo {eid}
  {arXiv:2310.04506}} (\bibinfo {year} {2023}{\natexlab{b}})},\ \Eprint
  {https://arxiv.org/abs/2310.04506} {arXiv:2310.04506 [astro-ph.HE]}
  \BibitemShut {NoStop}%
\bibitem [{\citenamefont {{Shakura}}\ and\ \citenamefont
  {{Sunyaev}}(1973)}]{Shakura1973}%
  \BibitemOpen
  \bibfield  {author} {\bibinfo {author} {\bibfnamefont {N.~I.}\ \bibnamefont
  {{Shakura}}}\ and\ \bibinfo {author} {\bibfnamefont {R.~A.}\ \bibnamefont
  {{Sunyaev}}},\ }\href@noop {} {\bibfield  {journal} {\bibinfo  {journal}
  {\aap}\ }\textbf {\bibinfo {volume} {24}},\ \bibinfo {pages} {337} (\bibinfo
  {year} {1973})}\BibitemShut {NoStop}%
\bibitem [{\citenamefont {{Jim{\'e}nez-Vicente}}\ \emph
  {et~al.}(2012)\citenamefont {{Jim{\'e}nez-Vicente}}, \citenamefont
  {{Mediavilla}}, \citenamefont {{Mu{\~n}oz}},\ and\ \citenamefont
  {{Kochanek}}}]{2012ApJ...751..106J}%
  \BibitemOpen
  \bibfield  {author} {\bibinfo {author} {\bibfnamefont {J.}~\bibnamefont
  {{Jim{\'e}nez-Vicente}}}, \bibinfo {author} {\bibfnamefont {E.}~\bibnamefont
  {{Mediavilla}}}, \bibinfo {author} {\bibfnamefont {J.~A.}\ \bibnamefont
  {{Mu{\~n}oz}}},\ and\ \bibinfo {author} {\bibfnamefont {C.~S.}\ \bibnamefont
  {{Kochanek}}},\ }\href {https://doi.org/10.1088/0004-637X/751/2/106}
  {\bibfield  {journal} {\bibinfo  {journal} {\apj}\ }\textbf {\bibinfo
  {volume} {751}},\ \bibinfo {eid} {106} (\bibinfo {year} {2012})},\ \Eprint
  {https://arxiv.org/abs/1201.3187} {arXiv:1201.3187 [astro-ph.CO]}
  \BibitemShut {NoStop}%
\bibitem [{\citenamefont {{Jim{\'e}nez-Vicente}}\ \emph
  {et~al.}(2014)\citenamefont {{Jim{\'e}nez-Vicente}}, \citenamefont
  {{Mediavilla}}, \citenamefont {{Kochanek}}, \citenamefont {{Mu{\~n}oz}},
  \citenamefont {{Motta}}, \citenamefont {{Falco}},\ and\ \citenamefont
  {{Mosquera}}}]{2014ApJ...783...47J}%
  \BibitemOpen
  \bibfield  {author} {\bibinfo {author} {\bibfnamefont {J.}~\bibnamefont
  {{Jim{\'e}nez-Vicente}}}, \bibinfo {author} {\bibfnamefont {E.}~\bibnamefont
  {{Mediavilla}}}, \bibinfo {author} {\bibfnamefont {C.~S.}\ \bibnamefont
  {{Kochanek}}}, \bibinfo {author} {\bibfnamefont {J.~A.}\ \bibnamefont
  {{Mu{\~n}oz}}}, \bibinfo {author} {\bibfnamefont {V.}~\bibnamefont
  {{Motta}}}, \bibinfo {author} {\bibfnamefont {E.}~\bibnamefont {{Falco}}},\
  and\ \bibinfo {author} {\bibfnamefont {A.~M.}\ \bibnamefont {{Mosquera}}},\
  }\href {https://doi.org/10.1088/0004-637X/783/1/47} {\bibfield  {journal}
  {\bibinfo  {journal} {\apj}\ }\textbf {\bibinfo {volume} {783}},\ \bibinfo
  {eid} {47} (\bibinfo {year} {2014})},\ \Eprint
  {https://arxiv.org/abs/1401.2785} {arXiv:1401.2785 [astro-ph.CO]}
  \BibitemShut {NoStop}%
\bibitem [{\citenamefont {{Morgan}}\ \emph {et~al.}(2018)\citenamefont
  {{Morgan}}, \citenamefont {{Hyer}}, \citenamefont {{Bonvin}}, \citenamefont
  {{Mosquera}}, \citenamefont {{Cornachione}}, \citenamefont {{Courbin}},
  \citenamefont {{Kochanek}},\ and\ \citenamefont {{Falco}}}]{Morgan2018}%
  \BibitemOpen
  \bibfield  {author} {\bibinfo {author} {\bibfnamefont {C.~W.}\ \bibnamefont
  {{Morgan}}}, \bibinfo {author} {\bibfnamefont {G.~E.}\ \bibnamefont
  {{Hyer}}}, \bibinfo {author} {\bibfnamefont {V.}~\bibnamefont {{Bonvin}}},
  \bibinfo {author} {\bibfnamefont {A.~M.}\ \bibnamefont {{Mosquera}}},
  \bibinfo {author} {\bibfnamefont {M.}~\bibnamefont {{Cornachione}}}, \bibinfo
  {author} {\bibfnamefont {F.}~\bibnamefont {{Courbin}}}, \bibinfo {author}
  {\bibfnamefont {C.~S.}\ \bibnamefont {{Kochanek}}},\ and\ \bibinfo {author}
  {\bibfnamefont {E.~E.}\ \bibnamefont {{Falco}}},\ }\href
  {https://doi.org/10.3847/1538-4357/aaed3e} {\bibfield  {journal} {\bibinfo
  {journal} {\apj}\ }\textbf {\bibinfo {volume} {869}},\ \bibinfo {eid} {106}
  (\bibinfo {year} {2018})},\ \Eprint {https://arxiv.org/abs/1812.05639}
  {arXiv:1812.05639 [astro-ph.GA]} \BibitemShut {NoStop}%
\bibitem [{\citenamefont {{Huterer}}\ and\ \citenamefont
  {{Starkman}}(2003)}]{Huterer2003}%
  \BibitemOpen
  \bibfield  {author} {\bibinfo {author} {\bibfnamefont {D.}~\bibnamefont
  {{Huterer}}}\ and\ \bibinfo {author} {\bibfnamefont {G.}~\bibnamefont
  {{Starkman}}},\ }\href {https://doi.org/10.1103/PhysRevLett.90.031301}
  {\bibfield  {journal} {\bibinfo  {journal} {\prl}\ }\textbf {\bibinfo
  {volume} {90}},\ \bibinfo {eid} {031301} (\bibinfo {year} {2003})},\ \Eprint
  {https://arxiv.org/abs/astro-ph/0207517} {arXiv:astro-ph/0207517 [astro-ph]}
  \BibitemShut {NoStop}%
\bibitem [{\citenamefont {{V{\'e}ron-Cetty}}\ and\ \citenamefont
  {{V{\'e}ron}}(2010)}]{AGNcatalog}%
  \BibitemOpen
  \bibfield  {author} {\bibinfo {author} {\bibfnamefont {M.~P.}\ \bibnamefont
  {{V{\'e}ron-Cetty}}}\ and\ \bibinfo {author} {\bibfnamefont {P.}~\bibnamefont
  {{V{\'e}ron}}},\ }\href {https://doi.org/10.1051/0004-6361/201014188}
  {\bibfield  {journal} {\bibinfo  {journal} {\aap}\ }\textbf {\bibinfo
  {volume} {518}},\ \bibinfo {eid} {A10} (\bibinfo {year} {2010})}\BibitemShut
  {NoStop}%
\bibitem [{\citenamefont {{Pe{\~n}a-Herazo}}\ \emph {et~al.}(2022)\citenamefont
  {{Pe{\~n}a-Herazo}}, \citenamefont {{Massaro}}, \citenamefont {{Chavushyan}},
  \citenamefont {{Masetti}}, \citenamefont {{Paggi}},\ and\ \citenamefont
  {{Capetti}}}]{TurinSyCAT}%
  \BibitemOpen
  \bibfield  {author} {\bibinfo {author} {\bibfnamefont {H.~A.}\ \bibnamefont
  {{Pe{\~n}a-Herazo}}}, \bibinfo {author} {\bibfnamefont {F.}~\bibnamefont
  {{Massaro}}}, \bibinfo {author} {\bibfnamefont {V.}~\bibnamefont
  {{Chavushyan}}}, \bibinfo {author} {\bibfnamefont {N.}~\bibnamefont
  {{Masetti}}}, \bibinfo {author} {\bibfnamefont {A.}~\bibnamefont {{Paggi}}},\
  and\ \bibinfo {author} {\bibfnamefont {A.}~\bibnamefont {{Capetti}}},\ }\href
  {https://doi.org/10.1051/0004-6361/202038752} {\bibfield  {journal} {\bibinfo
   {journal} {\aap}\ }\textbf {\bibinfo {volume} {659}},\ \bibinfo {eid} {A32}
  (\bibinfo {year} {2022})}\BibitemShut {NoStop}%
\bibitem [{\citenamefont {{Antonucci}}(1993)}]{Antonucci1993}%
  \BibitemOpen
  \bibfield  {author} {\bibinfo {author} {\bibfnamefont {R.}~\bibnamefont
  {{Antonucci}}},\ }\href {https://doi.org/10.1146/annurev.aa.31.090193.002353}
  {\bibfield  {journal} {\bibinfo  {journal} {\araa}\ }\textbf {\bibinfo
  {volume} {31}},\ \bibinfo {pages} {473} (\bibinfo {year} {1993})}\BibitemShut
  {NoStop}%
\bibitem [{\citenamefont {{Baptista}}\ \emph {et~al.}(1995)\citenamefont
  {{Baptista}}, \citenamefont {{Horne}}, \citenamefont {{Hilditch}},
  \citenamefont {{Mason}},\ and\ \citenamefont {{Drew}}}]{Baptista1995}%
  \BibitemOpen
  \bibfield  {author} {\bibinfo {author} {\bibfnamefont {R.}~\bibnamefont
  {{Baptista}}}, \bibinfo {author} {\bibfnamefont {K.}~\bibnamefont {{Horne}}},
  \bibinfo {author} {\bibfnamefont {R.~W.}\ \bibnamefont {{Hilditch}}},
  \bibinfo {author} {\bibfnamefont {K.~O.}\ \bibnamefont {{Mason}}},\ and\
  \bibinfo {author} {\bibfnamefont {J.~E.}\ \bibnamefont {{Drew}}},\ }\href
  {https://doi.org/10.1086/175970} {\bibfield  {journal} {\bibinfo  {journal}
  {\apj}\ }\textbf {\bibinfo {volume} {448}},\ \bibinfo {pages} {395} (\bibinfo
  {year} {1995})}\BibitemShut {NoStop}%
\bibitem [{\citenamefont {{Linnell}}\ \emph {et~al.}(2008)\citenamefont
  {{Linnell}}, \citenamefont {{Godon}}, \citenamefont {{Hubeny}}, \citenamefont
  {{Sion}},\ and\ \citenamefont {{Szkody}}}]{Linnell2008}%
  \BibitemOpen
  \bibfield  {author} {\bibinfo {author} {\bibfnamefont {A.~P.}\ \bibnamefont
  {{Linnell}}}, \bibinfo {author} {\bibfnamefont {P.}~\bibnamefont {{Godon}}},
  \bibinfo {author} {\bibfnamefont {I.}~\bibnamefont {{Hubeny}}}, \bibinfo
  {author} {\bibfnamefont {E.~M.}\ \bibnamefont {{Sion}}},\ and\ \bibinfo
  {author} {\bibfnamefont {P.}~\bibnamefont {{Szkody}}},\ }\href
  {https://doi.org/10.1086/592104} {\bibfield  {journal} {\bibinfo  {journal}
  {\apj}\ }\textbf {\bibinfo {volume} {688}},\ \bibinfo {pages} {568} (\bibinfo
  {year} {2008})},\ \Eprint {https://arxiv.org/abs/0807.3920} {arXiv:0807.3920
  [astro-ph]} \BibitemShut {NoStop}%
\bibitem [{\citenamefont {{Linnell}}\ \emph {et~al.}(2010)\citenamefont
  {{Linnell}}, \citenamefont {{Godon}}, \citenamefont {{Hubeny}}, \citenamefont
  {{Sion}},\ and\ \citenamefont {{Szkody}}}]{Linnell2010}%
  \BibitemOpen
  \bibfield  {author} {\bibinfo {author} {\bibfnamefont {A.~P.}\ \bibnamefont
  {{Linnell}}}, \bibinfo {author} {\bibfnamefont {P.}~\bibnamefont {{Godon}}},
  \bibinfo {author} {\bibfnamefont {I.}~\bibnamefont {{Hubeny}}}, \bibinfo
  {author} {\bibfnamefont {E.~M.}\ \bibnamefont {{Sion}}},\ and\ \bibinfo
  {author} {\bibfnamefont {P.}~\bibnamefont {{Szkody}}},\ }\href
  {https://doi.org/10.1088/0004-637X/719/1/271} {\bibfield  {journal} {\bibinfo
   {journal} {\apj}\ }\textbf {\bibinfo {volume} {719}},\ \bibinfo {pages}
  {271} (\bibinfo {year} {2010})},\ \Eprint {https://arxiv.org/abs/1006.2832}
  {arXiv:1006.2832 [astro-ph.SR]} \BibitemShut {NoStop}%
\bibitem [{\citenamefont {{Novikov}}\ and\ \citenamefont
  {{Thorne}}(1973)}]{NovikovThorne}%
  \BibitemOpen
  \bibfield  {author} {\bibinfo {author} {\bibfnamefont {I.~D.}\ \bibnamefont
  {{Novikov}}}\ and\ \bibinfo {author} {\bibfnamefont {K.~S.}\ \bibnamefont
  {{Thorne}}},\ }in\ \href@noop {} {\emph {\bibinfo {booktitle} {Black Holes
  (Les Astres Occlus)}}}\ (\bibinfo {year} {1973})\ pp.\ \bibinfo {pages}
  {343--450}\BibitemShut {NoStop}%
\bibitem [{\citenamefont {{Mo{\'s}cibrodzka}}\ and\ \citenamefont
  {{Gammie}}(2018)}]{ipole}%
  \BibitemOpen
  \bibfield  {author} {\bibinfo {author} {\bibfnamefont {M.}~\bibnamefont
  {{Mo{\'s}cibrodzka}}}\ and\ \bibinfo {author} {\bibfnamefont {C.~F.}\
  \bibnamefont {{Gammie}}},\ }\href {https://doi.org/10.1093/mnras/stx3162}
  {\bibfield  {journal} {\bibinfo  {journal} {\mnras}\ }\textbf {\bibinfo
  {volume} {475}},\ \bibinfo {pages} {43} (\bibinfo {year} {2018})},\ \Eprint
  {https://arxiv.org/abs/1712.03057} {arXiv:1712.03057 [astro-ph.HE]}
  \BibitemShut {NoStop}%
\bibitem [{\citenamefont {{Peterson}}(2006)}]{Peterson2006}%
  \BibitemOpen
  \bibfield  {author} {\bibinfo {author} {\bibfnamefont {B.~M.}\ \bibnamefont
  {{Peterson}}},\ }in\ \href {https://doi.org/10.1007/3-540-34621-X_3} {\emph
  {\bibinfo {booktitle} {Physics of Active Galactic Nuclei at all Scales}}},\
  Vol.\ \bibinfo {volume} {693},\ \bibinfo {editor} {edited by\ \bibinfo
  {editor} {\bibfnamefont {D.}~\bibnamefont {{Alloin}}}}\ (\bibinfo
  {publisher} {{Springer}},\ \bibinfo {year} {2006})\ p.~\bibinfo {pages}
  {77},\ \bibinfo {note}
  {\url{https://ned.ipac.caltech.edu/level5/Sept16/Peterson/paper.pdf}}\BibitemShut
  {NoStop}%
\bibitem [{\citenamefont {{Kinemuchi}}\ \emph {et~al.}(2020)\citenamefont
  {{Kinemuchi}}, \citenamefont {{Hall}}, \citenamefont {{McGreer}} \emph
  {et~al.}}]{SDSSRM}%
  \BibitemOpen
  \bibfield  {author} {\bibinfo {author} {\bibfnamefont {K.}~\bibnamefont
  {{Kinemuchi}}}, \bibinfo {author} {\bibfnamefont {P.~B.}\ \bibnamefont
  {{Hall}}}, \bibinfo {author} {\bibfnamefont {I.}~\bibnamefont {{McGreer}}},
  \emph {et~al.},\ }\href {https://doi.org/10.3847/1538-4365/aba43f} {\bibfield
   {journal} {\bibinfo  {journal} {\apjs}\ }\textbf {\bibinfo {volume} {250}},\
  \bibinfo {eid} {10} (\bibinfo {year} {2020})},\ \Eprint
  {https://arxiv.org/abs/2007.05160} {arXiv:2007.05160 [astro-ph.GA]}
  \BibitemShut {NoStop}%
\bibitem [{\citenamefont {{Shen}}\ \emph {et~al.}(2023)\citenamefont {{Shen}},
  \citenamefont {{Grier}}, \citenamefont {{Horne}} \emph
  {et~al.}}]{2023arXiv230501014S}%
  \BibitemOpen
  \bibfield  {author} {\bibinfo {author} {\bibfnamefont {Y.}~\bibnamefont
  {{Shen}}}, \bibinfo {author} {\bibfnamefont {C.~J.}\ \bibnamefont {{Grier}}},
  \bibinfo {author} {\bibfnamefont {K.}~\bibnamefont {{Horne}}}, \emph
  {et~al.},\ }\href {https://doi.org/10.48550/arXiv.2305.01014} {\bibfield
  {journal} {\bibinfo  {journal} {arXiv e-prints}\ ,\ \bibinfo {eid}
  {arXiv:2305.01014}} (\bibinfo {year} {2023})},\ \Eprint
  {https://arxiv.org/abs/2305.01014} {arXiv:2305.01014 [astro-ph.GA]}
  \BibitemShut {NoStop}%
\bibitem [{\citenamefont {{Yu}}\ \emph {et~al.}(2023)\citenamefont {{Yu}},
  \citenamefont {{Martini}}, \citenamefont {{Penton}} \emph {et~al.}}]{OzDES}%
  \BibitemOpen
  \bibfield  {author} {\bibinfo {author} {\bibfnamefont {Z.}~\bibnamefont
  {{Yu}}}, \bibinfo {author} {\bibfnamefont {P.}~\bibnamefont {{Martini}}},
  \bibinfo {author} {\bibfnamefont {A.}~\bibnamefont {{Penton}}}, \emph
  {et~al.},\ }\href {https://doi.org/10.1093/mnras/stad1224} {\bibfield
  {journal} {\bibinfo  {journal} {\mnras}\ }\textbf {\bibinfo {volume} {522}},\
  \bibinfo {pages} {4132} (\bibinfo {year} {2023})},\ \Eprint
  {https://arxiv.org/abs/2208.05491} {arXiv:2208.05491 [astro-ph.GA]}
  \BibitemShut {NoStop}%
\bibitem [{\citenamefont {{Bahcall}}\ \emph {et~al.}(1972)\citenamefont
  {{Bahcall}}, \citenamefont {{Kozlovsky}},\ and\ \citenamefont
  {{Salpeter}}}]{Bahcall1972}%
  \BibitemOpen
  \bibfield  {author} {\bibinfo {author} {\bibfnamefont {J.~N.}\ \bibnamefont
  {{Bahcall}}}, \bibinfo {author} {\bibfnamefont {B.-Z.}\ \bibnamefont
  {{Kozlovsky}}},\ and\ \bibinfo {author} {\bibfnamefont {E.~E.}\ \bibnamefont
  {{Salpeter}}},\ }\href {https://doi.org/10.1086/151300} {\bibfield  {journal}
  {\bibinfo  {journal} {\apj}\ }\textbf {\bibinfo {volume} {171}},\ \bibinfo
  {pages} {467} (\bibinfo {year} {1972})}\BibitemShut {NoStop}%
\bibitem [{\citenamefont {{Blandford}}\ and\ \citenamefont
  {{McKee}}(1982)}]{Blandford1982}%
  \BibitemOpen
  \bibfield  {author} {\bibinfo {author} {\bibfnamefont {R.~D.}\ \bibnamefont
  {{Blandford}}}\ and\ \bibinfo {author} {\bibfnamefont {C.~F.}\ \bibnamefont
  {{McKee}}},\ }\href {https://doi.org/10.1086/159843} {\bibfield  {journal}
  {\bibinfo  {journal} {\apj}\ }\textbf {\bibinfo {volume} {255}},\ \bibinfo
  {pages} {419} (\bibinfo {year} {1982})}\BibitemShut {NoStop}%
\bibitem [{\citenamefont {{Peterson}}(1993)}]{Peterson1993}%
  \BibitemOpen
  \bibfield  {author} {\bibinfo {author} {\bibfnamefont {B.~M.}\ \bibnamefont
  {{Peterson}}},\ }\href {https://doi.org/10.1086/133140} {\bibfield  {journal}
  {\bibinfo  {journal} {\pasp}\ }\textbf {\bibinfo {volume} {105}},\ \bibinfo
  {pages} {247} (\bibinfo {year} {1993})}\BibitemShut {NoStop}%
\bibitem [{\citenamefont {Kamionkowski}\ and\ \citenamefont
  {Riess}(2023)}]{Kamionkowski2023}%
  \BibitemOpen
  \bibfield  {author} {\bibinfo {author} {\bibfnamefont {M.}~\bibnamefont
  {Kamionkowski}}\ and\ \bibinfo {author} {\bibfnamefont {A.~G.}\ \bibnamefont
  {Riess}},\ }\href {https://doi.org/10.1146/annurev-nucl-111422-024107}
  {\bibfield  {journal} {\bibinfo  {journal} {Annual Review of Nuclear and
  Particle Science}\ }\textbf {\bibinfo {volume} {73}},\ \bibinfo {pages} {153}
  (\bibinfo {year} {2023})},\ \Eprint {https://arxiv.org/abs/2211.04492}
  {arXiv:2211.04492 [astro-ph.CO]} \BibitemShut {NoStop}%
\bibitem [{\citenamefont {{Gravity Collaboration}}\ \emph
  {et~al.}(2018)\citenamefont {{Gravity Collaboration}}, \citenamefont
  {{Sturm}}, \citenamefont {{Dexter}}, \citenamefont {{Pfuhl}} \emph
  {et~al.}}]{2018Natur.563..657G}%
  \BibitemOpen
  \bibfield  {author} {\bibinfo {author} {\bibnamefont {{Gravity
  Collaboration}}}, \bibinfo {author} {\bibfnamefont {E.}~\bibnamefont
  {{Sturm}}}, \bibinfo {author} {\bibfnamefont {J.}~\bibnamefont {{Dexter}}},
  \bibinfo {author} {\bibfnamefont {O.}~\bibnamefont {{Pfuhl}}}, \emph
  {et~al.},\ }\href {https://doi.org/10.1038/s41586-018-0731-9} {\bibfield
  {journal} {\bibinfo  {journal} {\nat}\ }\textbf {\bibinfo {volume} {563}},\
  \bibinfo {pages} {657} (\bibinfo {year} {2018})},\ \Eprint
  {https://arxiv.org/abs/1811.11195} {arXiv:1811.11195 [astro-ph.GA]}
  \BibitemShut {NoStop}%
\bibitem [{\citenamefont {{GRAVITY Collaboration}}\ \emph
  {et~al.}(2020)\citenamefont {{GRAVITY Collaboration}}, \citenamefont
  {{Amorim}}, \citenamefont {{Baub{\"o}ck}}, \citenamefont {{Brandner}} \emph
  {et~al.}}]{2020A&A...643A.154G}%
  \BibitemOpen
  \bibfield  {author} {\bibinfo {author} {\bibnamefont {{GRAVITY
  Collaboration}}}, \bibinfo {author} {\bibfnamefont {A.}~\bibnamefont
  {{Amorim}}}, \bibinfo {author} {\bibfnamefont {M.}~\bibnamefont
  {{Baub{\"o}ck}}}, \bibinfo {author} {\bibfnamefont {W.}~\bibnamefont
  {{Brandner}}}, \emph {et~al.},\ }\href
  {https://doi.org/10.1051/0004-6361/202039067} {\bibfield  {journal} {\bibinfo
   {journal} {\aap}\ }\textbf {\bibinfo {volume} {643}},\ \bibinfo {eid} {A154}
  (\bibinfo {year} {2020})},\ \Eprint {https://arxiv.org/abs/2009.08463}
  {arXiv:2009.08463 [astro-ph.GA]} \BibitemShut {NoStop}%
\bibitem [{\citenamefont {{GRAVITY Collaboration}}\ \emph
  {et~al.}(2021)\citenamefont {{GRAVITY Collaboration}}, \citenamefont
  {{Amorim}}, \citenamefont {{Baub{\"o}ck}}, \citenamefont {{Brandner}} \emph
  {et~al.}}]{2021A&A...648A.117G}%
  \BibitemOpen
  \bibfield  {author} {\bibinfo {author} {\bibnamefont {{GRAVITY
  Collaboration}}}, \bibinfo {author} {\bibfnamefont {A.}~\bibnamefont
  {{Amorim}}}, \bibinfo {author} {\bibfnamefont {M.}~\bibnamefont
  {{Baub{\"o}ck}}}, \bibinfo {author} {\bibfnamefont {W.}~\bibnamefont
  {{Brandner}}}, \emph {et~al.},\ }\href
  {https://doi.org/10.1051/0004-6361/202040061} {\bibfield  {journal} {\bibinfo
   {journal} {\aap}\ }\textbf {\bibinfo {volume} {648}},\ \bibinfo {eid} {A117}
  (\bibinfo {year} {2021})},\ \Eprint {https://arxiv.org/abs/2102.00068}
  {arXiv:2102.00068 [astro-ph.GA]} \BibitemShut {NoStop}%
\bibitem [{\citenamefont {{GRAVITY Collaboration}}\ \emph
  {et~al.}(2024)\citenamefont {{GRAVITY Collaboration}}, \citenamefont
  {{Amorim}}, \citenamefont {{Bourdarot}}, \citenamefont {{Brandner}} \emph
  {et~al.}}]{2024arXiv240107676G}%
  \BibitemOpen
  \bibfield  {author} {\bibinfo {author} {\bibnamefont {{GRAVITY
  Collaboration}}}, \bibinfo {author} {\bibfnamefont {A.}~\bibnamefont
  {{Amorim}}}, \bibinfo {author} {\bibfnamefont {G.}~\bibnamefont
  {{Bourdarot}}}, \bibinfo {author} {\bibfnamefont {W.}~\bibnamefont
  {{Brandner}}}, \emph {et~al.},\ }\href
  {https://doi.org/10.48550/arXiv.2401.07676} {\bibfield  {journal} {\bibinfo
  {journal} {arXiv e-prints}\ ,\ \bibinfo {eid} {arXiv:2401.07676}} (\bibinfo
  {year} {2024})},\ \Eprint {https://arxiv.org/abs/2401.07676}
  {arXiv:2401.07676 [astro-ph.GA]} \BibitemShut {NoStop}%
\bibitem [{\citenamefont {{Abuter}}\ \emph {et~al.}(2024)\citenamefont
  {{Abuter}}, \citenamefont {{Allouche}}, \citenamefont {{Amorim}} \emph
  {et~al.}}]{2024arXiv240114567A}%
  \BibitemOpen
  \bibfield  {author} {\bibinfo {author} {\bibfnamefont {R.}~\bibnamefont
  {{Abuter}}}, \bibinfo {author} {\bibfnamefont {F.}~\bibnamefont
  {{Allouche}}}, \bibinfo {author} {\bibfnamefont {A.}~\bibnamefont
  {{Amorim}}}, \emph {et~al.},\ }\href
  {https://doi.org/10.48550/arXiv.2401.14567} {\bibfield  {journal} {\bibinfo
  {journal} {arXiv e-prints}\ ,\ \bibinfo {eid} {arXiv:2401.14567}} (\bibinfo
  {year} {2024})},\ \Eprint {https://arxiv.org/abs/2401.14567}
  {arXiv:2401.14567 [astro-ph.GA]} \BibitemShut {NoStop}%
\bibitem [{\citenamefont {{Koss}}\ \emph {et~al.}(2022)\citenamefont {{Koss}},
  \citenamefont {{Ricci}}, \citenamefont {{Trakhtenbrot}} \emph
  {et~al.}}]{BASS_DR2}%
  \BibitemOpen
  \bibfield  {author} {\bibinfo {author} {\bibfnamefont {M.~J.}\ \bibnamefont
  {{Koss}}}, \bibinfo {author} {\bibfnamefont {C.}~\bibnamefont {{Ricci}}},
  \bibinfo {author} {\bibfnamefont {B.}~\bibnamefont {{Trakhtenbrot}}}, \emph
  {et~al.},\ }\href {https://doi.org/10.3847/1538-4365/ac6c05} {\bibfield
  {journal} {\bibinfo  {journal} {\apjs}\ }\textbf {\bibinfo {volume} {261}},\
  \bibinfo {eid} {2} (\bibinfo {year} {2022})},\ \Eprint
  {https://arxiv.org/abs/2207.12432} {arXiv:2207.12432 [astro-ph.GA]}
  \BibitemShut {NoStop}%
\bibitem [{\citenamefont {{Dietrich}}\ \emph {et~al.}(1999)\citenamefont
  {{Dietrich}}, \citenamefont {{Wagner}}, \citenamefont {{Courvoisier}},
  \citenamefont {{Bock}},\ and\ \citenamefont {{North}}}]{Dietrich1999}%
  \BibitemOpen
  \bibfield  {author} {\bibinfo {author} {\bibfnamefont {M.}~\bibnamefont
  {{Dietrich}}}, \bibinfo {author} {\bibfnamefont {S.~J.}\ \bibnamefont
  {{Wagner}}}, \bibinfo {author} {\bibfnamefont {T.~J.~L.}\ \bibnamefont
  {{Courvoisier}}}, \bibinfo {author} {\bibfnamefont {H.}~\bibnamefont
  {{Bock}}},\ and\ \bibinfo {author} {\bibfnamefont {P.}~\bibnamefont
  {{North}}},\ }\href@noop {} {\bibfield  {journal} {\bibinfo  {journal}
  {\aap}\ }\textbf {\bibinfo {volume} {351}},\ \bibinfo {pages} {31} (\bibinfo
  {year} {1999})}\BibitemShut {NoStop}%
\bibitem [{\citenamefont {{Murray}}\ and\ \citenamefont
  {{Chiang}}(1997)}]{Murray1997}%
  \BibitemOpen
  \bibfield  {author} {\bibinfo {author} {\bibfnamefont {N.}~\bibnamefont
  {{Murray}}}\ and\ \bibinfo {author} {\bibfnamefont {J.}~\bibnamefont
  {{Chiang}}},\ }\href {https://doi.org/10.1086/303443} {\bibfield  {journal}
  {\bibinfo  {journal} {\apj}\ }\textbf {\bibinfo {volume} {474}},\ \bibinfo
  {pages} {91} (\bibinfo {year} {1997})}\BibitemShut {NoStop}%
\bibitem [{\citenamefont {{Riebe}}\ \emph {et~al.}(2013)\citenamefont
  {{Riebe}}, \citenamefont {{Partl}}, \citenamefont {{Enke}}, \citenamefont
  {{Forero-Romero}}, \citenamefont {{Gottl{\"o}ber}}, \citenamefont {{Klypin}},
  \citenamefont {{Lemson}}, \citenamefont {{Prada}}, \citenamefont {{Primack}},
  \citenamefont {{Steinmetz}},\ and\ \citenamefont
  {{Turchaninov}}}]{MultiDark}%
  \BibitemOpen
  \bibfield  {author} {\bibinfo {author} {\bibfnamefont {K.}~\bibnamefont
  {{Riebe}}}, \bibinfo {author} {\bibfnamefont {A.~M.}\ \bibnamefont
  {{Partl}}}, \bibinfo {author} {\bibfnamefont {H.}~\bibnamefont {{Enke}}},
  \bibinfo {author} {\bibfnamefont {J.}~\bibnamefont {{Forero-Romero}}},
  \bibinfo {author} {\bibfnamefont {S.}~\bibnamefont {{Gottl{\"o}ber}}},
  \bibinfo {author} {\bibfnamefont {A.}~\bibnamefont {{Klypin}}}, \bibinfo
  {author} {\bibfnamefont {G.}~\bibnamefont {{Lemson}}}, \bibinfo {author}
  {\bibfnamefont {F.}~\bibnamefont {{Prada}}}, \bibinfo {author} {\bibfnamefont
  {J.~R.}\ \bibnamefont {{Primack}}}, \bibinfo {author} {\bibfnamefont
  {M.}~\bibnamefont {{Steinmetz}}},\ and\ \bibinfo {author} {\bibfnamefont
  {V.}~\bibnamefont {{Turchaninov}}},\ }\href
  {https://doi.org/10.1002/asna.201211900} {\bibfield  {journal} {\bibinfo
  {journal} {Astronomische Nachrichten}\ }\textbf {\bibinfo {volume} {334}},\
  \bibinfo {pages} {691} (\bibinfo {year} {2013})}\BibitemShut {NoStop}%
\bibitem [{\citenamefont {Riess}\ \emph {et~al.}(2022)\citenamefont {Riess},
  \citenamefont {Yuan}, \citenamefont {Macri} \emph {et~al.}}]{Riess_2022}%
  \BibitemOpen
  \bibfield  {author} {\bibinfo {author} {\bibfnamefont {A.~G.}\ \bibnamefont
  {Riess}}, \bibinfo {author} {\bibfnamefont {W.}~\bibnamefont {Yuan}},
  \bibinfo {author} {\bibfnamefont {L.~M.}\ \bibnamefont {Macri}}, \emph
  {et~al.},\ }\href {https://doi.org/10.3847/2041-8213/ac5c5b} {\bibfield
  {journal} {\bibinfo  {journal} {The Astrophysical Journal Letters}\ }\textbf
  {\bibinfo {volume} {934}},\ \bibinfo {pages} {L7} (\bibinfo {year}
  {2022})}\BibitemShut {NoStop}%
\bibitem [{\citenamefont {{Planck Collaboration}}\ \emph
  {et~al.}(2020)\citenamefont {{Planck Collaboration}}, \citenamefont
  {{Aghanim, N.}}, \citenamefont {{Akrami, Y.}}, \citenamefont {{Ashdown, M.}}
  \emph {et~al.}}]{Planck2018}%
  \BibitemOpen
  \bibfield  {author} {\bibinfo {author} {\bibnamefont {{Planck
  Collaboration}}}, \bibinfo {author} {\bibnamefont {{Aghanim, N.}}}, \bibinfo
  {author} {\bibnamefont {{Akrami, Y.}}}, \bibinfo {author} {\bibnamefont
  {{Ashdown, M.}}}, \emph {et~al.},\ }\href
  {https://doi.org/10.1051/0004-6361/201833910} {\bibfield  {journal} {\bibinfo
   {journal} {A\&A}\ }\textbf {\bibinfo {volume} {641}},\ \bibinfo {pages} {A6}
  (\bibinfo {year} {2020})}\BibitemShut {NoStop}%
\bibitem [{\citenamefont {{Sergeev}}\ \emph {et~al.}(2005)\citenamefont
  {{Sergeev}}, \citenamefont {{Doroshenko}}, \citenamefont {{Golubinskiy}},
  \citenamefont {{Merkulova}},\ and\ \citenamefont {{Sergeeva}}}]{Sergeev2005}%
  \BibitemOpen
  \bibfield  {author} {\bibinfo {author} {\bibfnamefont {S.~G.}\ \bibnamefont
  {{Sergeev}}}, \bibinfo {author} {\bibfnamefont {V.~T.}\ \bibnamefont
  {{Doroshenko}}}, \bibinfo {author} {\bibfnamefont {Y.~V.}\ \bibnamefont
  {{Golubinskiy}}}, \bibinfo {author} {\bibfnamefont {N.~I.}\ \bibnamefont
  {{Merkulova}}},\ and\ \bibinfo {author} {\bibfnamefont {E.~A.}\ \bibnamefont
  {{Sergeeva}}},\ }\href {https://doi.org/10.1086/427820} {\bibfield  {journal}
  {\bibinfo  {journal} {\apj}\ }\textbf {\bibinfo {volume} {622}},\ \bibinfo
  {pages} {129} (\bibinfo {year} {2005})}\BibitemShut {NoStop}%
\bibitem [{\citenamefont {{Cackett}}\ \emph {et~al.}(2007)\citenamefont
  {{Cackett}}, \citenamefont {{Horne}},\ and\ \citenamefont
  {{Winkler}}}]{Cackett2007}%
  \BibitemOpen
  \bibfield  {author} {\bibinfo {author} {\bibfnamefont {E.~M.}\ \bibnamefont
  {{Cackett}}}, \bibinfo {author} {\bibfnamefont {K.}~\bibnamefont {{Horne}}},\
  and\ \bibinfo {author} {\bibfnamefont {H.}~\bibnamefont {{Winkler}}},\ }\href
  {https://doi.org/10.1111/j.1365-2966.2007.12098.x} {\bibfield  {journal}
  {\bibinfo  {journal} {\mnras}\ }\textbf {\bibinfo {volume} {380}},\ \bibinfo
  {pages} {669} (\bibinfo {year} {2007})},\ \Eprint
  {https://arxiv.org/abs/0706.1464} {arXiv:0706.1464 [astro-ph]} \BibitemShut
  {NoStop}%
\bibitem [{\citenamefont {{Neustadt}}\ and\ \citenamefont
  {{Kochanek}}(2022)}]{Neustadt2022MNRAS}%
  \BibitemOpen
  \bibfield  {author} {\bibinfo {author} {\bibfnamefont {J.~M.~M.}\
  \bibnamefont {{Neustadt}}}\ and\ \bibinfo {author} {\bibfnamefont {C.~S.}\
  \bibnamefont {{Kochanek}}},\ }\href {https://doi.org/10.1093/mnras/stac888}
  {\bibfield  {journal} {\bibinfo  {journal} {\mnras}\ }\textbf {\bibinfo
  {volume} {513}},\ \bibinfo {pages} {1046} (\bibinfo {year} {2022})},\ \Eprint
  {https://arxiv.org/abs/2201.10565} {arXiv:2201.10565 [astro-ph.GA]}
  \BibitemShut {NoStop}%
\bibitem [{\citenamefont {{Shappee}}\ \emph {et~al.}(2014)\citenamefont
  {{Shappee}}, \citenamefont {{Prieto}}, \citenamefont {{Grupe}} \emph
  {et~al.}}]{Shappee2014}%
  \BibitemOpen
  \bibfield  {author} {\bibinfo {author} {\bibfnamefont {B.~J.}\ \bibnamefont
  {{Shappee}}}, \bibinfo {author} {\bibfnamefont {J.~L.}\ \bibnamefont
  {{Prieto}}}, \bibinfo {author} {\bibfnamefont {D.}~\bibnamefont {{Grupe}}},
  \emph {et~al.},\ }\href {https://doi.org/10.1088/0004-637X/788/1/48}
  {\bibfield  {journal} {\bibinfo  {journal} {\apj}\ }\textbf {\bibinfo
  {volume} {788}},\ \bibinfo {eid} {48} (\bibinfo {year} {2014})},\ \Eprint
  {https://arxiv.org/abs/1310.2241} {arXiv:1310.2241 [astro-ph.HE]}
  \BibitemShut {NoStop}%
\bibitem [{\citenamefont {{Fausnaugh}}\ \emph {et~al.}(2016)\citenamefont
  {{Fausnaugh}}, \citenamefont {{Denney}}, \citenamefont {{Barth}} \emph
  {et~al.}}]{Fausnaugh2016}%
  \BibitemOpen
  \bibfield  {author} {\bibinfo {author} {\bibfnamefont {M.~M.}\ \bibnamefont
  {{Fausnaugh}}}, \bibinfo {author} {\bibfnamefont {K.~D.}\ \bibnamefont
  {{Denney}}}, \bibinfo {author} {\bibfnamefont {A.~J.}\ \bibnamefont
  {{Barth}}}, \emph {et~al.},\ }\href
  {https://doi.org/10.3847/0004-637X/821/1/56} {\bibfield  {journal} {\bibinfo
  {journal} {\apj}\ }\textbf {\bibinfo {volume} {821}},\ \bibinfo {eid} {56}
  (\bibinfo {year} {2016})},\ \Eprint {https://arxiv.org/abs/1510.05648}
  {arXiv:1510.05648 [astro-ph.GA]} \BibitemShut {NoStop}%
\bibitem [{\citenamefont {{Werner}}\ \emph {et~al.}(2022)\citenamefont
  {{Werner}}, \citenamefont {{{\v{R}}{\'\i}pa}}, \citenamefont {{M{\"u}nz}}
  \emph {et~al.}}]{QUVIK}%
  \BibitemOpen
  \bibfield  {author} {\bibinfo {author} {\bibfnamefont {N.}~\bibnamefont
  {{Werner}}}, \bibinfo {author} {\bibfnamefont {J.}~\bibnamefont
  {{{\v{R}}{\'\i}pa}}}, \bibinfo {author} {\bibfnamefont {F.}~\bibnamefont
  {{M{\"u}nz}}}, \emph {et~al.},\ }in\ \href
  {https://doi.org/10.1117/12.2629531} {\emph {\bibinfo {booktitle} {Space
  Telescopes and Instrumentation 2022: Ultraviolet to Gamma Ray}}},\ \bibinfo
  {series} {Society of Photo-Optical Instrumentation Engineers (SPIE)
  Conference Series}, Vol.\ \bibinfo {volume} {12181},\ \bibinfo {editor}
  {edited by\ \bibinfo {editor} {\bibfnamefont {J.-W.~A.}\ \bibnamefont {{den
  Herder}}}, \bibinfo {editor} {\bibfnamefont {S.}~\bibnamefont {{Nikzad}}},\
  and\ \bibinfo {editor} {\bibfnamefont {K.}~\bibnamefont {{Nakazawa}}}}\
  (\bibinfo {year} {2022})\ p.\ \bibinfo {pages} {121810B},\ \Eprint
  {https://arxiv.org/abs/2207.05485} {arXiv:2207.05485 [astro-ph.IM]}
  \BibitemShut {NoStop}%
\bibitem [{\citenamefont {{Stone}}\ \emph {et~al.}(2019)\citenamefont
  {{Stone}}, \citenamefont {{Kesden}}, \citenamefont {{Cheng}},\ and\
  \citenamefont {{van Velzen}}}]{Stone2019}%
  \BibitemOpen
  \bibfield  {author} {\bibinfo {author} {\bibfnamefont {N.~C.}\ \bibnamefont
  {{Stone}}}, \bibinfo {author} {\bibfnamefont {M.}~\bibnamefont {{Kesden}}},
  \bibinfo {author} {\bibfnamefont {R.~M.}\ \bibnamefont {{Cheng}}},\ and\
  \bibinfo {author} {\bibfnamefont {S.}~\bibnamefont {{van Velzen}}},\ }\href
  {https://doi.org/10.1007/s10714-019-2510-9} {\bibfield  {journal} {\bibinfo
  {journal} {General Relativity and Gravitation}\ }\textbf {\bibinfo {volume}
  {51}},\ \bibinfo {eid} {30} (\bibinfo {year} {2019})},\ \Eprint
  {https://arxiv.org/abs/1801.10180} {arXiv:1801.10180 [astro-ph.HE]}
  \BibitemShut {NoStop}%
\bibitem [{\citenamefont {{Gezari}}(2021)}]{Gezari2021}%
  \BibitemOpen
  \bibfield  {author} {\bibinfo {author} {\bibfnamefont {S.}~\bibnamefont
  {{Gezari}}},\ }\href {https://doi.org/10.1146/annurev-astro-111720-030029}
  {\bibfield  {journal} {\bibinfo  {journal} {\araa}\ }\textbf {\bibinfo
  {volume} {59}},\ \bibinfo {pages} {21} (\bibinfo {year} {2021})},\ \Eprint
  {https://arxiv.org/abs/2104.14580} {arXiv:2104.14580 [astro-ph.HE]}
  \BibitemShut {NoStop}%
\bibitem [{\citenamefont {{Piran}}\ \emph {et~al.}(2015)\citenamefont
  {{Piran}}, \citenamefont {{Svirski}}, \citenamefont {{Krolik}}, \citenamefont
  {{Cheng}},\ and\ \citenamefont {{Shiokawa}}}]{Piran2015}%
  \BibitemOpen
  \bibfield  {author} {\bibinfo {author} {\bibfnamefont {T.}~\bibnamefont
  {{Piran}}}, \bibinfo {author} {\bibfnamefont {G.}~\bibnamefont {{Svirski}}},
  \bibinfo {author} {\bibfnamefont {J.}~\bibnamefont {{Krolik}}}, \bibinfo
  {author} {\bibfnamefont {R.~M.}\ \bibnamefont {{Cheng}}},\ and\ \bibinfo
  {author} {\bibfnamefont {H.}~\bibnamefont {{Shiokawa}}},\ }\href
  {https://doi.org/10.1088/0004-637X/806/2/164} {\bibfield  {journal} {\bibinfo
   {journal} {\apj}\ }\textbf {\bibinfo {volume} {806}},\ \bibinfo {eid} {164}
  (\bibinfo {year} {2015})},\ \Eprint {https://arxiv.org/abs/1502.05792}
  {arXiv:1502.05792 [astro-ph.HE]} \BibitemShut {NoStop}%
\bibitem [{\citenamefont {{Metzger}}\ and\ \citenamefont
  {{Stone}}(2016)}]{Metzger2016}%
  \BibitemOpen
  \bibfield  {author} {\bibinfo {author} {\bibfnamefont {B.~D.}\ \bibnamefont
  {{Metzger}}}\ and\ \bibinfo {author} {\bibfnamefont {N.~C.}\ \bibnamefont
  {{Stone}}},\ }\href {https://doi.org/10.1093/mnras/stw1394} {\bibfield
  {journal} {\bibinfo  {journal} {\mnras}\ }\textbf {\bibinfo {volume} {461}},\
  \bibinfo {pages} {948} (\bibinfo {year} {2016})},\ \Eprint
  {https://arxiv.org/abs/1506.03453} {arXiv:1506.03453 [astro-ph.HE]}
  \BibitemShut {NoStop}%
\bibitem [{\citenamefont {{Steinberg}}\ and\ \citenamefont
  {{Stone}}(2022)}]{Steinberg2022}%
  \BibitemOpen
  \bibfield  {author} {\bibinfo {author} {\bibfnamefont {E.}~\bibnamefont
  {{Steinberg}}}\ and\ \bibinfo {author} {\bibfnamefont {N.~C.}\ \bibnamefont
  {{Stone}}},\ }\href {https://doi.org/10.48550/arXiv.2206.10641} {\bibfield
  {journal} {\bibinfo  {journal} {arXiv e-prints}\ ,\ \bibinfo {eid}
  {arXiv:2206.10641}} (\bibinfo {year} {2022})},\ \Eprint
  {https://arxiv.org/abs/2206.10641} {arXiv:2206.10641 [astro-ph.HE]}
  \BibitemShut {NoStop}%
\bibitem [{\citenamefont {{Luminet}}(1979)}]{luminet1979}%
  \BibitemOpen
  \bibfield  {author} {\bibinfo {author} {\bibfnamefont {J.~P.}\ \bibnamefont
  {{Luminet}}},\ }\href@noop {} {\bibfield  {journal} {\bibinfo  {journal}
  {\aap}\ }\textbf {\bibinfo {volume} {75}},\ \bibinfo {pages} {228} (\bibinfo
  {year} {1979})}\BibitemShut {NoStop}%
\bibitem [{\citenamefont {{Gralla}}\ \emph {et~al.}(2019)\citenamefont
  {{Gralla}}, \citenamefont {{Holz}},\ and\ \citenamefont
  {{Wald}}}]{Gralla2019}%
  \BibitemOpen
  \bibfield  {author} {\bibinfo {author} {\bibfnamefont {S.~E.}\ \bibnamefont
  {{Gralla}}}, \bibinfo {author} {\bibfnamefont {D.~E.}\ \bibnamefont
  {{Holz}}},\ and\ \bibinfo {author} {\bibfnamefont {R.~M.}\ \bibnamefont
  {{Wald}}},\ }\href {https://doi.org/10.1103/PhysRevD.100.024018} {\bibfield
  {journal} {\bibinfo  {journal} {\prd}\ }\textbf {\bibinfo {volume} {100}},\
  \bibinfo {eid} {024018} (\bibinfo {year} {2019})},\ \Eprint
  {https://arxiv.org/abs/1906.00873} {arXiv:1906.00873 [astro-ph.HE]}
  \BibitemShut {NoStop}%
\bibitem [{\citenamefont {{Johnson}}\ \emph {et~al.}(2020)\citenamefont
  {{Johnson}}, \citenamefont {{Lupsasca}}, \citenamefont {{Strominger}} \emph
  {et~al.}}]{Johnson2020}%
  \BibitemOpen
  \bibfield  {author} {\bibinfo {author} {\bibfnamefont {M.~D.}\ \bibnamefont
  {{Johnson}}}, \bibinfo {author} {\bibfnamefont {A.}~\bibnamefont
  {{Lupsasca}}}, \bibinfo {author} {\bibfnamefont {A.}~\bibnamefont
  {{Strominger}}}, \emph {et~al.},\ }\href
  {https://doi.org/10.1126/sciadv.aaz1310} {\bibfield  {journal} {\bibinfo
  {journal} {Science Advances}\ }\textbf {\bibinfo {volume} {6}},\ \bibinfo
  {pages} {1310} (\bibinfo {year} {2020})},\ \Eprint
  {https://arxiv.org/abs/1907.04329} {arXiv:1907.04329 [astro-ph.IM]}
  \BibitemShut {NoStop}%
\bibitem [{\citenamefont {{Gralla}}\ \emph {et~al.}(2020)\citenamefont
  {{Gralla}}, \citenamefont {{Lupsasca}},\ and\ \citenamefont
  {{Marrone}}}]{gralla2020b}%
  \BibitemOpen
  \bibfield  {author} {\bibinfo {author} {\bibfnamefont {S.~E.}\ \bibnamefont
  {{Gralla}}}, \bibinfo {author} {\bibfnamefont {A.}~\bibnamefont
  {{Lupsasca}}},\ and\ \bibinfo {author} {\bibfnamefont {D.~P.}\ \bibnamefont
  {{Marrone}}},\ }\href {https://doi.org/10.1103/PhysRevD.102.124004}
  {\bibfield  {journal} {\bibinfo  {journal} {\prd}\ }\textbf {\bibinfo
  {volume} {102}},\ \bibinfo {eid} {124004} (\bibinfo {year} {2020})},\ \Eprint
  {https://arxiv.org/abs/2008.03879} {arXiv:2008.03879 [gr-qc]} \BibitemShut
  {NoStop}%
\bibitem [{\citenamefont {Staelens}\ \emph {et~al.}(2023)\citenamefont
  {Staelens}, \citenamefont {Mayerson}, \citenamefont {Bacchini}, \citenamefont
  {Ripperda},\ and\ \citenamefont {K\"uchler}}]{Staelens:2023jgr}%
  \BibitemOpen
  \bibfield  {author} {\bibinfo {author} {\bibfnamefont {S.}~\bibnamefont
  {Staelens}}, \bibinfo {author} {\bibfnamefont {D.~R.}\ \bibnamefont
  {Mayerson}}, \bibinfo {author} {\bibfnamefont {F.}~\bibnamefont {Bacchini}},
  \bibinfo {author} {\bibfnamefont {B.}~\bibnamefont {Ripperda}},\ and\
  \bibinfo {author} {\bibfnamefont {L.}~\bibnamefont {K\"uchler}},\ }\href
  {https://doi.org/10.1103/PhysRevD.107.124026} {\bibfield  {journal} {\bibinfo
   {journal} {Phys. Rev. D}\ }\textbf {\bibinfo {volume} {107}},\ \bibinfo
  {pages} {124026} (\bibinfo {year} {2023})},\ \Eprint
  {https://arxiv.org/abs/2303.02111} {arXiv:2303.02111 [gr-qc]} \BibitemShut
  {NoStop}%
\bibitem [{\citenamefont {{C{\'a}rdenas-Avenda{\~n}o}}\ and\ \citenamefont
  {{Held}}(2023)}]{Cardenas-Avendano:2023obg}%
  \BibitemOpen
  \bibfield  {author} {\bibinfo {author} {\bibfnamefont {A.}~\bibnamefont
  {{C{\'a}rdenas-Avenda{\~n}o}}}\ and\ \bibinfo {author} {\bibfnamefont
  {A.}~\bibnamefont {{Held}}},\ }\href
  {https://doi.org/10.48550/arXiv.2312.06590} {\bibfield  {journal} {\bibinfo
  {journal} {arXiv e-prints}\ ,\ \bibinfo {eid} {arXiv:2312.06590}} (\bibinfo
  {year} {2023})},\ \Eprint {https://arxiv.org/abs/2312.06590}
  {arXiv:2312.06590 [gr-qc]} \BibitemShut {NoStop}%
\bibitem [{\citenamefont {{C{\'a}rdenas-Avenda{\~n}o}}\ and\ \citenamefont
  {{Lupsasca}}(2023)}]{Cardenas-Avendano:2023dzo}%
  \BibitemOpen
  \bibfield  {author} {\bibinfo {author} {\bibfnamefont {A.}~\bibnamefont
  {{C{\'a}rdenas-Avenda{\~n}o}}}\ and\ \bibinfo {author} {\bibfnamefont
  {A.}~\bibnamefont {{Lupsasca}}},\ }\href
  {https://doi.org/10.1103/PhysRevD.108.064043} {\bibfield  {journal} {\bibinfo
   {journal} {\prd}\ }\textbf {\bibinfo {volume} {108}},\ \bibinfo {eid}
  {064043} (\bibinfo {year} {2023})},\ \Eprint
  {https://arxiv.org/abs/2305.12956} {arXiv:2305.12956 [gr-qc]} \BibitemShut
  {NoStop}%
\bibitem [{\citenamefont {{Lupsasca}}\ \emph {et~al.}(2024)\citenamefont
  {{Lupsasca}}, \citenamefont {{Mayerson}}, \citenamefont {{Ripperda}},\ and\
  \citenamefont {{Staelens}}}]{Lupsasca:2024wkp}%
  \BibitemOpen
  \bibfield  {author} {\bibinfo {author} {\bibfnamefont {A.}~\bibnamefont
  {{Lupsasca}}}, \bibinfo {author} {\bibfnamefont {D.~R.}\ \bibnamefont
  {{Mayerson}}}, \bibinfo {author} {\bibfnamefont {B.}~\bibnamefont
  {{Ripperda}}},\ and\ \bibinfo {author} {\bibfnamefont {S.}~\bibnamefont
  {{Staelens}}},\ }\href {https://doi.org/10.48550/arXiv.2402.01290} {\bibfield
   {journal} {\bibinfo  {journal} {arXiv e-prints}\ ,\ \bibinfo {eid}
  {arXiv:2402.01290}} (\bibinfo {year} {2024})},\ \Eprint
  {https://arxiv.org/abs/2402.01290} {arXiv:2402.01290 [gr-qc]} \BibitemShut
  {NoStop}%
\bibitem [{\citenamefont {{Gammie}}(1999)}]{Gammie1999}%
  \BibitemOpen
  \bibfield  {author} {\bibinfo {author} {\bibfnamefont {C.~F.}\ \bibnamefont
  {{Gammie}}},\ }\href {https://doi.org/10.1086/312207} {\bibfield  {journal}
  {\bibinfo  {journal} {\apjl}\ }\textbf {\bibinfo {volume} {522}},\ \bibinfo
  {pages} {L57} (\bibinfo {year} {1999})},\ \Eprint
  {https://arxiv.org/abs/astro-ph/9906223} {arXiv:astro-ph/9906223 [astro-ph]}
  \BibitemShut {NoStop}%
\bibitem [{\citenamefont {{Agol}}\ and\ \citenamefont
  {{Krolik}}(2000)}]{Agol2000}%
  \BibitemOpen
  \bibfield  {author} {\bibinfo {author} {\bibfnamefont {E.}~\bibnamefont
  {{Agol}}}\ and\ \bibinfo {author} {\bibfnamefont {J.~H.}\ \bibnamefont
  {{Krolik}}},\ }\href {https://doi.org/10.1086/308177} {\bibfield  {journal}
  {\bibinfo  {journal} {\apj}\ }\textbf {\bibinfo {volume} {528}},\ \bibinfo
  {pages} {161} (\bibinfo {year} {2000})},\ \Eprint
  {https://arxiv.org/abs/astro-ph/9908049} {arXiv:astro-ph/9908049 [astro-ph]}
  \BibitemShut {NoStop}%
\bibitem [{\citenamefont {{Hadar}}\ \emph {et~al.}(2021)\citenamefont
  {{Hadar}}, \citenamefont {{Johnson}}, \citenamefont {{Lupsasca}},\ and\
  \citenamefont {{Wong}}}]{Hadar2021}%
  \BibitemOpen
  \bibfield  {author} {\bibinfo {author} {\bibfnamefont {S.}~\bibnamefont
  {{Hadar}}}, \bibinfo {author} {\bibfnamefont {M.~D.}\ \bibnamefont
  {{Johnson}}}, \bibinfo {author} {\bibfnamefont {A.}~\bibnamefont
  {{Lupsasca}}},\ and\ \bibinfo {author} {\bibfnamefont {G.~N.}\ \bibnamefont
  {{Wong}}},\ }\href {https://doi.org/10.1103/PhysRevD.103.104038} {\bibfield
  {journal} {\bibinfo  {journal} {\prd}\ }\textbf {\bibinfo {volume} {103}},\
  \bibinfo {eid} {104038} (\bibinfo {year} {2021})},\ \Eprint
  {https://arxiv.org/abs/2010.03683} {arXiv:2010.03683 [gr-qc]} \BibitemShut
  {NoStop}%
\bibitem [{\citenamefont {{Braun}}\ \emph {et~al.}(2019)\citenamefont
  {{Braun}}, \citenamefont {{Bonaldi}}, \citenamefont {{Bourke}}, \citenamefont
  {{Keane}},\ and\ \citenamefont {{Wagg}}}]{SKA}%
  \BibitemOpen
  \bibfield  {author} {\bibinfo {author} {\bibfnamefont {R.}~\bibnamefont
  {{Braun}}}, \bibinfo {author} {\bibfnamefont {A.}~\bibnamefont {{Bonaldi}}},
  \bibinfo {author} {\bibfnamefont {T.}~\bibnamefont {{Bourke}}}, \bibinfo
  {author} {\bibfnamefont {E.}~\bibnamefont {{Keane}}},\ and\ \bibinfo {author}
  {\bibfnamefont {J.}~\bibnamefont {{Wagg}}},\ }\href
  {https://doi.org/10.48550/arXiv.1912.12699} {\bibfield  {journal} {\bibinfo
  {journal} {arXiv e-prints}\ ,\ \bibinfo {eid} {arXiv:1912.12699}} (\bibinfo
  {year} {2019})},\ \Eprint {https://arxiv.org/abs/1912.12699}
  {arXiv:1912.12699 [astro-ph.IM]} \BibitemShut {NoStop}%
\bibitem [{\citenamefont {{Angel}}\ \emph {et~al.}(2022)\citenamefont
  {{Angel}}, \citenamefont {{Bender}}, \citenamefont {{Berkson}} \emph
  {et~al.}}]{LFAST}%
  \BibitemOpen
  \bibfield  {author} {\bibinfo {author} {\bibfnamefont {R.}~\bibnamefont
  {{Angel}}}, \bibinfo {author} {\bibfnamefont {C.}~\bibnamefont {{Bender}}},
  \bibinfo {author} {\bibfnamefont {J.}~\bibnamefont {{Berkson}}}, \emph
  {et~al.},\ }in\ \href {https://doi.org/10.1117/12.2629655} {\emph {\bibinfo
  {booktitle} {Ground-based and Airborne Telescopes IX}}},\ \bibinfo {series}
  {Society of Photo-Optical Instrumentation Engineers (SPIE) Conference
  Series}, Vol.\ \bibinfo {volume} {12182},\ \bibinfo {editor} {edited by\
  \bibinfo {editor} {\bibfnamefont {H.~K.}\ \bibnamefont {{Marshall}}},
  \bibinfo {editor} {\bibfnamefont {J.}~\bibnamefont {{Spyromilio}}},\ and\
  \bibinfo {editor} {\bibfnamefont {T.}~\bibnamefont {{Usuda}}}}\ (\bibinfo
  {year} {2022})\ p.\ \bibinfo {pages} {121821U}\BibitemShut {NoStop}%
\bibitem [{\citenamefont {{Chael}}\ \emph {et~al.}(2018)\citenamefont
  {{Chael}}, \citenamefont {{Johnson}}, \citenamefont {{Bouman}}, \citenamefont
  {{Blackburn}}, \citenamefont {{Akiyama}},\ and\ \citenamefont
  {{Narayan}}}]{Chael2018}%
  \BibitemOpen
  \bibfield  {author} {\bibinfo {author} {\bibfnamefont {A.~A.}\ \bibnamefont
  {{Chael}}}, \bibinfo {author} {\bibfnamefont {M.~D.}\ \bibnamefont
  {{Johnson}}}, \bibinfo {author} {\bibfnamefont {K.~L.}\ \bibnamefont
  {{Bouman}}}, \bibinfo {author} {\bibfnamefont {L.~L.}\ \bibnamefont
  {{Blackburn}}}, \bibinfo {author} {\bibfnamefont {K.}~\bibnamefont
  {{Akiyama}}},\ and\ \bibinfo {author} {\bibfnamefont {R.}~\bibnamefont
  {{Narayan}}},\ }\href {https://doi.org/10.3847/1538-4357/aab6a8} {\bibfield
  {journal} {\bibinfo  {journal} {\apj}\ }\textbf {\bibinfo {volume} {857}},\
  \bibinfo {eid} {23} (\bibinfo {year} {2018})},\ \Eprint
  {https://arxiv.org/abs/1803.07088} {arXiv:1803.07088 [astro-ph.IM]}
  \BibitemShut {NoStop}%
\bibitem [{\citenamefont {Bely}(2003)}]{bely2003}%
  \BibitemOpen
  \bibfield  {author} {\bibinfo {author} {\bibfnamefont {P.~Y.}\ \bibnamefont
  {Bely}},\ }\href@noop {} {\emph {\bibinfo {title} {The Design and
  Construction of Large Optical Telescopes}}},\ \bibinfo {edition} {1st}\ ed.,\
  Lecture Notes in Physics\ (\bibinfo  {publisher} {Springer},\ \bibinfo {year}
  {2003})\ pp.\ \bibinfo {pages} {XXIV, 508},\ \bibinfo {note} {hardcover ISBN:
  978-0-387-95512-4}\BibitemShut {NoStop}%
\bibitem [{\citenamefont {{Milster}}\ \emph {et~al.}(2020)\citenamefont
  {{Milster}}, \citenamefont {{Sik Kim}}, \citenamefont {{Wang}},\ and\
  \citenamefont {{Purvin}}}]{Milster2020}%
  \BibitemOpen
  \bibfield  {author} {\bibinfo {author} {\bibfnamefont {T.~D.}\ \bibnamefont
  {{Milster}}}, \bibinfo {author} {\bibfnamefont {Y.}~\bibnamefont {{Sik
  Kim}}}, \bibinfo {author} {\bibfnamefont {Z.}~\bibnamefont {{Wang}}},\ and\
  \bibinfo {author} {\bibfnamefont {K.}~\bibnamefont {{Purvin}}},\ }\href
  {https://doi.org/10.1364/AO.394124} {\bibfield  {journal} {\bibinfo
  {journal} {\ao}\ }\textbf {\bibinfo {volume} {59}},\ \bibinfo {pages} {7900}
  (\bibinfo {year} {2020})}\BibitemShut {NoStop}%
\bibitem [{\citenamefont {Kim}\ \emph {et~al.}(2020)\citenamefont {Kim},
  \citenamefont {Wang},\ and\ \citenamefont {Milster}}]{Kim:20}%
  \BibitemOpen
  \bibfield  {author} {\bibinfo {author} {\bibfnamefont {Y.}~\bibnamefont
  {Kim}}, \bibinfo {author} {\bibfnamefont {Z.}~\bibnamefont {Wang}},\ and\
  \bibinfo {author} {\bibfnamefont {T.}~\bibnamefont {Milster}},\ }in\ \href
  {https://doi.org/10.1364/FIO.2020.JTu7A.1} {\emph {\bibinfo {booktitle}
  {Frontiers in Optics / Laser Science}}}\ (\bibinfo  {publisher} {Optica
  Publishing Group},\ \bibinfo {year} {2020})\ p.\ \bibinfo {pages}
  {JTu7A.1}\BibitemShut {NoStop}%
\bibitem [{\citenamefont {Chen}\ \emph {et~al.}()\citenamefont {Chen},
  \citenamefont {Dunsky}, \citenamefont {Huang},\ and\ \citenamefont {{Van
  Tilburg}}}]{Chen2024}%
  \BibitemOpen
  \bibfield  {author} {\bibinfo {author} {\bibfnamefont {I.-K.}\ \bibnamefont
  {Chen}}, \bibinfo {author} {\bibfnamefont {D.}~\bibnamefont {Dunsky}},
  \bibinfo {author} {\bibfnamefont {J.}~\bibnamefont {Huang}},\ and\ \bibinfo
  {author} {\bibfnamefont {K.}~\bibnamefont {{Van Tilburg}}},\ }\href@noop {}
  {}\bibinfo {note} {In preparation}\BibitemShut {NoStop}%
\bibitem [{\citenamefont {Galassi}\ \emph {et~al.}(2019)\citenamefont
  {Galassi}, \citenamefont {Davies}, \citenamefont {Theiler}, \citenamefont
  {Gough}, \citenamefont {Jungman}, \citenamefont {Alken}, \citenamefont
  {Booth}, \citenamefont {Rossi},\ and\ \citenamefont {Ulerich}}]{GSL}%
  \BibitemOpen
  \bibfield  {author} {\bibinfo {author} {\bibfnamefont {M.~C.}\ \bibnamefont
  {Galassi}}, \bibinfo {author} {\bibfnamefont {J.}~\bibnamefont {Davies}},
  \bibinfo {author} {\bibfnamefont {J.}~\bibnamefont {Theiler}}, \bibinfo
  {author} {\bibfnamefont {B.}~\bibnamefont {Gough}}, \bibinfo {author}
  {\bibfnamefont {G.}~\bibnamefont {Jungman}}, \bibinfo {author} {\bibfnamefont
  {P.}~\bibnamefont {Alken}}, \bibinfo {author} {\bibfnamefont
  {M.}~\bibnamefont {Booth}}, \bibinfo {author} {\bibfnamefont
  {F.}~\bibnamefont {Rossi}},\ and\ \bibinfo {author} {\bibfnamefont
  {R.}~\bibnamefont {Ulerich}},\ }\href@noop {} {\emph {\bibinfo {title} {{GNU
  Scientific Library}}}}\ (\bibinfo  {publisher} {Network Theory, Ltd.},\
  \bibinfo {year} {2019})\BibitemShut {NoStop}%
\bibitem [{\citenamefont {Frigo}\ and\ \citenamefont {Johnson}(2005)}]{FFTW}%
  \BibitemOpen
  \bibfield  {author} {\bibinfo {author} {\bibfnamefont {M.}~\bibnamefont
  {Frigo}}\ and\ \bibinfo {author} {\bibfnamefont {S.}~\bibnamefont
  {Johnson}},\ }\href {https://doi.org/10.1109/JPROC.2004.840301} {\bibfield
  {journal} {\bibinfo  {journal} {IEEE Proc.}\ }\textbf {\bibinfo {volume}
  {93}},\ \bibinfo {pages} {216} (\bibinfo {year} {2005})}\BibitemShut
  {NoStop}%
\bibitem [{\citenamefont {{Hahn}}(2005)}]{CUBA}%
  \BibitemOpen
  \bibfield  {author} {\bibinfo {author} {\bibfnamefont {T.}~\bibnamefont
  {{Hahn}}},\ }\href {https://doi.org/10.1016/j.cpc.2005.01.010} {\bibfield
  {journal} {\bibinfo  {journal} {Computer Physics Communications}\ }\textbf
  {\bibinfo {volume} {168}},\ \bibinfo {pages} {78} (\bibinfo {year} {2005})},\
  \Eprint {https://arxiv.org/abs/hep-ph/0404043} {arXiv:hep-ph/0404043
  [hep-ph]} \BibitemShut {NoStop}%
\bibitem [{\citenamefont {{Ochsenbein}}\ \emph {et~al.}(2000)\citenamefont
  {{Ochsenbein}}, \citenamefont {{Bauer}},\ and\ \citenamefont
  {{Marcout}}}]{vizier}%
  \BibitemOpen
  \bibfield  {author} {\bibinfo {author} {\bibfnamefont {F.}~\bibnamefont
  {{Ochsenbein}}}, \bibinfo {author} {\bibfnamefont {P.}~\bibnamefont
  {{Bauer}}},\ and\ \bibinfo {author} {\bibfnamefont {J.}~\bibnamefont
  {{Marcout}}},\ }\href {https://doi.org/10.26093/cds/vizier} {\bibfield
  {journal} {\bibinfo  {journal} {\aaps}\ }\textbf {\bibinfo {volume} {143}},\
  \bibinfo {pages} {23} (\bibinfo {year} {2000})},\ \Eprint
  {https://arxiv.org/abs/astro-ph/0002122} {arXiv:astro-ph/0002122 [astro-ph]}
  \BibitemShut {NoStop}%
\bibitem [{\citenamefont {{Afshordi}}\ and\ \citenamefont
  {{Paczy{\'n}ski}}(2003)}]{Afshordi2003}%
  \BibitemOpen
  \bibfield  {author} {\bibinfo {author} {\bibfnamefont {N.}~\bibnamefont
  {{Afshordi}}}\ and\ \bibinfo {author} {\bibfnamefont {B.}~\bibnamefont
  {{Paczy{\'n}ski}}},\ }\href {https://doi.org/10.1086/375559} {\bibfield
  {journal} {\bibinfo  {journal} {\apj}\ }\textbf {\bibinfo {volume} {592}},\
  \bibinfo {pages} {354} (\bibinfo {year} {2003})},\ \Eprint
  {https://arxiv.org/abs/astro-ph/0202409} {arXiv:astro-ph/0202409 [astro-ph]}
  \BibitemShut {NoStop}%
\bibitem [{\citenamefont {{Shafee}}\ \emph {et~al.}(2008)\citenamefont
  {{Shafee}}, \citenamefont {{Narayan}},\ and\ \citenamefont
  {{McClintock}}}]{Shafee2008}%
  \BibitemOpen
  \bibfield  {author} {\bibinfo {author} {\bibfnamefont {R.}~\bibnamefont
  {{Shafee}}}, \bibinfo {author} {\bibfnamefont {R.}~\bibnamefont
  {{Narayan}}},\ and\ \bibinfo {author} {\bibfnamefont {J.~E.}\ \bibnamefont
  {{McClintock}}},\ }\href {https://doi.org/10.1086/527346} {\bibfield
  {journal} {\bibinfo  {journal} {\apj}\ }\textbf {\bibinfo {volume} {676}},\
  \bibinfo {pages} {549} (\bibinfo {year} {2008})},\ \Eprint
  {https://arxiv.org/abs/0705.2244} {arXiv:0705.2244 [astro-ph]} \BibitemShut
  {NoStop}%
\bibitem [{\citenamefont {{Beckwith}}\ \emph {et~al.}(2008)\citenamefont
  {{Beckwith}}, \citenamefont {{Hawley}},\ and\ \citenamefont
  {{Krolik}}}]{Beckwith2008}%
  \BibitemOpen
  \bibfield  {author} {\bibinfo {author} {\bibfnamefont {K.}~\bibnamefont
  {{Beckwith}}}, \bibinfo {author} {\bibfnamefont {J.~F.}\ \bibnamefont
  {{Hawley}}},\ and\ \bibinfo {author} {\bibfnamefont {J.~H.}\ \bibnamefont
  {{Krolik}}},\ }\href {https://doi.org/10.1111/j.1365-2966.2008.13710.x}
  {\bibfield  {journal} {\bibinfo  {journal} {\mnras}\ }\textbf {\bibinfo
  {volume} {390}},\ \bibinfo {pages} {21} (\bibinfo {year} {2008})},\ \Eprint
  {https://arxiv.org/abs/0801.2974} {arXiv:0801.2974 [astro-ph]} \BibitemShut
  {NoStop}%
\bibitem [{\citenamefont {{Hollywood}}\ and\ \citenamefont
  {{Melia}}(1997)}]{hollywood1997}%
  \BibitemOpen
  \bibfield  {author} {\bibinfo {author} {\bibfnamefont {J.~M.}\ \bibnamefont
  {{Hollywood}}}\ and\ \bibinfo {author} {\bibfnamefont {F.}~\bibnamefont
  {{Melia}}},\ }\href {https://doi.org/10.1086/313036} {\bibfield  {journal}
  {\bibinfo  {journal} {\apjs}\ }\textbf {\bibinfo {volume} {112}},\ \bibinfo
  {pages} {423} (\bibinfo {year} {1997})}\BibitemShut {NoStop}%
\bibitem [{\citenamefont {{Beckwith}}\ and\ \citenamefont
  {{Done}}(2005)}]{beckwith2005}%
  \BibitemOpen
  \bibfield  {author} {\bibinfo {author} {\bibfnamefont {K.}~\bibnamefont
  {{Beckwith}}}\ and\ \bibinfo {author} {\bibfnamefont {C.}~\bibnamefont
  {{Done}}},\ }\href {https://doi.org/10.1111/j.1365-2966.2005.08980.x}
  {\bibfield  {journal} {\bibinfo  {journal} {\mnras}\ }\textbf {\bibinfo
  {volume} {359}},\ \bibinfo {pages} {1217} (\bibinfo {year} {2005})},\ \Eprint
  {https://arxiv.org/abs/astro-ph/0411339} {arXiv:astro-ph/0411339 [astro-ph]}
  \BibitemShut {NoStop}%
\bibitem [{\citenamefont {{Gralla}}\ and\ \citenamefont
  {{Lupsasca}}(2020)}]{gralla2020}%
  \BibitemOpen
  \bibfield  {author} {\bibinfo {author} {\bibfnamefont {S.~E.}\ \bibnamefont
  {{Gralla}}}\ and\ \bibinfo {author} {\bibfnamefont {A.}~\bibnamefont
  {{Lupsasca}}},\ }\href {https://doi.org/10.1103/PhysRevD.101.044031}
  {\bibfield  {journal} {\bibinfo  {journal} {\prd}\ }\textbf {\bibinfo
  {volume} {101}},\ \bibinfo {eid} {044031} (\bibinfo {year} {2020})},\ \Eprint
  {https://arxiv.org/abs/1910.12873} {arXiv:1910.12873 [gr-qc]} \BibitemShut
  {NoStop}%
\bibitem [{\citenamefont {{Mendes}}\ and\ \citenamefont
  {{Pavlis}}(2004)}]{mendes2004high}%
  \BibitemOpen
  \bibfield  {author} {\bibinfo {author} {\bibfnamefont {V.~B.}\ \bibnamefont
  {{Mendes}}}\ and\ \bibinfo {author} {\bibfnamefont {E.~C.}\ \bibnamefont
  {{Pavlis}}},\ }\href {https://doi.org/10.1029/2004GL020308} {\bibfield
  {journal} {\bibinfo  {journal} {\grl}\ }\textbf {\bibinfo {volume} {31}},\
  \bibinfo {eid} {L14602} (\bibinfo {year} {2004})}\BibitemShut {NoStop}%
\bibitem [{\citenamefont {{Hulley}}\ and\ \citenamefont
  {{Pavlis}}(2007)}]{hulley2007ray}%
  \BibitemOpen
  \bibfield  {author} {\bibinfo {author} {\bibfnamefont {G.~C.}\ \bibnamefont
  {{Hulley}}}\ and\ \bibinfo {author} {\bibfnamefont {E.~C.}\ \bibnamefont
  {{Pavlis}}},\ }\href {https://doi.org/10.1029/2006JB004834} {\bibfield
  {journal} {\bibinfo  {journal} {Journal of Geophysical Research (Solid
  Earth)}\ }\textbf {\bibinfo {volume} {112}},\ \bibinfo {eid} {B06417}
  (\bibinfo {year} {2007})}\BibitemShut {NoStop}%
\bibitem [{\citenamefont {Stuhl}(2021)}]{stuhl2021atmospheric}%
  \BibitemOpen
  \bibfield  {author} {\bibinfo {author} {\bibfnamefont {B.~K.}\ \bibnamefont
  {Stuhl}},\ }\href@noop {} {\bibfield  {journal} {\bibinfo  {journal} {Optics
  Express}\ }\textbf {\bibinfo {volume} {29}},\ \bibinfo {pages} {13706}
  (\bibinfo {year} {2021})}\BibitemShut {NoStop}%
\bibitem [{\citenamefont {Hase}\ \emph {et~al.}(2012)\citenamefont {Hase},
  \citenamefont {Behrend}, \citenamefont {Ma}, \citenamefont {Petrachenko},
  \citenamefont {Schuh},\ and\ \citenamefont {Whitney}}]{hase2012emerging}%
  \BibitemOpen
  \bibfield  {author} {\bibinfo {author} {\bibfnamefont {H.}~\bibnamefont
  {Hase}}, \bibinfo {author} {\bibfnamefont {D.}~\bibnamefont {Behrend}},
  \bibinfo {author} {\bibfnamefont {C.}~\bibnamefont {Ma}}, \bibinfo {author}
  {\bibfnamefont {B.}~\bibnamefont {Petrachenko}}, \bibinfo {author}
  {\bibfnamefont {H.}~\bibnamefont {Schuh}},\ and\ \bibinfo {author}
  {\bibfnamefont {A.}~\bibnamefont {Whitney}},\ }in\ \href@noop {} {\emph
  {\bibinfo {booktitle} {IVS 2012 General Meeting Proceedings,
  NASA/CP-2012-217504}}}\ (\bibinfo {year} {2012})\ pp.\ \bibinfo {pages}
  {8--12}\BibitemShut {NoStop}%
\bibitem [{\citenamefont {Behrend}\ \emph {et~al.}(2019)\citenamefont
  {Behrend}, \citenamefont {Petrachenko}, \citenamefont {Ruszczyk},\ and\
  \citenamefont {El{\'o}segui}}]{behrend2019roll}%
  \BibitemOpen
  \bibfield  {author} {\bibinfo {author} {\bibfnamefont {D.}~\bibnamefont
  {Behrend}}, \bibinfo {author} {\bibfnamefont {B.}~\bibnamefont
  {Petrachenko}}, \bibinfo {author} {\bibfnamefont {C.}~\bibnamefont
  {Ruszczyk}},\ and\ \bibinfo {author} {\bibfnamefont {P.}~\bibnamefont
  {El{\'o}segui}},\ }in\ \href@noop {} {\emph {\bibinfo {booktitle} {poster,
  24rd European VLBI group for geodesy and astrometry working meeting}}}\
  (\bibinfo {year} {2019})\BibitemShut {NoStop}%
\bibitem [{\citenamefont {Niell}\ \emph {et~al.}(2018)\citenamefont {Niell},
  \citenamefont {Barrett}, \citenamefont {Burns}, \citenamefont {Cappallo},
  \citenamefont {Corey}, \citenamefont {Derome}, \citenamefont {Eckert},
  \citenamefont {Elosegui}, \citenamefont {McWhirter}, \citenamefont {Poirier}
  \emph {et~al.}}]{niell2018demonstration}%
  \BibitemOpen
  \bibfield  {author} {\bibinfo {author} {\bibfnamefont {A.}~\bibnamefont
  {Niell}}, \bibinfo {author} {\bibfnamefont {J.}~\bibnamefont {Barrett}},
  \bibinfo {author} {\bibfnamefont {A.}~\bibnamefont {Burns}}, \bibinfo
  {author} {\bibfnamefont {R.}~\bibnamefont {Cappallo}}, \bibinfo {author}
  {\bibfnamefont {B.}~\bibnamefont {Corey}}, \bibinfo {author} {\bibfnamefont
  {M.}~\bibnamefont {Derome}}, \bibinfo {author} {\bibfnamefont
  {C.}~\bibnamefont {Eckert}}, \bibinfo {author} {\bibfnamefont
  {P.}~\bibnamefont {Elosegui}}, \bibinfo {author} {\bibfnamefont
  {R.}~\bibnamefont {McWhirter}}, \bibinfo {author} {\bibfnamefont
  {M.}~\bibnamefont {Poirier}}, \emph {et~al.},\ }\href@noop {} {\bibfield
  {journal} {\bibinfo  {journal} {Radio Science}\ }\textbf {\bibinfo {volume}
  {53}},\ \bibinfo {pages} {1269} (\bibinfo {year} {2018})}\BibitemShut
  {NoStop}%
\bibitem [{\citenamefont {Rioja}\ and\ \citenamefont
  {Dodson}(2020)}]{rioja2020precise}%
  \BibitemOpen
  \bibfield  {author} {\bibinfo {author} {\bibfnamefont {M.~J.}\ \bibnamefont
  {Rioja}}\ and\ \bibinfo {author} {\bibfnamefont {R.}~\bibnamefont {Dodson}},\
  }\href@noop {} {\bibfield  {journal} {\bibinfo  {journal} {The Astronomy and
  Astrophysics Review}\ }\textbf {\bibinfo {volume} {28}},\ \bibinfo {pages}
  {1} (\bibinfo {year} {2020})}\BibitemShut {NoStop}%
\bibitem [{\citenamefont {Inc.}(2023{\natexlab{a}})}]{mhm}%
  \BibitemOpen
  \bibfield  {author} {\bibinfo {author} {\bibfnamefont {M.~T.}\ \bibnamefont
  {Inc.}},\ }\href@noop {} {\bibinfo {title} {Mhm-2020 active hydrogen
  maser}},\ \bibinfo {howpublished}
  {\url{https://ww1.microchip.com/downloads/aemDocuments/documents/FTD/ProductDocuments/Brochures/MHM-2020-Active-Hydrogen-Maser-00003090.pdf}}
  (\bibinfo {year} {2023}{\natexlab{a}}),\ \bibinfo {note} {[Online; accessed
  11-Jan-2024]}\BibitemShut {NoStop}%
\bibitem [{\citenamefont {Quartzlock}(2023)}]{CH1-76A}%
  \BibitemOpen
  \bibfield  {author} {\bibinfo {author} {\bibnamefont {Quartzlock}},\
  }\href@noop {} {\bibinfo {title} {Ch1-76a passive hydrogen maser}},\ \bibinfo
  {howpublished}
  {\url{https://www.quartzlock.com/userfiles/downloads/datasheets/Passive_Hydrogen_Maser_CH1-76A.pdf}}
  (\bibinfo {year} {2023}),\ \bibinfo {note} {[Online; accessed
  11-Jan-2024]}\BibitemShut {NoStop}%
\bibitem [{\citenamefont {Akiyama}\ \emph {et~al.}(2019)\citenamefont {Akiyama}
  \emph {et~al.}}]{EHT_instrumentation}%
  \BibitemOpen
  \bibfield  {author} {\bibinfo {author} {\bibfnamefont {K.}~\bibnamefont
  {Akiyama}} \emph {et~al.} (\bibinfo {collaboration} {Event Horizon
  Telescope}),\ }\href {https://doi.org/10.3847/2041-8213/ab0c96} {\bibfield
  {journal} {\bibinfo  {journal} {Astrophys. J. Lett.}\ }\textbf {\bibinfo
  {volume} {875}},\ \bibinfo {pages} {L2} (\bibinfo {year} {2019})},\ \Eprint
  {https://arxiv.org/abs/1906.11239} {arXiv:1906.11239 [astro-ph.IM]}
  \BibitemShut {NoStop}%
\bibitem [{\citenamefont {Inc.}(2023{\natexlab{b}})}]{8040C}%
  \BibitemOpen
  \bibfield  {author} {\bibinfo {author} {\bibfnamefont {M.~T.}\ \bibnamefont
  {Inc.}},\ }\href@noop {} {\bibinfo {title} {8040c rubidium frequency
  standard}},\ \bibinfo {howpublished}
  {\url{https://ww1.microchip.com/downloads/aemDocuments/documents/FTD/ProductDocuments/Brochures/00003047.pdf}}
  (\bibinfo {year} {2023}{\natexlab{b}}),\ \bibinfo {note} {[Online; accessed
  11-Jan-2024]}\BibitemShut {NoStop}%
\bibitem [{\citenamefont {Corporation}(2023)}]{3352A}%
  \BibitemOpen
  \bibfield  {author} {\bibinfo {author} {\bibfnamefont {A.}~\bibnamefont
  {Corporation}},\ }\href@noop {} {\bibinfo {title} {3352a series rubidium
  frequency standard}},\ \bibinfo {howpublished}
  {\url{https://www.astronics.com/docs/default-source/Documents-ats/current-datasheets/rubidiums/3352a_0.pdf?sfvrsn=2
  }} (\bibinfo {year} {2023}),\ \bibinfo {note} {[Online; accessed
  11-Jan-2024]}\BibitemShut {NoStop}%
\bibitem [{\citenamefont {Inc.}(2023{\natexlab{c}})}]{5071A}%
  \BibitemOpen
  \bibfield  {author} {\bibinfo {author} {\bibfnamefont {M.~T.}\ \bibnamefont
  {Inc.}},\ }\href@noop {} {\bibinfo {title} {5071a primary frequency
  standard}},\ \bibinfo {howpublished}
  {\url{https://ww1.microchip.com/downloads/aemDocuments/documents/FTD/ProductDocuments/Brochures/5071A-Sell-Sheet-00002980.pdf}}
  (\bibinfo {year} {2023}{\natexlab{c}}),\ \bibinfo {note} {[Online; accessed
  11-Jan-2024]}\BibitemShut {NoStop}%
\bibitem [{\citenamefont {{Roslund}}\ \emph {et~al.}(2023)\citenamefont
  {{Roslund}}, \citenamefont {{Cing{\"o}z}}, \citenamefont {{Lunden}},
  \citenamefont {{Partridge}}, \citenamefont {{Kowligy}}, \citenamefont
  {{Roller}}, \citenamefont {{Sheredy}}, \citenamefont {{Skulason}},
  \citenamefont {{Song}}, \citenamefont {{Abo-Shaeer}},\ and\ \citenamefont
  {{Boyd}}}]{Roslund2023}%
  \BibitemOpen
  \bibfield  {author} {\bibinfo {author} {\bibfnamefont {J.~D.}\ \bibnamefont
  {{Roslund}}}, \bibinfo {author} {\bibfnamefont {A.}~\bibnamefont
  {{Cing{\"o}z}}}, \bibinfo {author} {\bibfnamefont {W.~D.}\ \bibnamefont
  {{Lunden}}}, \bibinfo {author} {\bibfnamefont {G.~B.}\ \bibnamefont
  {{Partridge}}}, \bibinfo {author} {\bibfnamefont {A.~S.}\ \bibnamefont
  {{Kowligy}}}, \bibinfo {author} {\bibfnamefont {F.}~\bibnamefont {{Roller}}},
  \bibinfo {author} {\bibfnamefont {D.~B.}\ \bibnamefont {{Sheredy}}}, \bibinfo
  {author} {\bibfnamefont {G.~E.}\ \bibnamefont {{Skulason}}}, \bibinfo
  {author} {\bibfnamefont {J.~P.}\ \bibnamefont {{Song}}}, \bibinfo {author}
  {\bibfnamefont {J.~R.}\ \bibnamefont {{Abo-Shaeer}}},\ and\ \bibinfo {author}
  {\bibfnamefont {M.~M.}\ \bibnamefont {{Boyd}}},\ }\href
  {https://doi.org/10.48550/arXiv.2308.12457} {\bibfield  {journal} {\bibinfo
  {journal} {arXiv e-prints}\ ,\ \bibinfo {eid} {arXiv:2308.12457}} (\bibinfo
  {year} {2023})},\ \Eprint {https://arxiv.org/abs/2308.12457}
  {arXiv:2308.12457 [physics.atom-ph]} \BibitemShut {NoStop}%
\bibitem [{\citenamefont {{Vanarsdale}}\ \emph {et~al.}(2023)\citenamefont
  {{Vanarsdale}}, \citenamefont {{Lombardi}}, \citenamefont {{Nelson}},
  \citenamefont {{Sherman}}, \citenamefont {{Yost}},\ and\ \citenamefont
  {{Brewer}}}]{fiber_transmit_time}%
  \BibitemOpen
  \bibfield  {author} {\bibinfo {author} {\bibfnamefont {J.}~\bibnamefont
  {{Vanarsdale}}}, \bibinfo {author} {\bibfnamefont {M.}~\bibnamefont
  {{Lombardi}}}, \bibinfo {author} {\bibfnamefont {G.}~\bibnamefont
  {{Nelson}}}, \bibinfo {author} {\bibfnamefont {J.}~\bibnamefont {{Sherman}}},
  \bibinfo {author} {\bibfnamefont {D.}~\bibnamefont {{Yost}}},\ and\ \bibinfo
  {author} {\bibfnamefont {S.}~\bibnamefont {{Brewer}}},\ }in\ \href@noop {}
  {\emph {\bibinfo {booktitle} {APS Division of Atomic, Molecular and Optical
  Physics Meeting Abstracts}}},\ \bibinfo {series} {APS Meeting Abstracts},
  Vol.\ \bibinfo {volume} {2023}\ (\bibinfo {year} {2023})\ p.\ \bibinfo
  {pages} {F01.135}\BibitemShut {NoStop}%
\bibitem [{\citenamefont {Koke}\ \emph {et~al.}(2021)\citenamefont {Koke},
  \citenamefont {Benkler}, \citenamefont {Kuhl},\ and\ \citenamefont
  {Grosche}}]{Koke_2021}%
  \BibitemOpen
  \bibfield  {author} {\bibinfo {author} {\bibfnamefont {S.}~\bibnamefont
  {Koke}}, \bibinfo {author} {\bibfnamefont {E.}~\bibnamefont {Benkler}},
  \bibinfo {author} {\bibfnamefont {A.}~\bibnamefont {Kuhl}},\ and\ \bibinfo
  {author} {\bibfnamefont {G.}~\bibnamefont {Grosche}},\ }\href
  {https://doi.org/10.1088/1367-2630/ac21a0} {\bibfield  {journal} {\bibinfo
  {journal} {New Journal of Physics}\ }\textbf {\bibinfo {volume} {23}},\
  \bibinfo {pages} {093024} (\bibinfo {year} {2021})}\BibitemShut {NoStop}%
\bibitem [{Note3()}]{Note3}%
  \BibitemOpen
  \bibinfo {note} {They also use a polarizer, which cuts the expected photon
  flux in half and which may account for some flux loss too. A polarizer does
  not change the SNR for intensity interferometry, so we ignore it in this
  estimate}\BibitemShut {NoStop}%
\bibitem [{\citenamefont {Marsili}\ \emph {et~al.}(2013)\citenamefont
  {Marsili}, \citenamefont {Verma}, \citenamefont {Stern}, \citenamefont
  {Harrington}, \citenamefont {Lita}, \citenamefont {Gerrits}, \citenamefont
  {Vayshenker}, \citenamefont {Baek}, \citenamefont {Shaw}, \citenamefont
  {Mirin},\ and\ \citenamefont {Nam}}]{high_eta}%
  \BibitemOpen
  \bibfield  {author} {\bibinfo {author} {\bibfnamefont {F.}~\bibnamefont
  {Marsili}}, \bibinfo {author} {\bibfnamefont {V.}~\bibnamefont {Verma}},
  \bibinfo {author} {\bibfnamefont {J.}~\bibnamefont {Stern}}, \bibinfo
  {author} {\bibfnamefont {S.}~\bibnamefont {Harrington}}, \bibinfo {author}
  {\bibfnamefont {A.}~\bibnamefont {Lita}}, \bibinfo {author} {\bibfnamefont
  {T.}~\bibnamefont {Gerrits}}, \bibinfo {author} {\bibfnamefont
  {I.}~\bibnamefont {Vayshenker}}, \bibinfo {author} {\bibfnamefont
  {B.}~\bibnamefont {Baek}}, \bibinfo {author} {\bibfnamefont {M.}~\bibnamefont
  {Shaw}}, \bibinfo {author} {\bibfnamefont {R.}~\bibnamefont {Mirin}},\ and\
  \bibinfo {author} {\bibfnamefont {S.}~\bibnamefont {Nam}},\ }\href
  {https://tsapps.nist.gov/publication/get_pdf.cfm?pub_id=911769} {\bibfield
  {journal} {\bibinfo  {journal} {Nature}\ } (\bibinfo {year}
  {2013})}\BibitemShut {NoStop}%
\bibitem [{\citenamefont {{National Institute of Standards and
  Technology}}(2024)}]{NISTQuantumDetectors}%
  \BibitemOpen
  \bibfield  {author} {\bibinfo {author} {\bibnamefont {{National Institute of
  Standards and Technology}}},\ }\href@noop {} {\bibinfo {title} {{Single
  Photon Detectors}}},\ \bibinfo {howpublished}
  {\url{https://www.nist.gov/pml/quantum-networks-nist/technologies-quantum-networks/single-photon-detectors}}
  (\bibinfo {year} {Accessed: 2024}),\ \bibinfo {note} {accessed:
  2024-03-08}\BibitemShut {NoStop}%
\bibitem [{\citenamefont {{Schneider}}\ \emph {et~al.}(1983)\citenamefont
  {{Schneider}}, \citenamefont {{Gunn}},\ and\ \citenamefont
  {{Hoessel}}}]{Schneider1983}%
  \BibitemOpen
  \bibfield  {author} {\bibinfo {author} {\bibfnamefont {D.~P.}\ \bibnamefont
  {{Schneider}}}, \bibinfo {author} {\bibfnamefont {J.~E.}\ \bibnamefont
  {{Gunn}}},\ and\ \bibinfo {author} {\bibfnamefont {J.~G.}\ \bibnamefont
  {{Hoessel}}},\ }\href {https://doi.org/10.1086/160602} {\bibfield  {journal}
  {\bibinfo  {journal} {\apj}\ }\textbf {\bibinfo {volume} {264}},\ \bibinfo
  {pages} {337} (\bibinfo {year} {1983})}\BibitemShut {NoStop}%
\end{thebibliography}%

\end{document}